\documentclass[useAMS,usenatbib,usegraphicx]{mn2e}
\usepackage{psfig}   
\usepackage{graphicx}
\usepackage{subfigure}

\title[Dynamical evolution of star forming regions]{Dynamical evolution of star forming regions}

\author[R.~J.~Parker, N.~J.~Wright, S.~P.~Goodwin \& M.~R.~Meyer]{
  Richard J.~Parker$^{1}$\thanks{E-mail: rparker@phys.ethz.ch},  Nicholas J.~Wright$^2$, Simon P.~Goodwin$^{3}$ and Michael R. Meyer$^1$
  \vspace*{0.1cm}\\
   $^1$ Institute for Astronomy, ETH Z{\"u}rich, Wolfgang-Pauli-Strasse 27, 8093, Z{\"u}rich, Switzerland \\
   $^2$ Centre for Astrophysics Research, Science and Technology Research Institute, University of Hertfordshire, Hatfield, AL10 9AB, UK\\
   $^3$ Department of Physics and Astronomy, University of Sheffield, Sheffield, S3 7RH, UK}

\begin{document}

\date{}
                             
\pagerange{\pageref{firstpage}--\pageref{lastpage}} \pubyear{2013}

\maketitle

\label{firstpage}

\begin{abstract}
We model the dynamical evolution of star forming regions with a wide range of initial properties. We follow the evolution of the regions' substructure using the $\mathcal{Q}$--parameter, we search 
for dynamical mass segregation using the $\Lambda_{\rm MSR}$ technique, and we also quantify the evolution of local density around stars as a function of mass using the $\Sigma_{\rm LDR}$ method.

The amount of dynamical mass segregation measured by $\Lambda_{\rm MSR}$ is generally only significant for subvirial and virialised, substructured 
regions -- which usually evolve to form bound clusters. The $\Sigma_{\rm LDR}$ method shows that massive stars attain higher local
densities than the median value in \emph{all} regions, even those that are supervirial and evolve to form (unbound) associations. 

We also introduce the $\mathcal{Q} - \Sigma_{\rm LDR}$ plot, which describes the evolution of spatial structure as a function of mass-weighted local density in a star forming region. Initially dense ($>$1000\,stars\,pc$^{-2}$), bound regions  
always have $\mathcal{Q}~>~1,~\Sigma_{\rm LDR} > 2$ after 5\,Myr, whereas dense  unbound regions always have $\mathcal{Q} < 1, \Sigma_{\rm LDR} > 2$ after 5\,Myr. 
Less dense regions ($<$100\,stars\,pc$^{-2}$) do not usually exhibit
$\Sigma_{\rm LDR} > 2$ values, and if relatively high local density
around massive stars arises purely from dynamics, then the
$\mathcal{Q} - \Sigma_{\rm LDR}$ plot can be used to estimate the initial density of a star forming region.
\end{abstract}


\section{Introduction}

Understanding the earliest phases of the dynamical evolution of stars
is important as their birth environments can impact planetary 
systems (through interactions with discs, or
through encounters with young planetary systems), as well as
stellar properties such as multiplicity.  Star
formation occurs most often in regions significantly denser than the
field in which interactions may be common\footnote{The typical field
  stellar density is around 0.1 stars pc$^{-3}$ \citep{Korchagin03} which is much
  lower than the densities of even loose associations, e.g. Taurus
  with roughly 5 stars pc$^{-3}$, and very much lower than clusters, e.g. the
  Orion Nebula Cluster with around 5000 stars pc$^{-3}$ \citep{King12a}.}
 \citep{Lada03,Gieles11,Kruijssen12b}.  Only a small fraction \citep[$\sim$10\,per
 cent,][]{Lada03} of stars remain in bound (open) clusters
 after 10~Myr, and so the vast majority of young stars disperse
 rapidly into the field.  It is interesting and important to know how
 and why most star forming regions dissolve rapidly, and what
 encounters stars may have had before this dissolution. Understanding this may place in context the exoplanet 
properties of nearby field stars.

Recent results from the \emph{Herschel} space telescope have shown that stars initially form in filamentary structures \citep[e.g.][]{Andre10}, which results in
hierarchical spatial distributions in star forming regions
\citep{Cartwright04,Sanchez09}. However, observations of  some young
($\sim$1\,Myr) clusters show them to be centrally concentrated, with
smooth radial profiles \citep[e.g.\,\,the Orion Nebular Cluster (ONC)
  and IC\,348 --][]{Hillenbrand98,Cartwright04}. Numerical studies
have shown that substructure can be erased  on very short timescales
\citep{Scally02,Goodwin04a,Allison10}, consistent with the hypothesis
that all star-forming regions form with substructure, and a certain
fraction dynamically evolve to attain smooth, centrally concentrated
profiles -- i.e.\,\,bound clusters,  whereas the remainder form
unbound associations that rapidly dissolve \citep{Kruijssen12b,Parker12d}.

If stars do form in substructured distributions, and this substructure
is erased in some star-forming regions, then in principle it may
be possible to compare observations of star clusters and associations at
different ages and  use measures of structure and kinematics to infer
the past, and potentially future, (dynamical) evolution of the system. 

As an example, the competitive accretion scenario of star formation
\citep{Zinnecker82,Bonnell01,Bonnell03,Bonnell08} predicts that the most massive
stars are over-concentrated at the centre of a region (primordial mass
segregation).  We would not expect a region to lose any primordial
mass segregation due to dynamical evolution. Whilst mass segregation
is observed in several clusters  (e.g.\,\, the ONC;
\citealp{Hillenbrand98}, \citealp{Allison09b}; NGC3603 \citep{Pang13};
and Trumpler~14; \citealp{Sana10}), it is not clear whether it is
primordial (i.e.\,\,an outcome of the star formation process).
Recently, \citet{Allison09b,Allison10} showed that mass segregation
can occur dynamically on very short timescales, negating the need for
the most massive stars to form at the centre of the region, as
previously thought  \citep{Bonnell98}. Furthermore, observations of
both high- and low-mass clusters (Berkeley~96 -- \citealp{Delgado13}; 
$\rho$~Oph -- \citealp*{Parker12c}) and associations (Taurus -- \citealp{Parker11b}; 
Cyg~OB2  -- \citealp{Wright13}) indicate that mass segregation
is not always present.

Assuming mass segregation is not always primordial, a combined measure
of the structure of a star forming region, and the amount of mass
segregation that can occur dynamically, could be a useful diagnostic
for tracing the dynamical evolution (if any) of the region. 

In this paper, we examine the dynamical evolution of $N$-body
simulations of star forming regions to ascertain how often, mass
segregation occurs (and quantify the amount)  as a function of the
initial bulk motion (virial state) of stars, and the amount of initial
substructure present in the region. We compare the evolution of
spatial structure as measured by the $\mathcal{Q}$--parameter
\citep{Cartwright04,Cartwright09a}, the occurence of mass
segregation as measured by the $\Lambda_{\rm MSR}$ minimum spanning
tree (MST) technique \citep{Allison09a}, and we follow the evolution of the mass-weighted local
density, $\Sigma - m$ \citep{Maschberger11}.

We will refer to the young substructured star-forming
complexes as \emph{regions}, and only  use the terminology `cluster'
or `association' when describing the regions at later times when they
have distinguishable morphologies.

The paper is organised as follows. We describe our $N$-body
simulations in Section~\ref{method}, in Section~\ref{tools} we
describe the algorithms used to quantify structure and mass
segregation, we present our results in Section~\ref{results}, we
provide a discussion in Section~\ref{discuss} and we conclude in
Section~\ref{conclude}.

\section{Method}
\label{method}

The regions we simulate have either 1500 members, which corresponds to a mass of $\sim 10^3$ M$_\odot$,  or 150 members, 
corresponding to a mass of $\sim 10^2$ M$_\odot$. For each set of initial conditions, 
we run an ensemble of 20 simulations, identical apart from the random number seed used to initialise the positions, masses and velocities of the stars. 

Our model regions are set up as fractals; observations of young unevolved star forming regions indicate a high level of substructure is present \citep[i.e. they do not  
have a radially smooth profile, e.g.][and references therein]{Cartwright04,Sanchez09,Schmeja11}. The fractal distribution provides a way of creating substructure on all scales. 
Note that we are not claiming that young star forming regions are fractal \citep[although they may be, e.g.][]{Elmegreen01}, but the fractal distribution is a relatively simple method of 
setting up substructure, as the level of substructure is described by just one parameter, the fractal dimension, $D$. In three dimensions, $D = 1.6$ indicates a 
highly substructured region, and $D = 3.0$ is a roughly uniform sphere.

We set up the fractals according to the method in \citet{Goodwin04a}. This begins by defining a cube of side $N_{\rm div}$ (we adopt $N_{\rm div} = 2.0$ 
throughout), inside of which the fractal is built. A first-generation parent is placed at the centre of the cube, which then spawns $N_{\rm div}^3$ subcubes, each containing a first generation 
child at its centre. The fractal is then built by determining which of the children themselves become parents, and spawn their own offspring. This is determined by the 
fractal dimension, $D$, where the probability that the child becomes a parent is given by $N_{\rm div}^{(D - 3)}$. For a lower fractal dimension fewer children 
mature and the final distribution contains more substructure. Any children that do not become parents in a given step are removed, along with all of their parents. 
A small amount of noise is then 
added to the positions of the remaining children, preventing the region from having a gridded appearance and the children become parents of the next generation. Each new parent 
then spawns $N_{\rm div}^3$ second-generation children in $N_{\rm div}^3$ sub-subcubes, with each second-generation child having a $N_{\rm div}^{(D - 3)}$ 
probability of becoming a second generation parent. This process is repeated until there are substantially more children than required. The children are pruned to produce a 
sphere from the cube and are then randomly removed (so maintaining the fractal dimension) until the required number of children is left. These children then become stars in the 
model. 

To determine the velocity structure of the cloud, children inherit their parent's velocity plus a random component that decreases with each generation of the fractal.  The children of the first 
generation are given random velocities from a Gaussian of mean zero. Each new generation inherits their parent's velocity plus an extra random component that becomes smaller with each 
generation. This results in a velocity structure in which nearby stars have similar velocities, but distant stars can have very different velocities. The velocity of every star is scaled to obtain the desired virial 
ratio of the region. In one set of simulations, we do not correlate the velocities 
according to position, and simply choose them randomly from a Gaussian of mean zero before scaling to the global virial ratio. 

We vary the initial global virial ratio, $\alpha_{\rm vir} =
T/|\Omega|$, where $T$ and $|\Omega|$ are the total kinetic energy and
total potential energy of the stars, respectively.   Note that a
virial ratio of $\alpha_{\rm vir} = 0.5$ does {\em not} necessarily mean that the
regions are in virial equilibrium.  Because of the spatial and
velocity substructure of the regions they are far from equilibrium and
will undergo a violent relaxation phase to attempt to attain virial
equilibrium and a smooth central profile (if they are bound).
Because we have correlated the velocities of the stars on local
scales, a substructured fractal with $\alpha_{\rm vir} = 0.5$ will violently relax in a similar way to a subvirial fractal. The main difference 
is that a subvirial fractal will collapse more quickly, and form a denser core, than a virial fractal \citep{Allison10}. Similarly, a supervirial fractal will expand on a global scale, but the pockets of substructure will not be supervirial. For this reason, we 
introduce the following terminology: a globally subvirial fractal
($\alpha_{\rm vir} = 0.3$) is `cool' because the stars are moving
slowly with respect to their `equilibrium' velocities, a globally virial fractal ($\alpha_{\rm vir} = 0.5$) is `tepid' because the stars are still able to 
interact in the substructure and it is bound,  and a globally
supervirial fractal ($\alpha_{\rm vir} = 1.5$) is `hot' and unbound.

The regions are set up with fractal dimensions of $D = 1.6$ (very clumpy), $D = 2.0$ and $D = 3.0$ (a roughly uniform sphere), in order to investigate 
the full parameter space. We reiterate that these initial conditions are based on observations of star forming regions, which appear to be filamentary \citep[e.g.][]{Andre10} and form stars with a hierarchical distribution \citep{Elmegreen01}. 
This substructured distribution of stars is also consistent with the outcome of hydrodynamical simulations of star formation \citep[e.g.][]{Schmeja06,Bate12,Girichidis12,Dale12a,Dale13}.

The regions contain 1500 or 150 stars each, have initial radii of 1\,pc with no primordial binaries or gas potential. We draw stellar masses from the recent fit to the field Initial Mass Function (IMF) by \citet{Maschberger13} which has a probability density function 
of the form:
\begin{equation}
p(m) \propto \left(\frac{m}{\mu}\right)^{-\alpha}\left(1 + \left(\frac{m}{\mu}\right)^{1 - \alpha}\right)^{-\beta}
\label{imf}.
\end{equation}
Eq.~\ref{imf} essentially combines the log-normal approximation for the IMF derived by \citet{Chabrier03,Chabrier05} with the \citet{Salpeter55} power-law slope for stars with mass $>$1\,M$_\odot$. Here, 
$\mu = 0.2$\,M$_\odot$ is the average stellar mass, $\alpha = 2.3$ is the Salpeter power-law exponent for higher mass stars, and $\beta = 1.4$ is the power-law exponent to describe the slope of the 
IMF for low-mass objects \citep*[which also deviates from the
  log-normal form;][]{Bastian10}. Finally, we sample from this IMF
within the mass range $m_{\rm low} = 0.01$\,M$_\odot$ to $m_{\rm up} =
50$\,M$_\odot$.

It is worth noting that the {\em average global} surface and volume
densities of all $N = 1500$ regions are very similar to each other (1500 stars
in a 1~pc radius sphere), and similarly for the $N=150$ star regions.
However, the {\em average} local surface and volume densities vary by
several orders of magnitude depending on the fractal dimension (degree
of substructure).  Highly substructured regions have their stars
concentrated in local `pockets' and have a filling factor much less
than unity (to some degree this is the definition of a fractal).
Thus, their local densities in these pockets may be considerable.

We run the simulations for 10\,Myr using the \texttt{kira} integrator in the Starlab package \citep{Zwart99,Zwart01}. We do not include stellar evolution in the simulations. A summary of the simulation parameter space is 
given in Table~\ref{cluster_setup}.

\begin{table}
\caption[bf]{A summary of the different star forming region properties adopted for the simulations.
The values in the columns are: the number of stars in each region ($N_{\rm stars}$), 
the typical mass of this region ($M_{\rm region}$),  the initial \emph{global} virial ratio of the region ($\alpha_{\rm vir}$), the initial fractal dimension ($D$) 
and whether or not the stellar velocities are correlated by distance.}
\begin{center}
\begin{tabular}{|c|c|c|c|c|}
\hline 
$N_{\rm stars}$ & $M_{\rm region}$  &  $\alpha_{\rm vir}$ & $D$ & Correlated velocities?\\
\hline
1500 & $\sim 10^3$\,M$_\odot$ & 0.3  & 1.6 & yes\\
1500 & $\sim 10^3$\,M$_\odot$ & 0.3  & 2.0 & yes\\
1500 & $\sim 10^3$\,M$_\odot$ & 0.3  & 3.0 & yes\\
\hline 
1500 & $\sim 10^3$\,M$_\odot$ & 0.5  & 1.6 & yes\\
1500 & $\sim 10^3$\,M$_\odot$ & 0.5  & 2.0 & yes\\
1500 & $\sim 10^3$\,M$_\odot$ & 0.5  & 3.0 & yes\\
\hline
1500 & $\sim 10^3$\,M$_\odot$ & 1.5  & 1.6 & yes\\
1500 & $\sim 10^3$\,M$_\odot$ & 1.5  & 2.0  & yes\\
1500 & $\sim 10^3$\,M$_\odot$ & 1.5  & 3.0 & yes\\
\hline
150 & $\sim 10^2$\,M$_\odot$ & 0.3  & 1.6 & yes\\
150 & $\sim 10^2$\,M$_\odot$ & 0.3  & 2.0 & yes\\
150 & $\sim 10^2$\,M$_\odot$ & 0.3  & 3.0 & yes\\
\hline 
150 & $\sim 10^2$\,M$_\odot$ & 0.5  & 1.6 & yes\\
150 & $\sim 10^2$\,M$_\odot$ & 0.5  & 2.0 & yes\\
150 & $\sim 10^2$\,M$_\odot$ & 0.5  & 3.0 & yes\\
\hline
150 & $\sim 10^2$\,M$_\odot$ & 1.5  & 1.6 & yes\\
150 & $\sim 10^2$\,M$_\odot$ & 1.5  & 2.0  & yes\\
150 & $\sim 10^2$\,M$_\odot$ & 1.5  & 3.0 & yes\\
\hline
1500 & $\sim 10^3$\,M$_\odot$ & 1.5  & 1.6 & no\\
\hline
\end{tabular}
\end{center}
\label{cluster_setup}
\end{table}

\section{Quantifying spatial structure and mass segregation}
\label{tools}

\subsection{Measuring spatial structure}

We determine the amount of structure in a star forming region by measuring the $\mathcal{Q}$-parameter. The $\mathcal{Q}$-parameter was pioneered by \citet{Cartwright04,Cartwright09a,Cartwright09b} and combines the normalised mean edge length of the minimum spanning tree of all the stars in the 
region, $\bar{m}$, with the normalised correlation length between all stars in the region, 
$\bar{s}$. The level of substructure is determined by the following equation:
\begin{equation}
\mathcal{Q} = \frac{\bar{m}}{\bar{s}}.
\end{equation}
A substructured association or region has $\mathcal{Q}<0.8$, whereas a smooth, centrally concentrated cluster has $\mathcal{Q}>0.8$. The $\mathcal{Q}$-parameter has the advantage of being independent of the density of the star forming region, and purely measures 
the level of substructure present. The original formulation of the $\mathcal{Q}$-parameter assumes the region is spherical, but can be modifed to take into account the effects of elongation \citep{Cartwright09a,Bastian09}. 

\subsection{Measuring mass segregation}

Mass segregation is a rather difficult thing to define.  It is usually
considered in the case of bound (spherical) clusters where a degree of
energy equipartition (primordial, dynamical, or both) results in the
most massive stars preferentially located in the cluster centre.  

However, here we take a more general definition of mass segregation
applicable to substructured regions and associations, as well as
clusters.  One way of viewing `mass segregation' is that the massive
stars are {\em closer to each other} than would be expected of random stars
(this is what is measured by the $\Lambda_{\rm MSR}$-parameter, see below).
Another view is that the massive stars are in {\em locally denser regions}
than would be expected of a typical star (this is what is measured by
the $\Sigma - m$ method, again see below for details).

Note that there are many other ways of defining mass segregation. For example, one can choose a cluster centre and measure the mass 
function as a function of radial distance \citep{Gouliermis04,Sabbi08}, use the mean square (Spitzer) radius of the cluster as a diagnostic for comparing stars with different mass ranges \citep*{Gouliermis09} or determine the distance of the most massive star(s) from the cluster centre 
compared to the average distance of low-mass stars to the cluster centre \citep{Kirk10}. It is also possible to quantify differences in luminosity between the centre and outskirts of a cluster \citep[e.g.][]{Carpenter97}. 

However, it is important to note that the $\Lambda_{\rm MSR}$ and the $\Sigma - m$ methods have the
significant advantage over other methods in that they do not require
the determination of a `centre' in a region, which is crucial for
analysing highly substructured regions.

\subsubsection{The $\Lambda_{\rm MSR}$ mass segregation ratio}

In order to quantify the amount of mass segregation present in a region, we first use the $\Lambda_{\rm MSR}$ method, introduced by \citet{Allison09a}.  This 
constructs a minimum spanning tree (MST) between a chosen subset of stars and then compares this MST to the average MST length of many 
random subsets. 

The MST
of a set of points is the path connecting all the points via the
shortest possible pathlength but which  contains no closed loops
\citep[e.g.][]{Prim57,Cartwright04}.

We use the algorithm of \citet{Prim57} to construct MSTs in our
dataset. We first make an ordered list of the separations  between all
possible pairs of stars. Stars are then connected together in `nodes',
starting with the shortest separations and  proceeding through the
list in order of increasing separation, forming new nodes if the
formation of the node does not result in a closed loop.

We find the MST of the $N_{\rm MST}$ stars in the chosen subset and
compare this to the MST of sets of $N_{\rm MST}$ random  stars in the
region. If the length of the MST of the chosen subset is shorter than
the average length of the MSTs for the  random stars then the subset
has a more concentrated distribution and is said to be mass segregated. Conversely, if the MST  length of the chosen subset is
longer than the average MST length, then the subset has a less
concentrated distribution, and is  said to be inversely mass
segregated \citep[see e.g.][]{Parker11b}. Alternatively, if the MST length of the chosen subset is
equal to the random MST length,  we can conclude that no mass
segregation is present.

By taking the ratio of the average (mean) random MST length to the subset MST
length, a quantitative measure of the degree of  mass segregation
(normal or inverse) can be obtained. We first determine the subset MST
length, $l_{\rm subset}$. We then  determine the average length of
sets of $N_{\rm MST}$ random stars each time, $\langle l_{\rm average}
\rangle$. There is a dispersion  associated with the average length of
random MSTs, which is roughly Gaussian and can be quantified as the
standard deviation  of the lengths  $\langle l_{\rm average} \rangle
\pm \sigma_{\rm average}$. However, we conservatively estimate the lower (upper) uncertainty 
as the MST length which lies 1/6 (5/6) of the way through an ordered list of all the random lengths (corresponding to a 66 per cent deviation from 
the median value, $\langle l_{\rm average} \rangle$). This determination 
prevents a single outlying object from heavily influencing the uncertainty. 
We can now define the `mass  segregation ratio' 
($\Lambda_{\rm MSR}$) as the ratio between the average random MST pathlength 
and that of a chosen subset, or mass range of objects:
\begin{equation}
\Lambda_{\rm MSR} = {\frac{\langle l_{\rm average} \rangle}{l_{\rm subset}}} ^{+ {\sigma_{\rm 5/6}}/{l_{\rm subset}}}_{- {\sigma_{\rm 1/6}}/{l_{\rm subset}}}.
\end{equation}
A $\Lambda_{\rm MSR}$ of $\sim$ 1 shows that the stars in the chosen
subset are distributed in the same way as all the other  stars,
whereas $\Lambda_{\rm MSR} > 1$ indicates mass segregation and
$\Lambda_{\rm MSR} < 1$ indicates inverse mass segregation,
i.e.\,\,the chosen subset is more sparsely distributed than the other stars.

There are several subtle variations of $\Lambda_{\rm MSR}$. \citet*{Olczak11} propose using the geometric mean to reduce the spread in uncertainties, 
and \citet{Maschberger11} propose using the median MST length to reduce the effects of outliers from influencing the results. However, in the subsequent 
analysis we will adopt the original $\Lambda_{\rm MSR}$  from Allison. 

\subsection{The $\Sigma - m$ method and the $\Sigma_{\rm LDR}$ local density ratio} 
 
Recently, \citet{Maschberger11} proposed a method to analyse mass segregation which measures the distribution of local stellar surface density, $\Sigma$, as a function of stellar mass.
We calculate the local stellar surface density following the prescription of \citet{Casertano85}, modified to account for the analysis in projection. For an individual star the local stellar surface density is given by
\begin{equation}
\Sigma = \frac{N - 1} {\pi r_{N}^2},
\end{equation}
where $r_{N}$ is the distance to the $N^{\rm th}$ nearest neighbouring star (we adopt $N = 10$ throughout this work).

If there is mass segregation, massive stars are concentrated in the central, dense regions and thus should have higher values of $\Sigma$.
This can be seen in a plot of $\Sigma$ versus mass, showing all stars and highlighting outliers.
Trends in the $\Sigma - m$ plot can be shown by the moving average (or median) of a subset, $\tilde{\Sigma}_\mathrm{subset} $, compared to the average (median) of the whole sample, $\tilde{\Sigma}_\mathrm{all}$.
The signature of mass segregation is then  $\tilde{\Sigma}_\mathrm{subset} >  \tilde{\Sigma}_\mathrm{all}$, and of inverse mass segregation  $\tilde{\Sigma}_\mathrm{subset} <  \tilde{\Sigma}_\mathrm{all}$.
The statistical significance of mass segregation can be established with a two-sample Kolmogorov-Smirnov (KS) test of the $\Sigma$ values of the subset against the $\Sigma$ values of the rest. 

Note that the $\Sigma - m$ method shares similarities with the two-dimensional convex hull method proposed by \citet{Moeckel09b}, in that as well as measuring the degree of mass segregation, it also 
provides information on the density within a region.

Whilst this method has the advantage of not being biased by outliers \citep[see the discussion in][]{Maschberger11}, it does lead to the artificial effect of placing each (massive) star in its own bin (defined by the local density of that star), and does not 
always reflect the \emph{spatial} distribution of a particular subset of stars. What it does effectively measure is the \emph{local density} distribution of a subset, which we will see in Section~\ref{results} is not necessarily 
the spatial distribution.

In this paper we will divide $\tilde{\Sigma}_\mathrm{subset}$ by $\tilde{\Sigma}_\mathrm{all}$ to define a `local density ratio', $\Sigma_{\rm LDR}$:
\begin{equation}
\Sigma_{\rm LDR} = \frac{\tilde{\Sigma}_\mathrm{subset}}{\tilde{\Sigma}_\mathrm{all}}
\end{equation} 
The significance of this measure of the local density of a subset of stars compared to the cluster will still be defined by the 
KS test between the $\Sigma$ values of the subset against the $\Sigma$ values of the rest. We will detail the number of stars used to determine $\tilde{\Sigma}_\mathrm{subset}$ 
in the following sections of the paper. 

The differences between $\Lambda_{\rm MSR}$ and $\Sigma_{\rm LDR}$ might seem subtle, but (as
we shall see) become important.  The $\Lambda_{\rm MSR}$ method measures the
collective concentration of massive stars (ie. are they close to
each other?).  The $\Sigma_{\rm LDR}$ method measures the relative local
densities of massive stars (ie. are they in dense regions?), but does
not consider how close the massive stars are to each other.  Therefore
it would be quite possible (and we find it so) to have significant 
`mass segregation' found by one method but not the other. For this reason, we do not refer to  $\Sigma - m$ as measuring `mass segregation' in the remainder of the paper.

\section{Results}
\label{results}

In this section we first examine the evolution of typical examples of
a substructured, subvirial (cool) star forming region and a
substructured, supervirial (warm) region, before  comparing the
average evolution of all of the models in our chosen parameter
space. We use the $\Lambda_{\rm MSR}$ measure of mass segregation, the $\Sigma_{\rm LDR}$ measure of local density, as well
as a measure  of the spatial structure of the region ($\mathcal{Q}$) described in
Section~\ref{tools}, and follow the evolution of these quantities over
time.

\subsection{Evolution of a substructured, subvirial region}

\begin{figure*}
  \begin{center}
\setlength{\subfigcapskip}{10pt}
\vspace*{-0.3cm}
\hspace*{-1.cm}\subfigure[Morphology, 0\,Myr]{\label{cool-a}\rotatebox{270}{\includegraphics[scale=0.26]{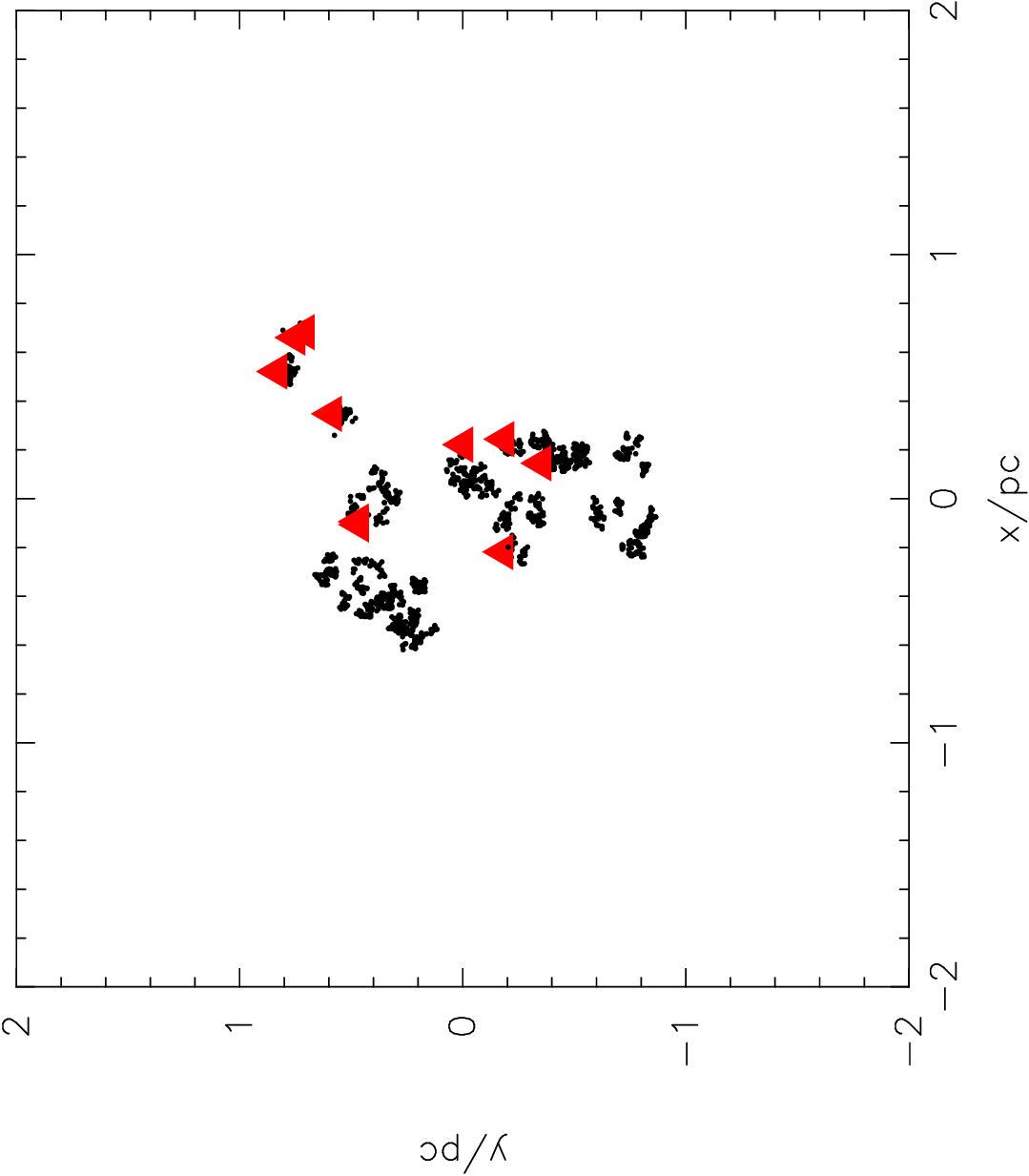}}}
\hspace*{0.3cm} 
\subfigure[Morphology, 5\,Myr]{\label{cool-b}\rotatebox{270}{\includegraphics[scale=0.26]{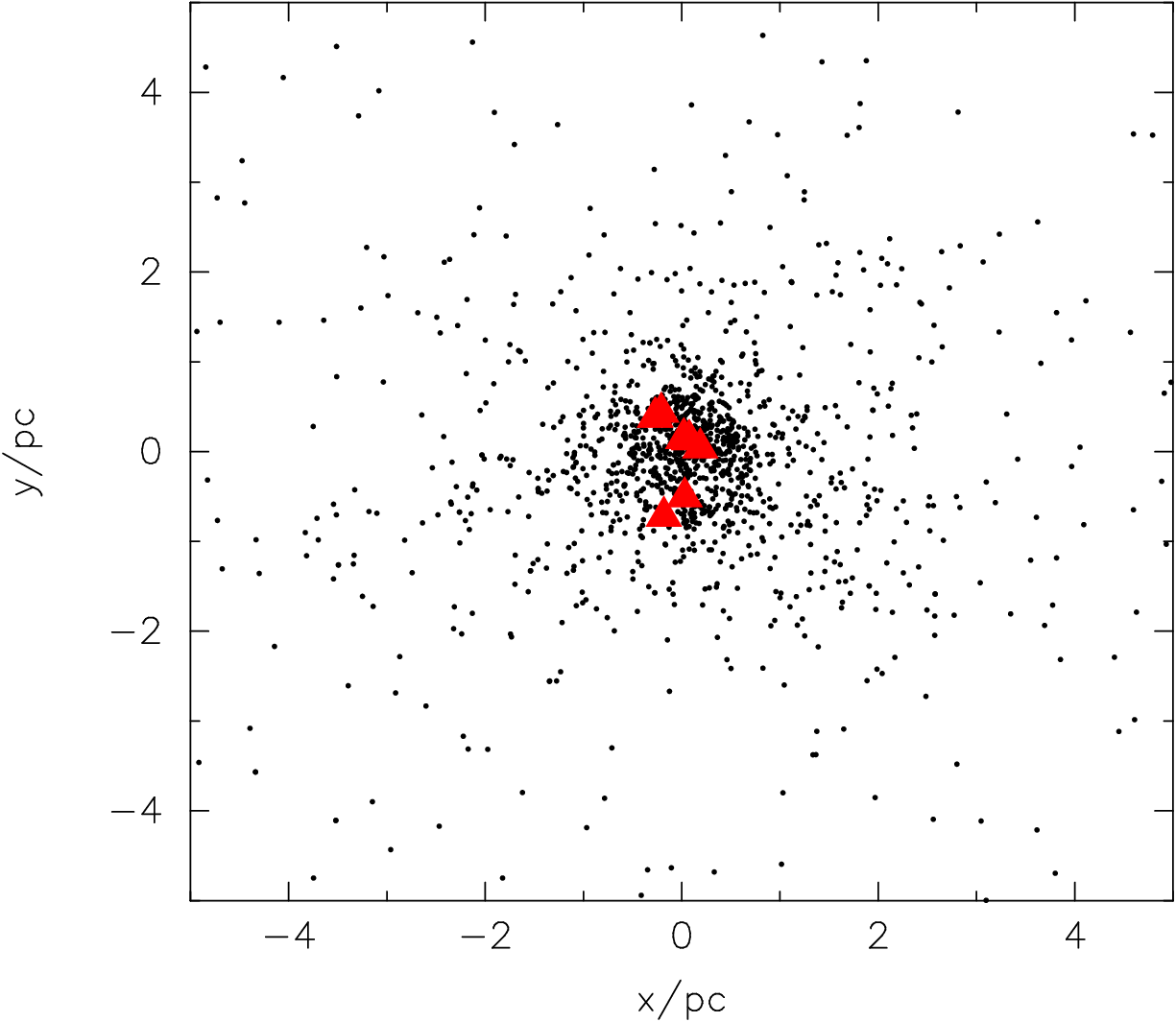}}}
\hspace*{0.3cm} 
\subfigure[Morphology, 10\,Myr]{\label{cool-c}\rotatebox{270}{\includegraphics[scale=0.26]{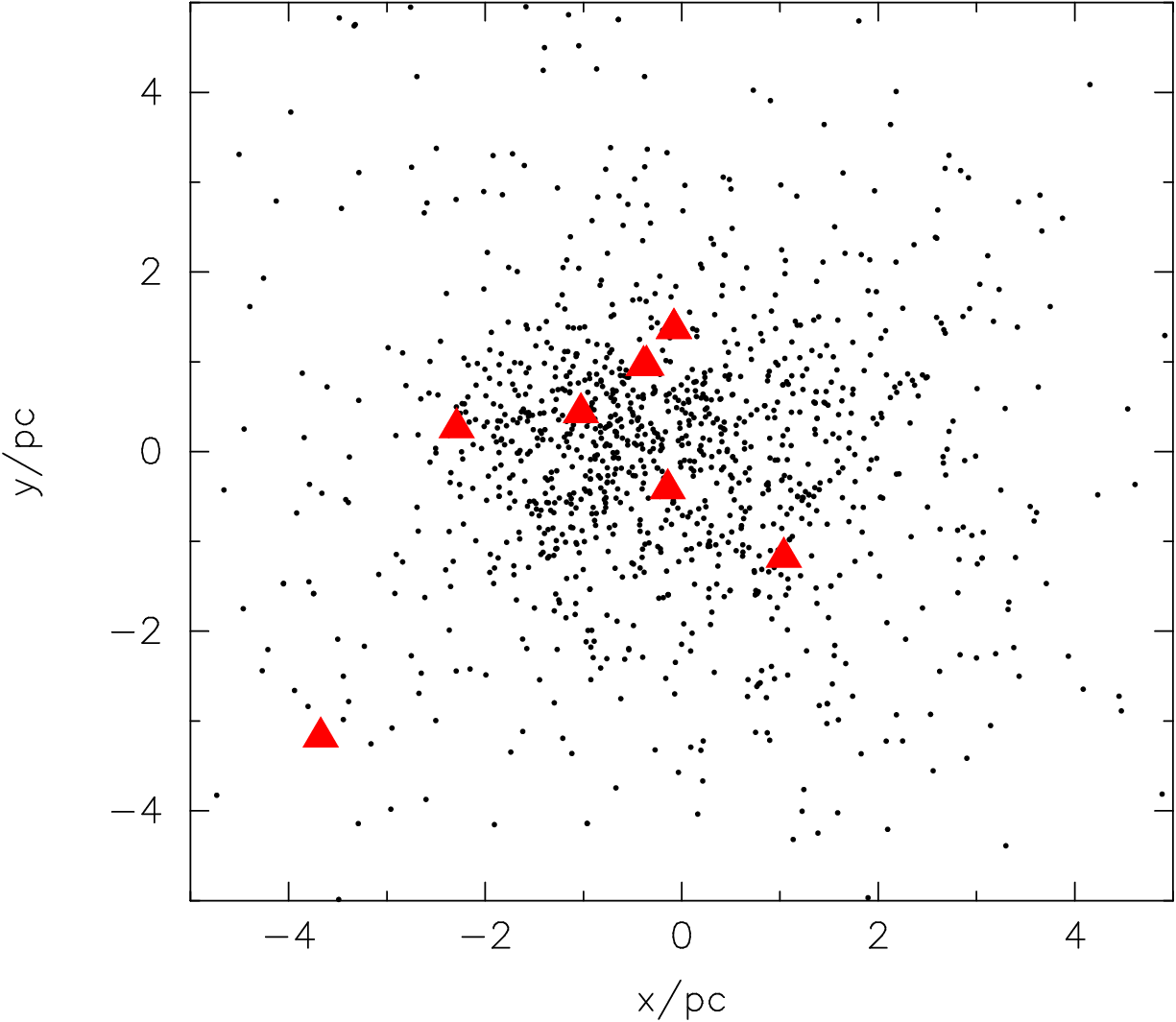}}}
\hspace*{-1.cm}
\subfigure[$\Lambda_{\rm MSR}$, 0\,Myr]{\label{cool-d}\rotatebox{270}{\includegraphics[scale=0.26]{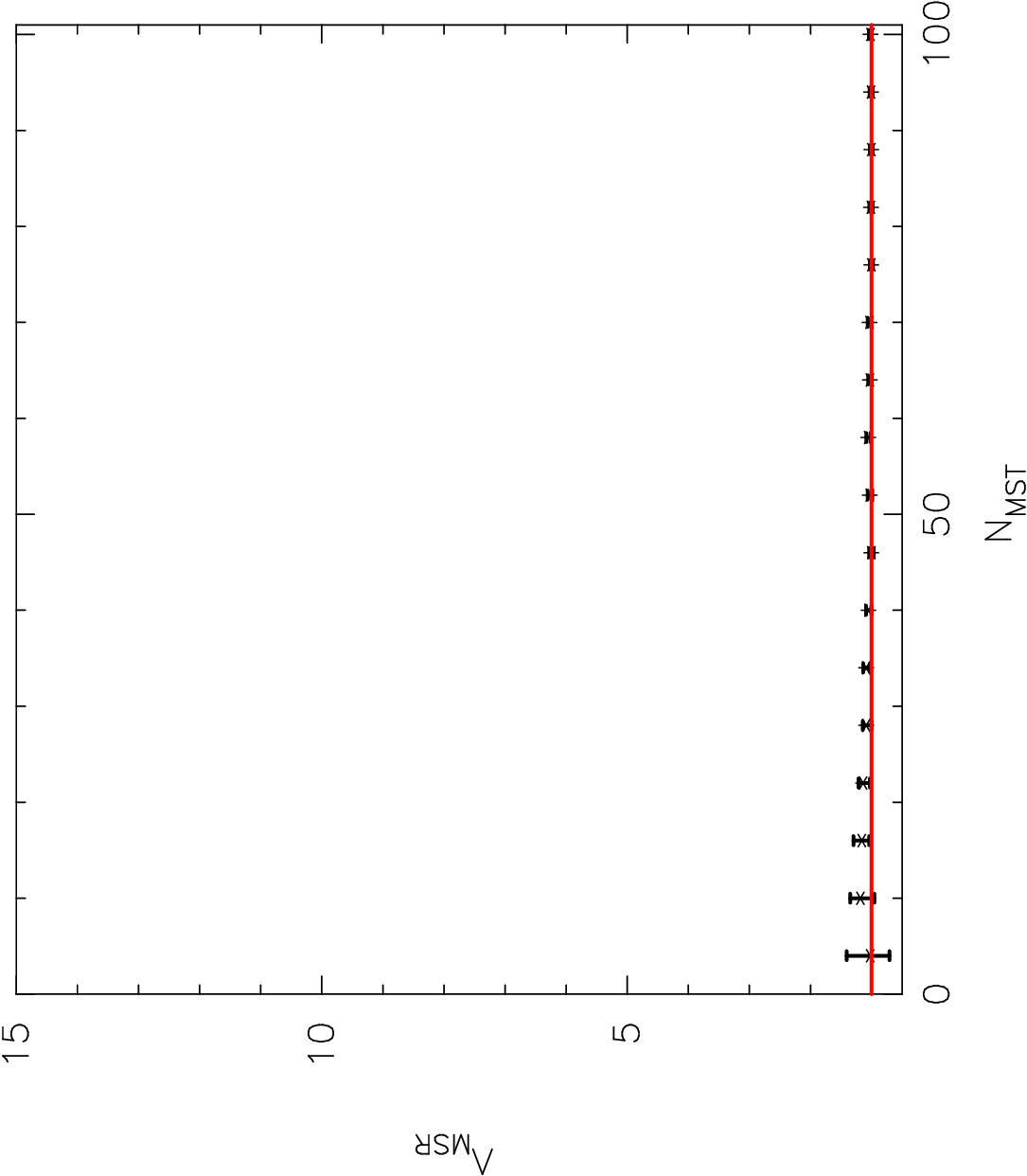}}}
\hspace*{0.3cm} 
\subfigure[$\Lambda_{\rm MSR}$, 5\,Myr]{\label{cool-e}\rotatebox{270}{\includegraphics[scale=0.26]{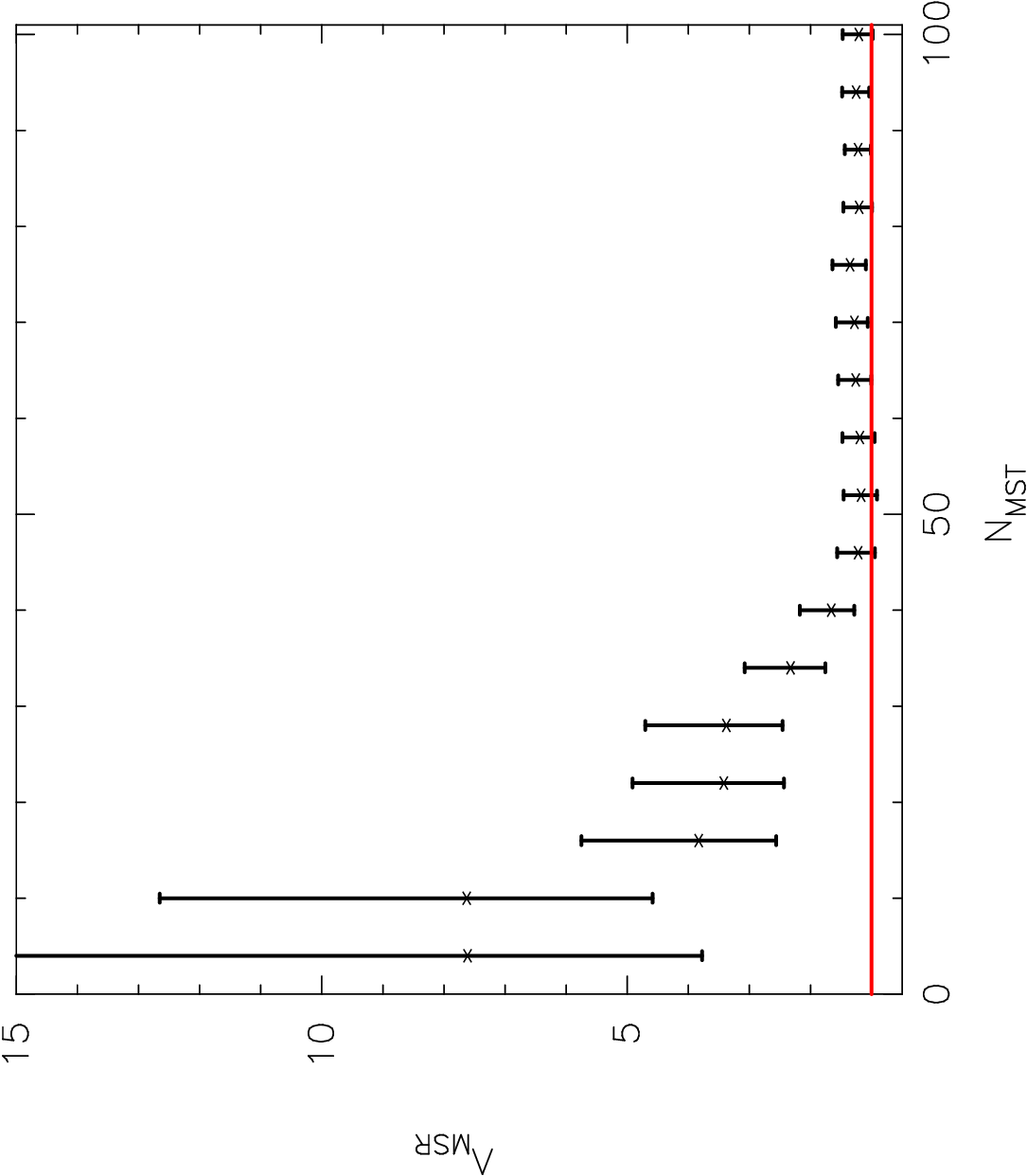}}}
\hspace*{0.3cm} 
\subfigure[$\Lambda_{\rm MSR}$, 10\,Myr]{\label{cool-f}\rotatebox{270}{\includegraphics[scale=0.26]{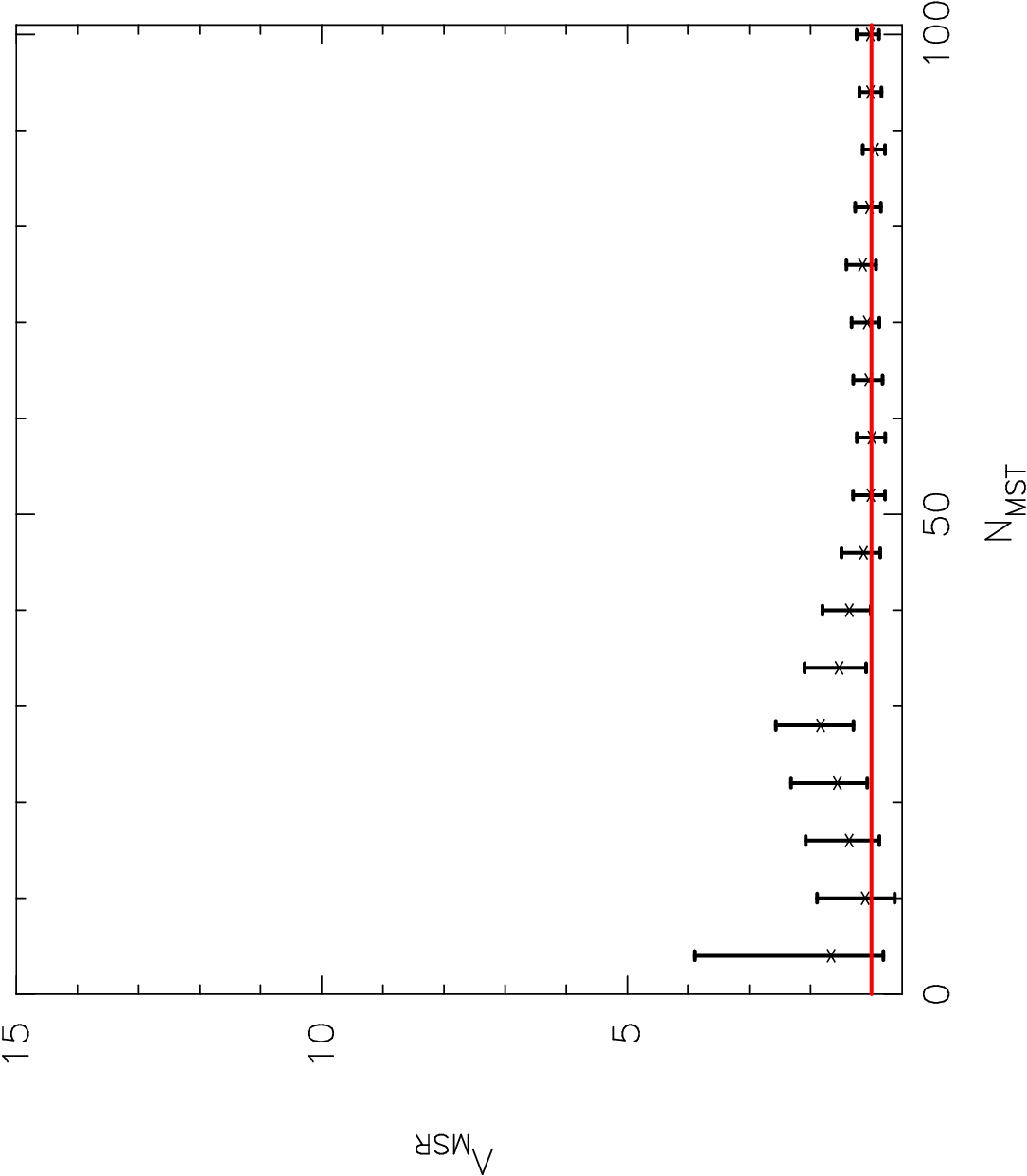}}}
\hspace*{-1.cm}
\subfigure[$\Sigma - m$, 0\,Myr]{\label{cool-g}\rotatebox{270}{\includegraphics[scale=0.26]{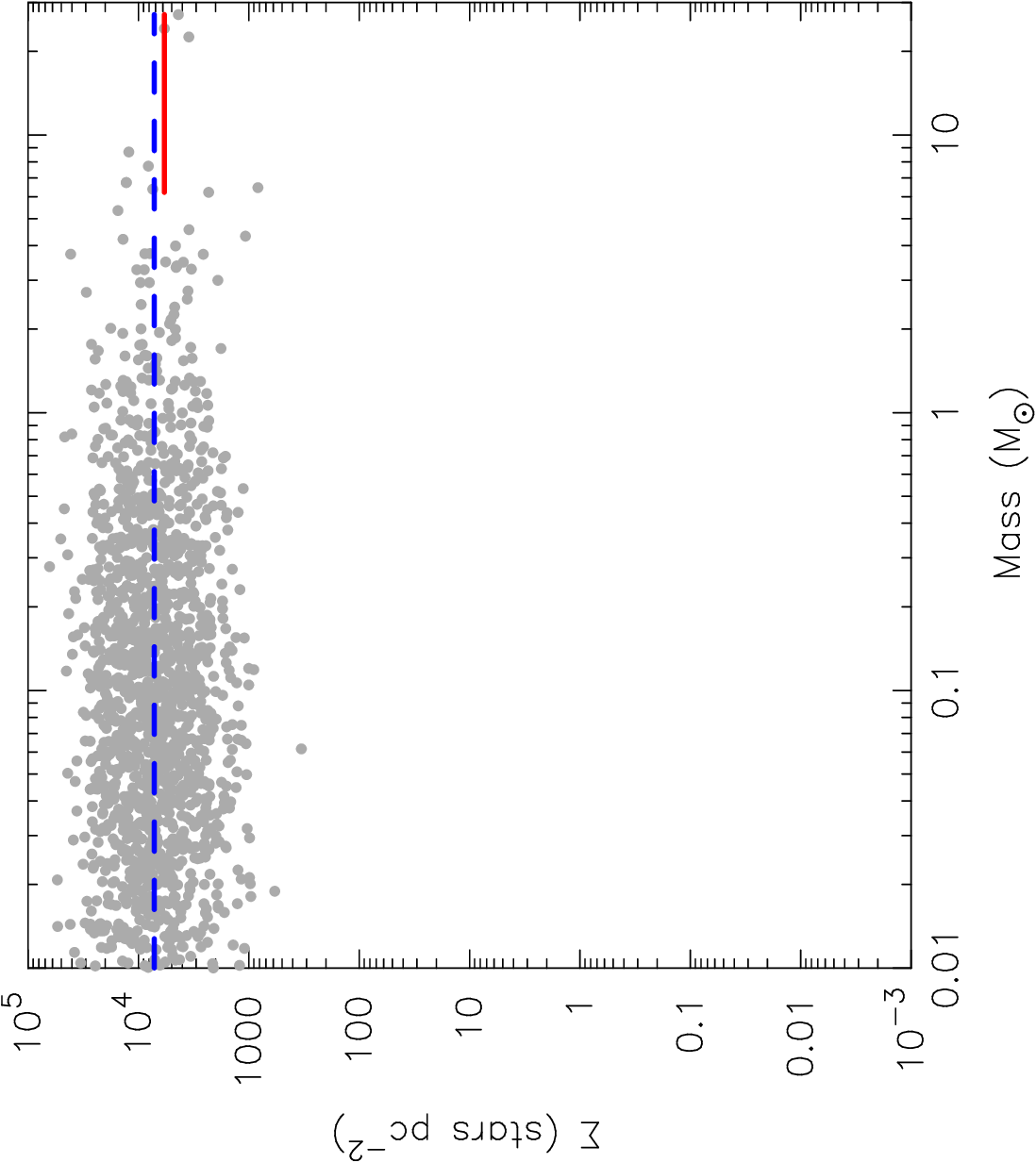}}}
\hspace*{0.3cm} 
\subfigure[$\Sigma - m$, 5\,Myr]{\label{cool-h}\rotatebox{270}{\includegraphics[scale=0.26]{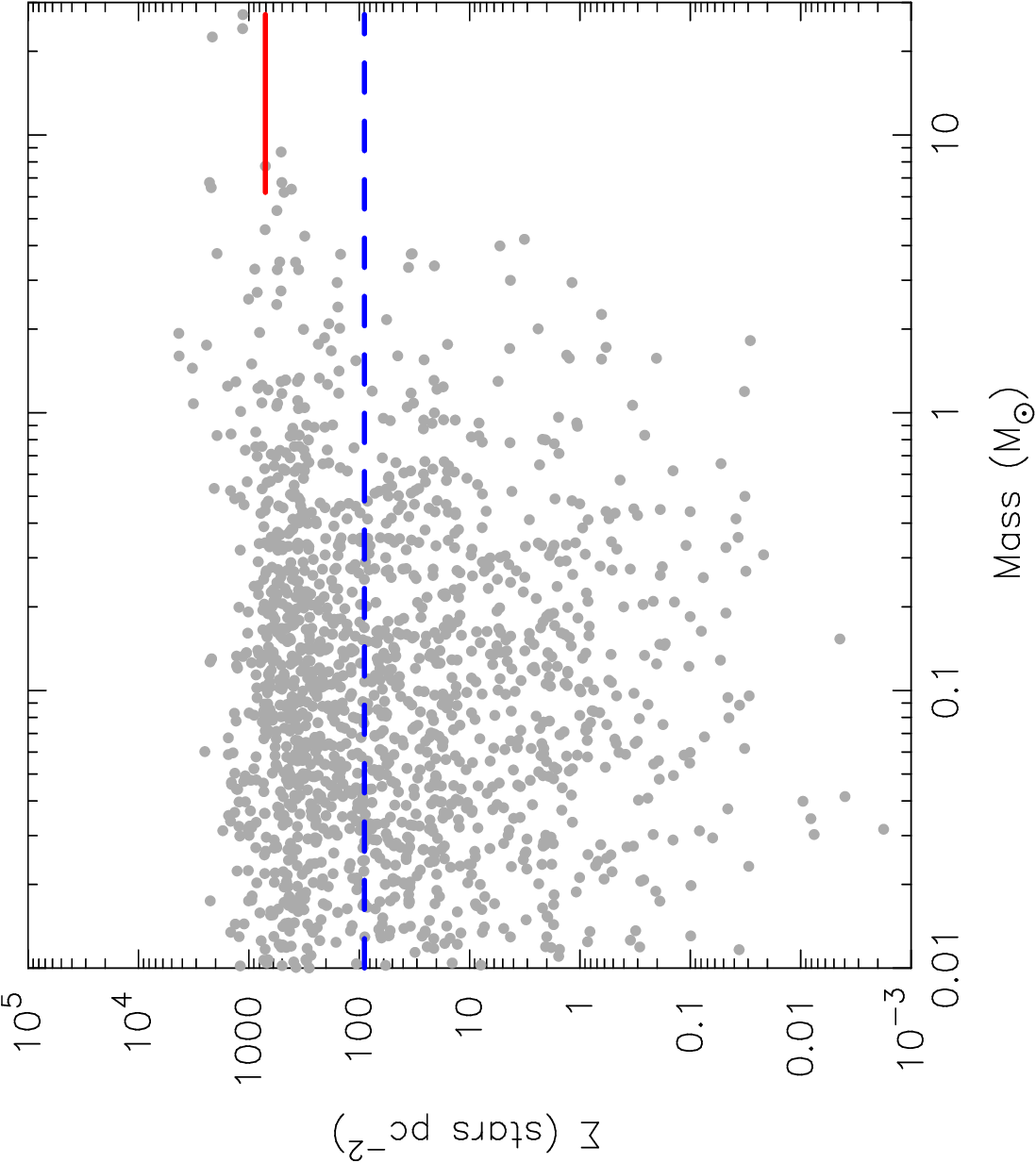}}}
\hspace*{0.3cm} 
\subfigure[$\Sigma - m$, 10\,Myr]{\label{cool-i}\rotatebox{270}{\includegraphics[scale=0.26]{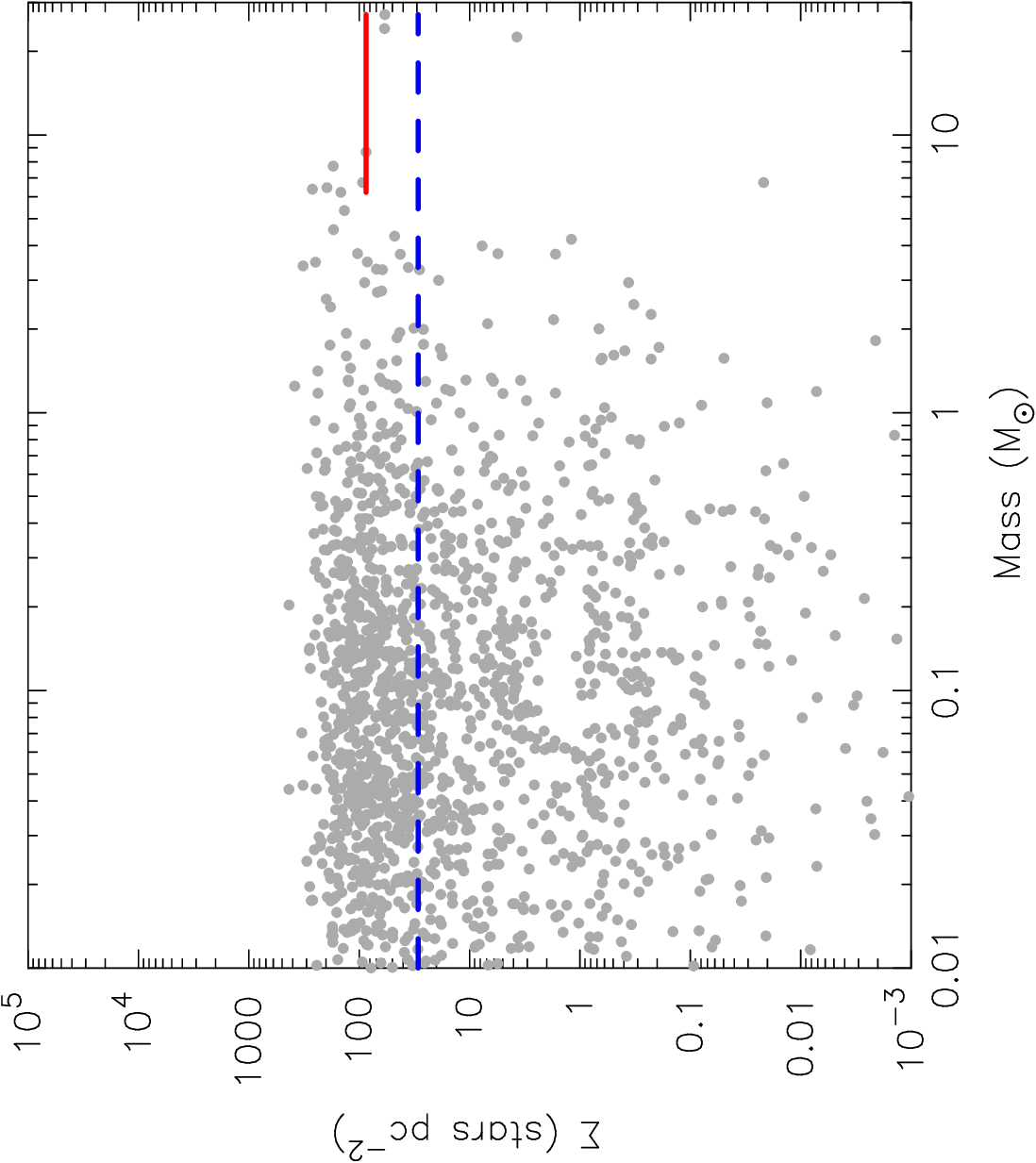}}}
\hspace*{-1.cm}
\subfigure[Evolution of $\Lambda_{\rm MSR}$]{\label{cool-j}\rotatebox{270}{\includegraphics[scale=0.26]{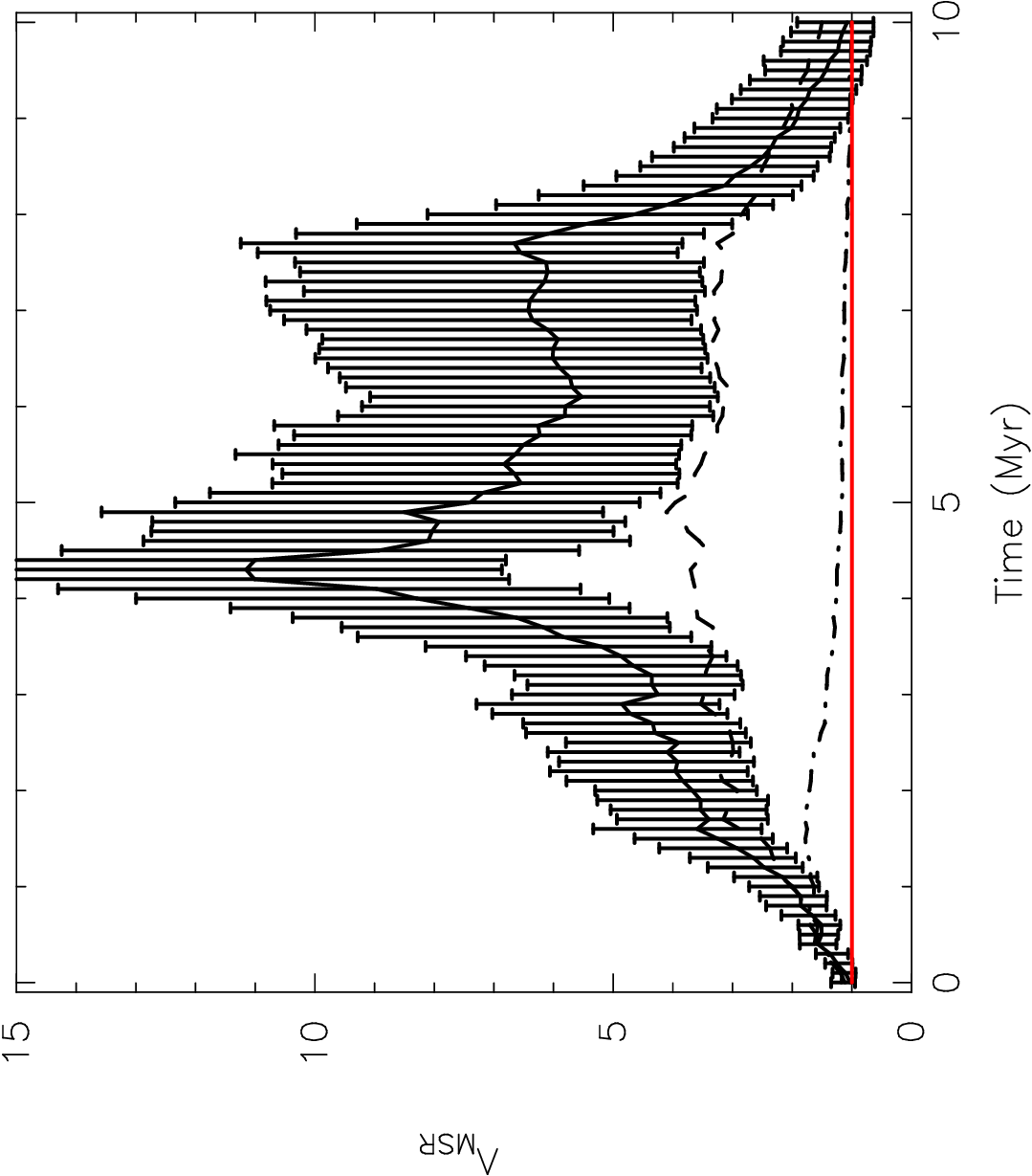}}}
\hspace*{0.3cm} 
\subfigure[Evolution of $\Sigma_{\rm LDR}$]{\label{cool-k}\rotatebox{270}{\includegraphics[scale=0.26]{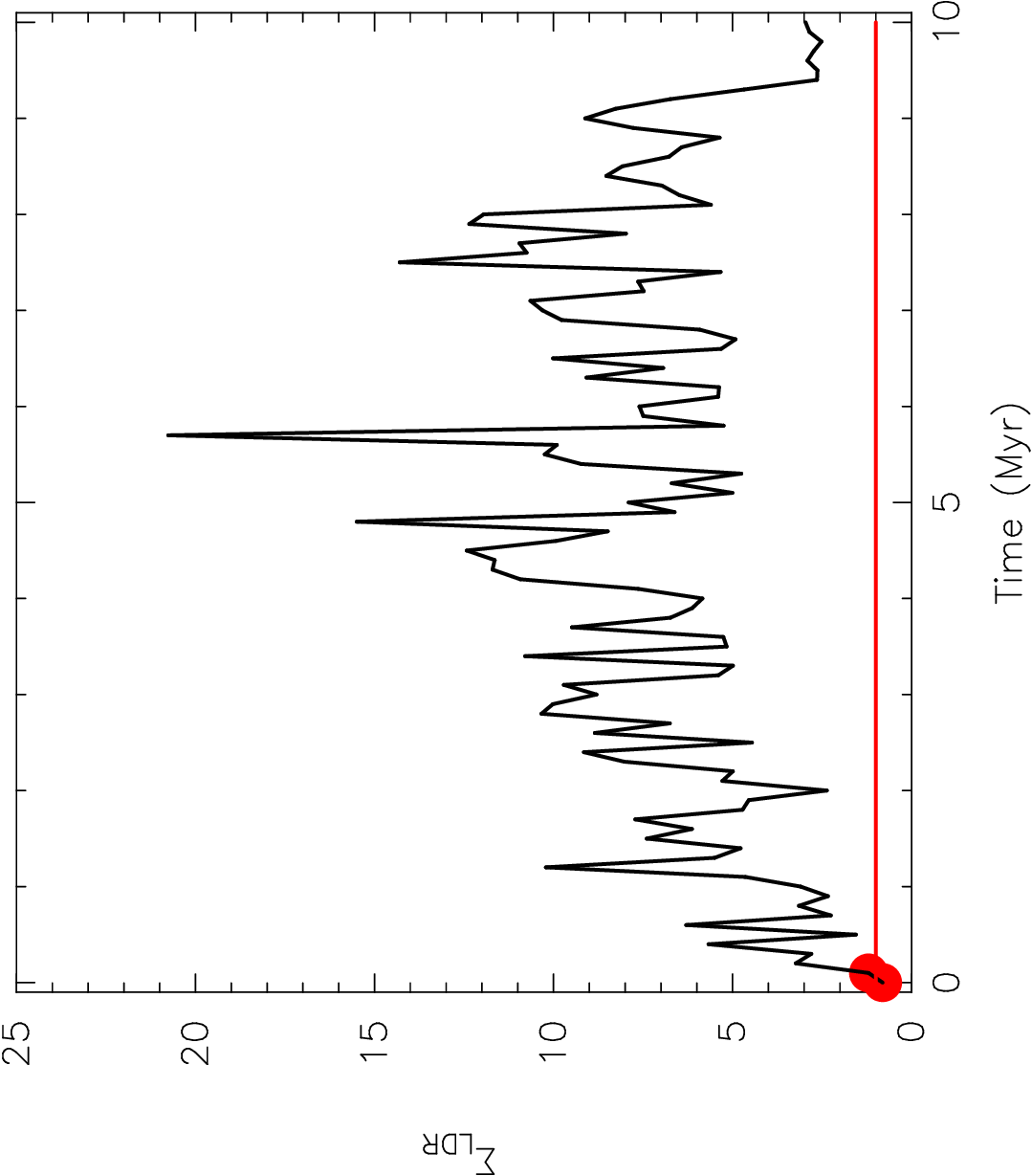}}}
\hspace*{0.3cm} 
\subfigure[Evolution of $\mathcal{Q}$--parameter]{\label{cool-l}\rotatebox{270}{\includegraphics[scale=0.26]{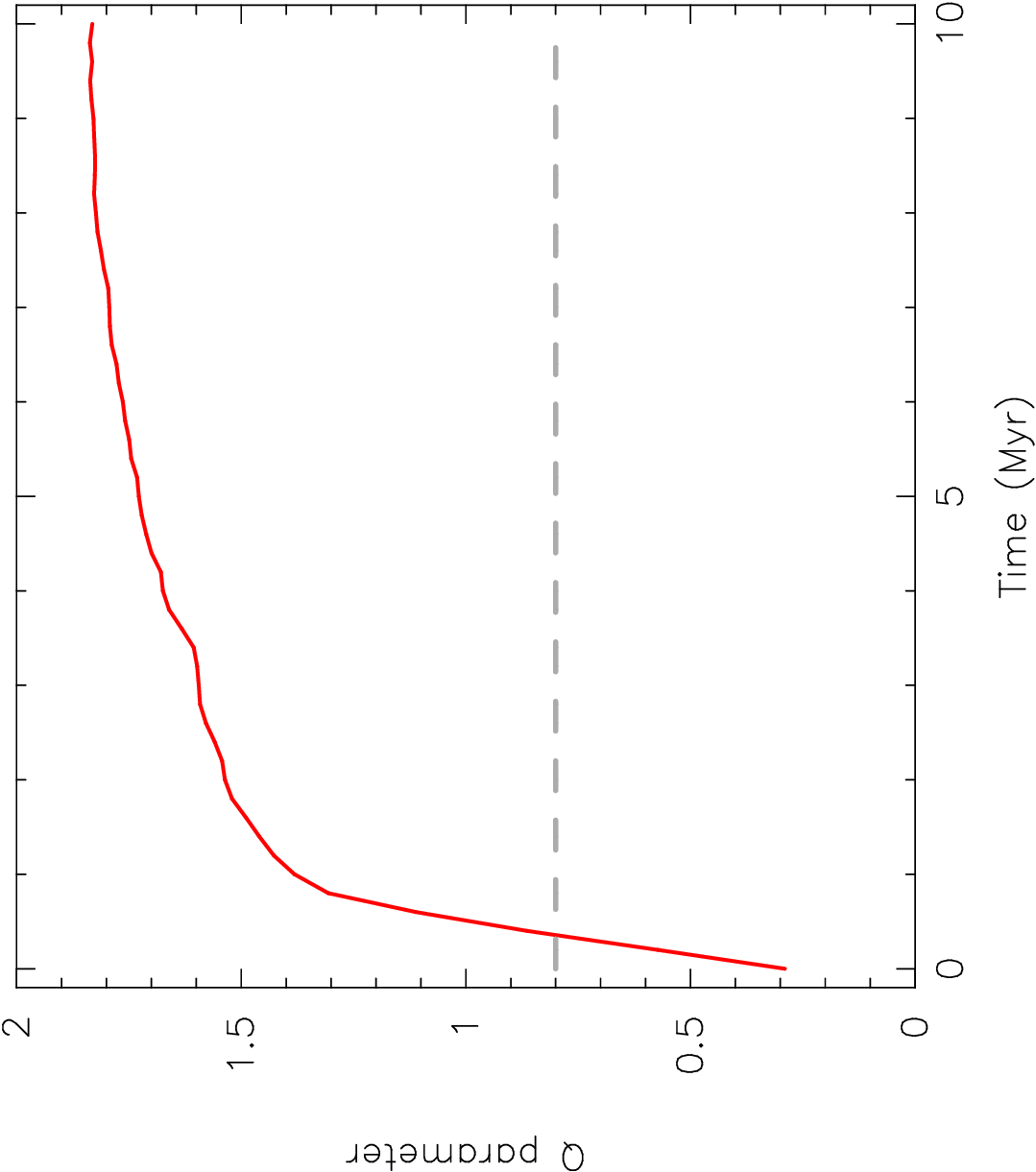}}}
\caption[bf]{Evolution of a subvirial ($\alpha_{\rm vir} = 0.3$), substructured star forming region ($D = 1.6$) with $N = 1500$ stars. We show the morphology at 0, 5 and 10\,Myr (a -- c), the mass segregation ratio, $\Lambda_{\rm MSR}$ of the 
$N_{\rm MST}$ most massive stars stars at 0, 5 and 10\,Myr (d -- f), and the surface density as a function of stellar mass, $\Sigma - m$ (g -- i) at 0, 5 and 10\,Myr. We also show the evolution 
of $\Lambda_{\rm MSR}$ in panel (j) -- the solid, dashed and dot-dashed lines are the values of $\Lambda_{\rm MSR}$ for the 10, 20 and 50 most massive stars respectively; the ratio of the surface density of the 10 most massive stars to the median surface density of all stars in the region ($\Sigma_{\rm LDR} = \tilde{\Sigma}_{10}/\tilde{\Sigma}_{\rm all}$ -- panel k) and the evolution of the $\mathcal{Q}$--parameter  (panel l), 
with time.  }
\label{cool_indiv}
  \end{center}
\end{figure*}

In Fig.~\ref{cool_indiv} we show the evolution of a
`typical'\footnote{By `typical' we mean that it shows the basic
  dynamics that occur in such a system, as we shall describe later
  different realisations of statistically the same initial conditions
  can result in very different behaviour \citep[see][]{Allison10}.} 
$N = 1500$ stars, subvirial ($\alpha_{\rm vir} = 0.3$), substructured ($D = 1.6$)
region. In panels (a) -- (c) we show the morphology at 0, 5 and
10\,Myr respectively. This region  undergoes violent relaxation and
collapses to form a bound, spherical cluster. The most massive stars
(shown by the red triangles) are randomly placed in the fractal initially but 
after 5\,Myr they are all in the cluster centre and have dynamically mass segregated.  

As the cluster relaxes, the most massive stars interact with each other, often forming unstable hierarchical 
multiple systems \citep{Allison11} -- indeed, the binary fraction of O--type stars in the simulation rises from 
zero to $\sim$70\,per cent after 10\,Myr. The formation of massive star binaries can in some cases destroy the cluster, 
as e.g.\,\,a 30\,M$_\odot$--30\,M$_\odot$ binary with 100\,au semimajor axis has a binding energy comparable to that of the whole cluster (10$^{41}$J), 
and this process is apparent after 10\,Myr (panel (c)).  

We can examine this behaviour using the the $\Lambda_{\rm MSR}$
measure of mass segregation. In panel (d) the mass segregation ratio
as a function of the number of stars in the minimum spanning tree,
$N_{\rm MST}$ is shown at 0\,Myr (i.e.\,\,before any dynamical
evolution has occured). The region rapidly mass segregates due to its
initial subvirial velocities and high level of substructure \citep[as
  demonstrated by][]{Allison10}, with significant mass segregation
down to the 40${\rm th}$ most massive star after 5\,Myr (panel
(e)). For example, if we focus on the  a subset of stars more massive
than the $N_{\rm MST} = 10^{\rm th}$ most massive star, the cluster
has a mass segregation ratio  $\Lambda_{\rm MSR} =
7.6^{+5.1}_{-2.9}$. However, as the higher-order massive star multiple
system in the centre decays \citep[see][]{Allison11}  
the amount of mass segregation (as measured by $\Lambda_{\rm MSR}$)
decreases to the point at which is not significant at 10\,Myr
(panel (f)). 


In panel (g) we show the
local surface density of every star in the region as a function of the
star's mass (the grey points). The median stellar surface density for
the entire region   is indicated by the blue dashed line in
Fig.~\ref{cool-g}. Initially, the median surface density for all stars
in the region is $\tilde{\Sigma}_{\rm all} = 7187$\,stars\,pc$^{-2}$,
whereas the 10 most massive stars have a median surface density
$\tilde{\Sigma}_{10} = 5829$\,stars\,pc$^{-2}$ as shown by the solid
red line.  This difference, however, is not significant -- a two
dimensional KS test returns a p-value of 0.57 that the two subsets
share the same parent distribution. However, after 5\,Myr (panel h)
the most massive stars have a median surface density of
$\tilde{\Sigma}_{10} = 708$\,stars\,pc$^{-2}$ compared to the  cluster
median value of $\tilde{\Sigma}_{\rm all} = 90$\,stars
\,pc$^{-2}$. Now the KS test between these two subsets returns a
p-value of $< 10^{-5}$, indicating that the massive stars are in areas of significantly higher local density than the average stars in the region. 
After 10\,Myr the 10 most massive stars again have a
higher surface density $\tilde{\Sigma}_{10} = 87$\,stars\,pc$^{-2}$,
whereas the median surface density for the whole cluster is
$\tilde{\Sigma}_{\rm all} = 29$\,stars\,pc$^{-2}$.
 
We show the full evolution of $\Lambda_{\rm MSR}$ in
Fig.~\ref{cool-j}. The value of $\Lambda_{\rm MSR}$ for the 10 most
massive stars is shown by the solid line (with error bars), and
$\Lambda_{\rm MSR}$ for the 20 and 50 most massive stars is shown by
the dashed and dot-dashed lines,  respectively. We plot $\Lambda_{\rm
  MSR} = 1$, i.e.\,\,no mass segregation, shown by the horizontal red
line. The region has dynamically mass segregated on a significant
level after 1.5\,Myr, and the value of $\Lambda_{\rm MSR}$ reaches a
maximum of $\Lambda_{\rm MSR} = 11.2^{+6.2}_{-4.3}$ after 4.3\,Myr.

In Fig.~\ref{cool-k} we show the evolution of the surface density of
the 10 most massive stars divided by the median surface density (the
$\Sigma_{\rm LDR}$ local density ratio, $\Sigma_{\rm LDR} =
\tilde{\Sigma}_{10}/\tilde{\Sigma}_{\rm all}$) in the region as a
function of time.  We plot a filled red circle at times when the
difference between the most massive stars and the whole region is not
significant (in this case only the first few snapshots in the $N$-body
simulation). We plot $\Sigma_{\rm LDR} = 1$, (i.e.\,\,no difference in local density as a function of mass) shown by the horizontal  red line. The $\Sigma_{\rm LDR}$
ratio quickly becomes significantly higher than unity, and is above 5
for the majority of the cluster's lifetime. 

Finally, we plot the evolution of the spatial structure (as measured
by the $\mathcal{Q}$--parameter) in Fig.~\ref{cool-l}. The boundary
between substructured and centrally concentrated morphologies
($\mathcal{Q} = 0.8$) is shown by the grey dashed line. The initial
substructure is rapidly erased  (as demonstrated in \citealp{Goodwin04a,Parker12d})
and the region becomes a smooth, centrally concentrated cluster after
only 1\,Myr (on a similar timescale to the mass  segregation).

\subsection{Evolution of a substructured, supervirial region}

\begin{figure*}
  \begin{center}
\setlength{\subfigcapskip}{10pt}
\vspace*{-0.3cm}
\hspace*{-1.cm}\subfigure[Morphology, 0\,Myr]{\label{hot-a}\rotatebox{270}{\includegraphics[scale=0.26]{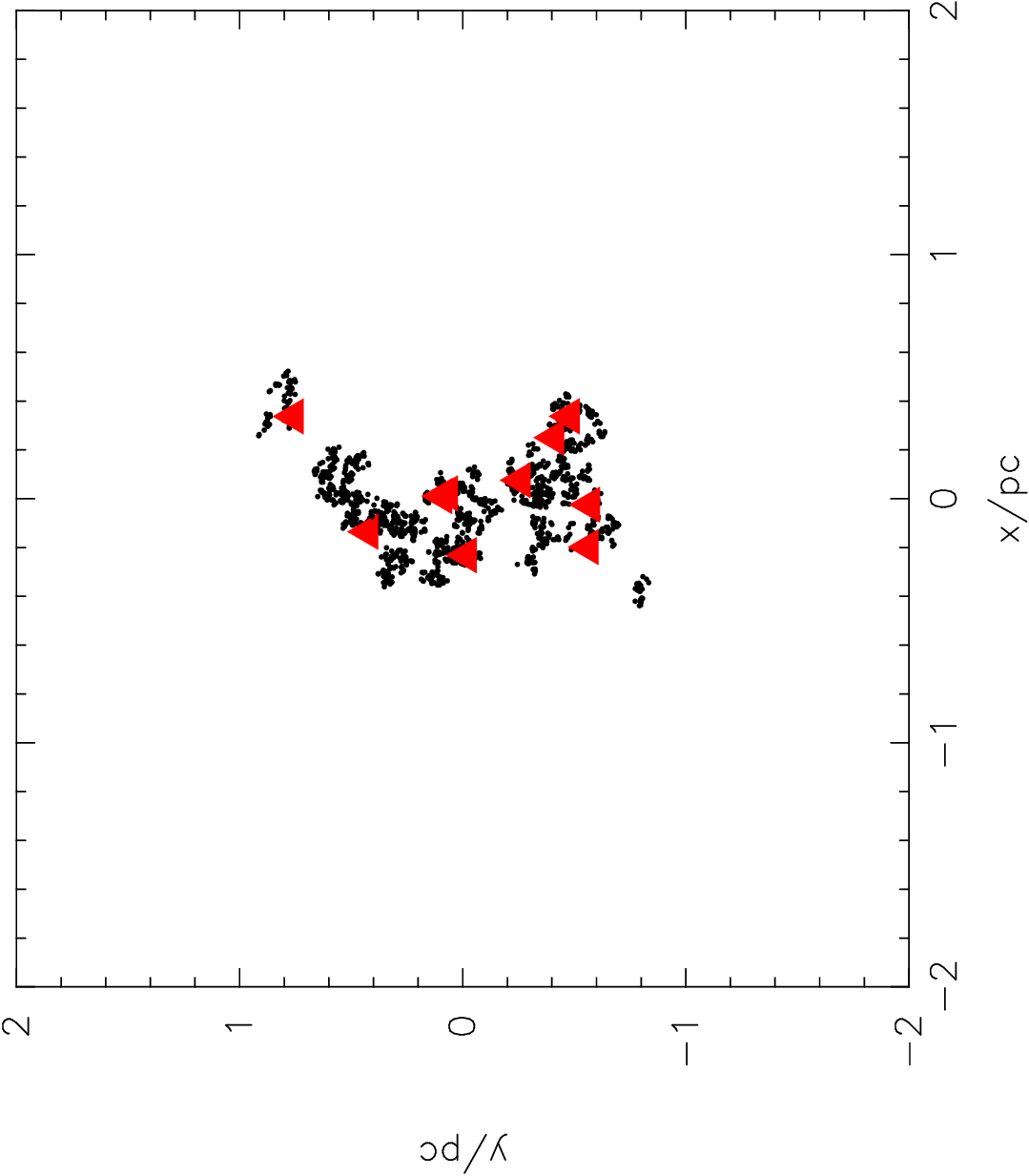}}}
\hspace*{0.3cm} 
\subfigure[Morphology, 5\,Myr]{\label{hot-b}\rotatebox{270}{\includegraphics[scale=0.26]{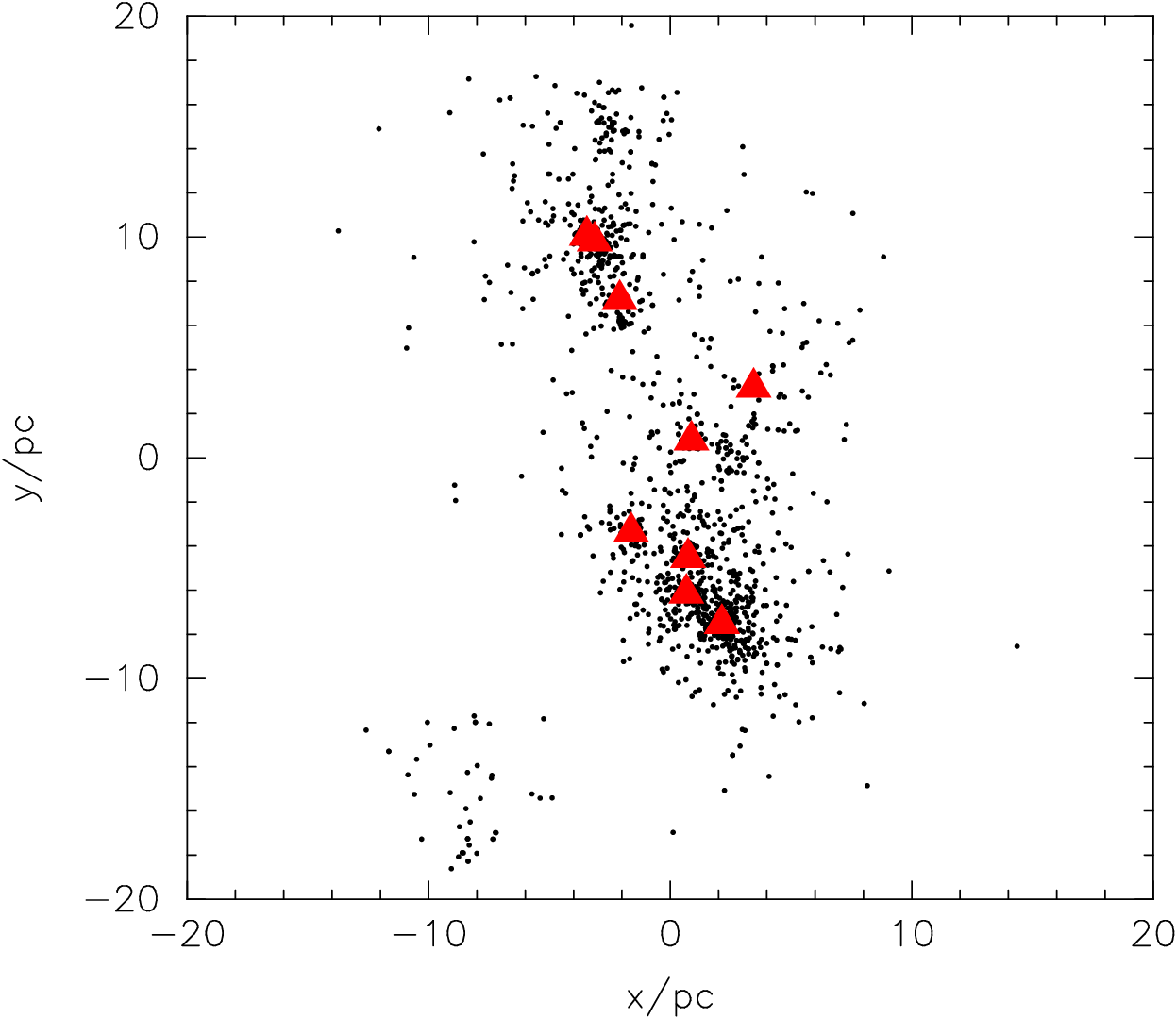}}}
\hspace*{0.3cm} 
\subfigure[Morphology, 10\,Myr]{\label{hot-c}\rotatebox{270}{\includegraphics[scale=0.26]{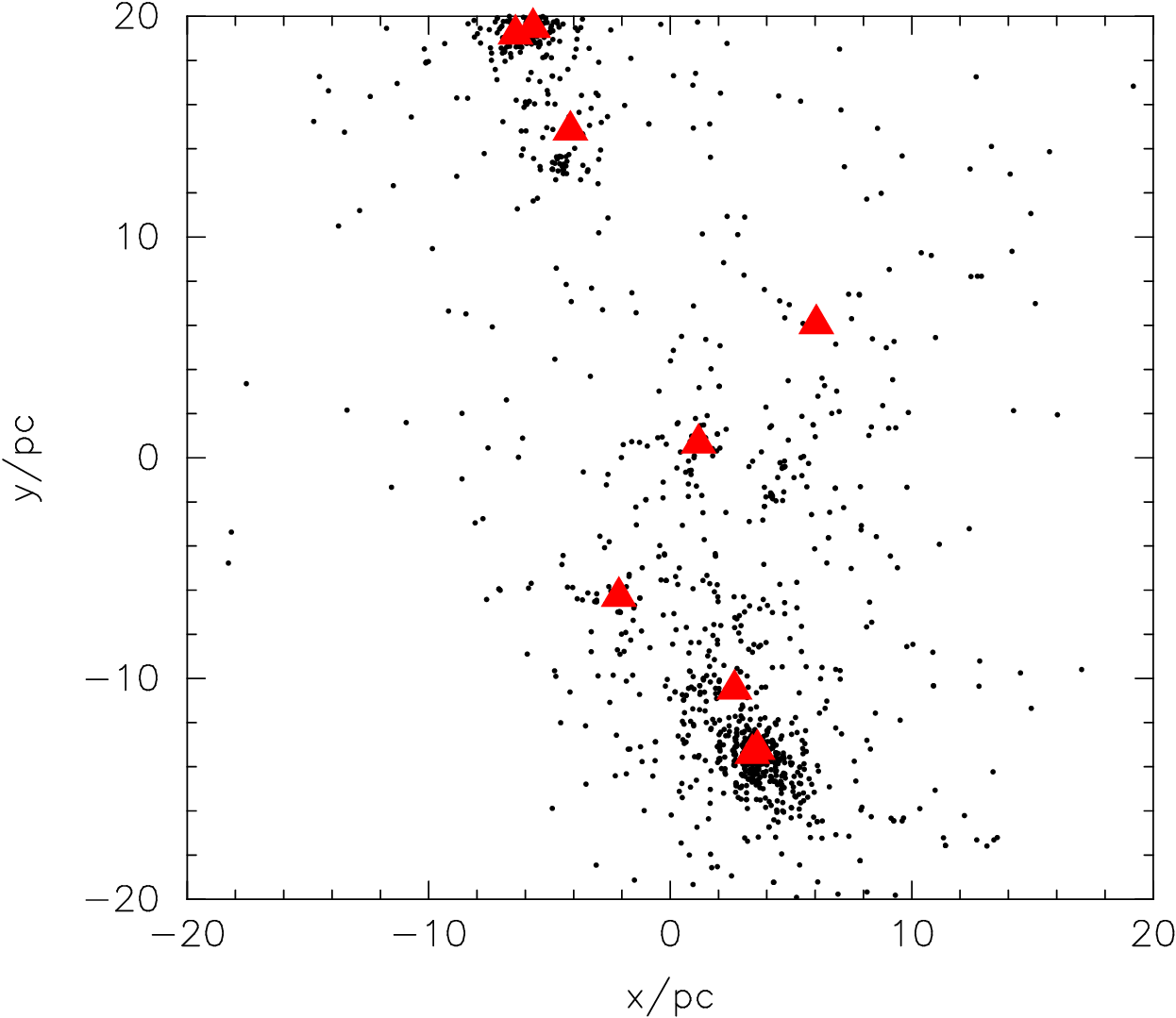}}}
\hspace*{-1.cm}
\subfigure[$\Lambda_{\rm MSR}$, 0\,Myr]{\label{hot-d}\rotatebox{270}{\includegraphics[scale=0.26]{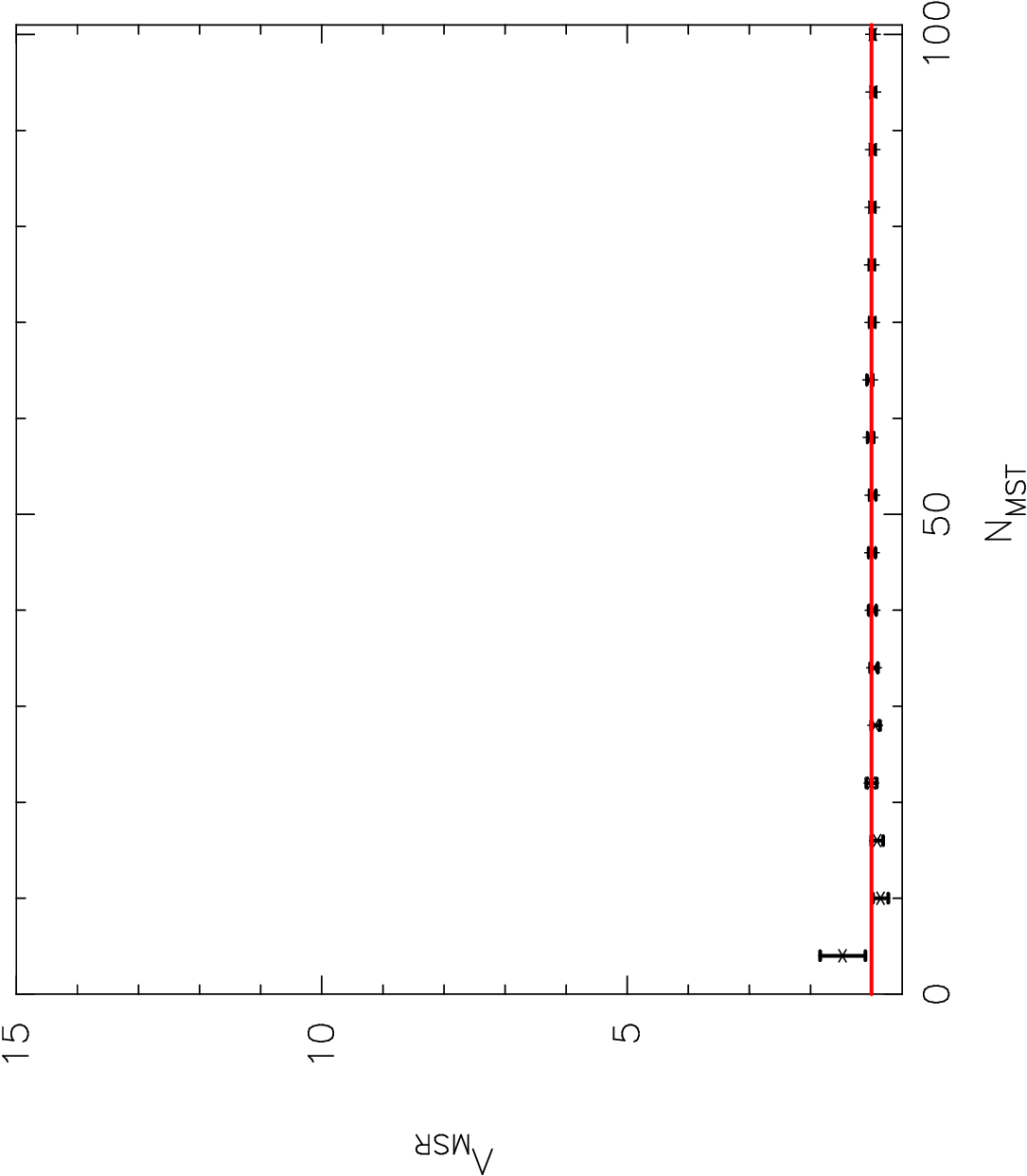}}}
\hspace*{0.3cm} 
\subfigure[$\Lambda_{\rm MSR}$, 5\,Myr]{\label{hot-e}\rotatebox{270}{\includegraphics[scale=0.26]{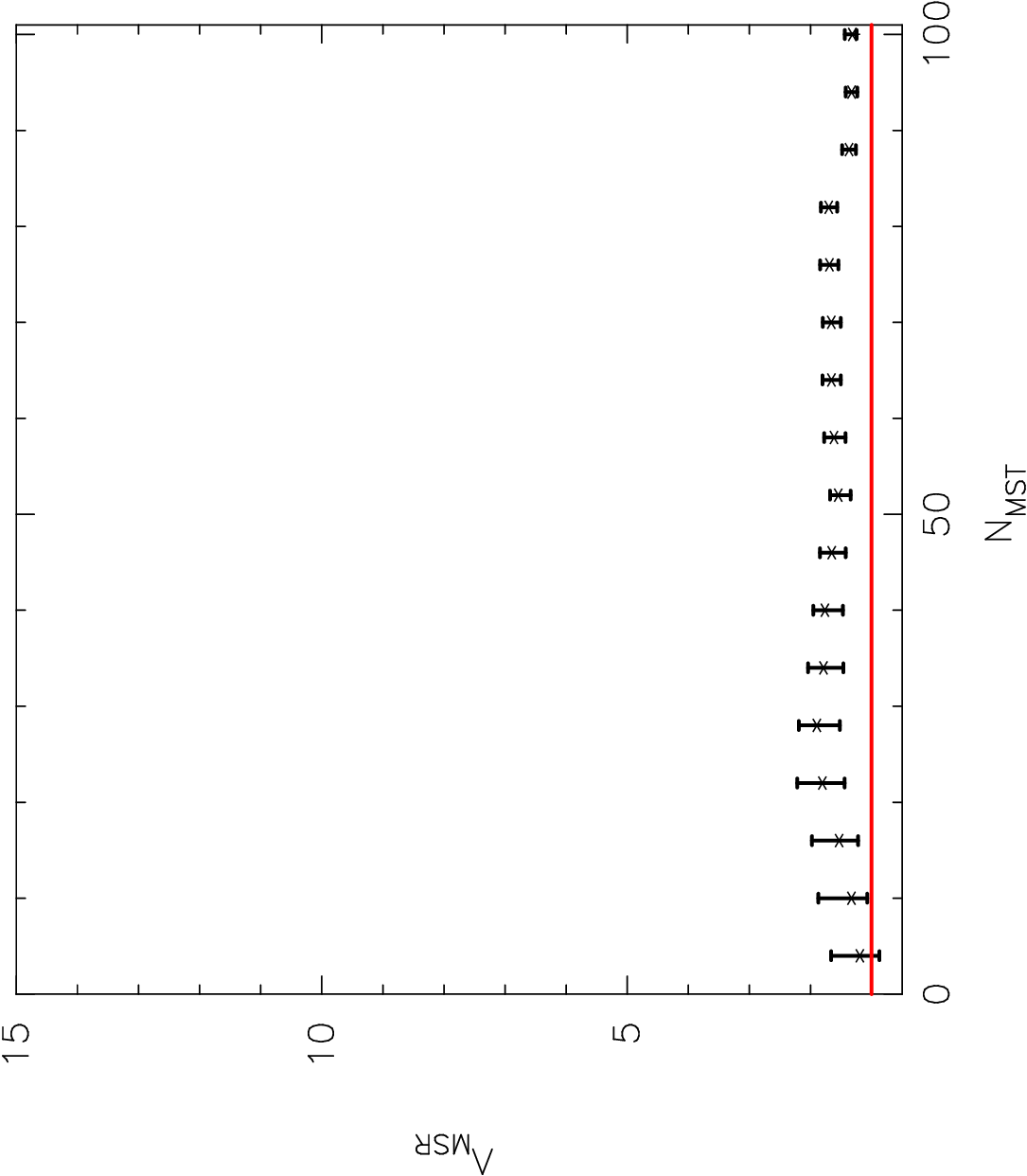}}}
\hspace*{0.3cm} 
\subfigure[$\Lambda_{\rm MSR}$, 10\,Myr]{\label{hot-f}\rotatebox{270}{\includegraphics[scale=0.26]{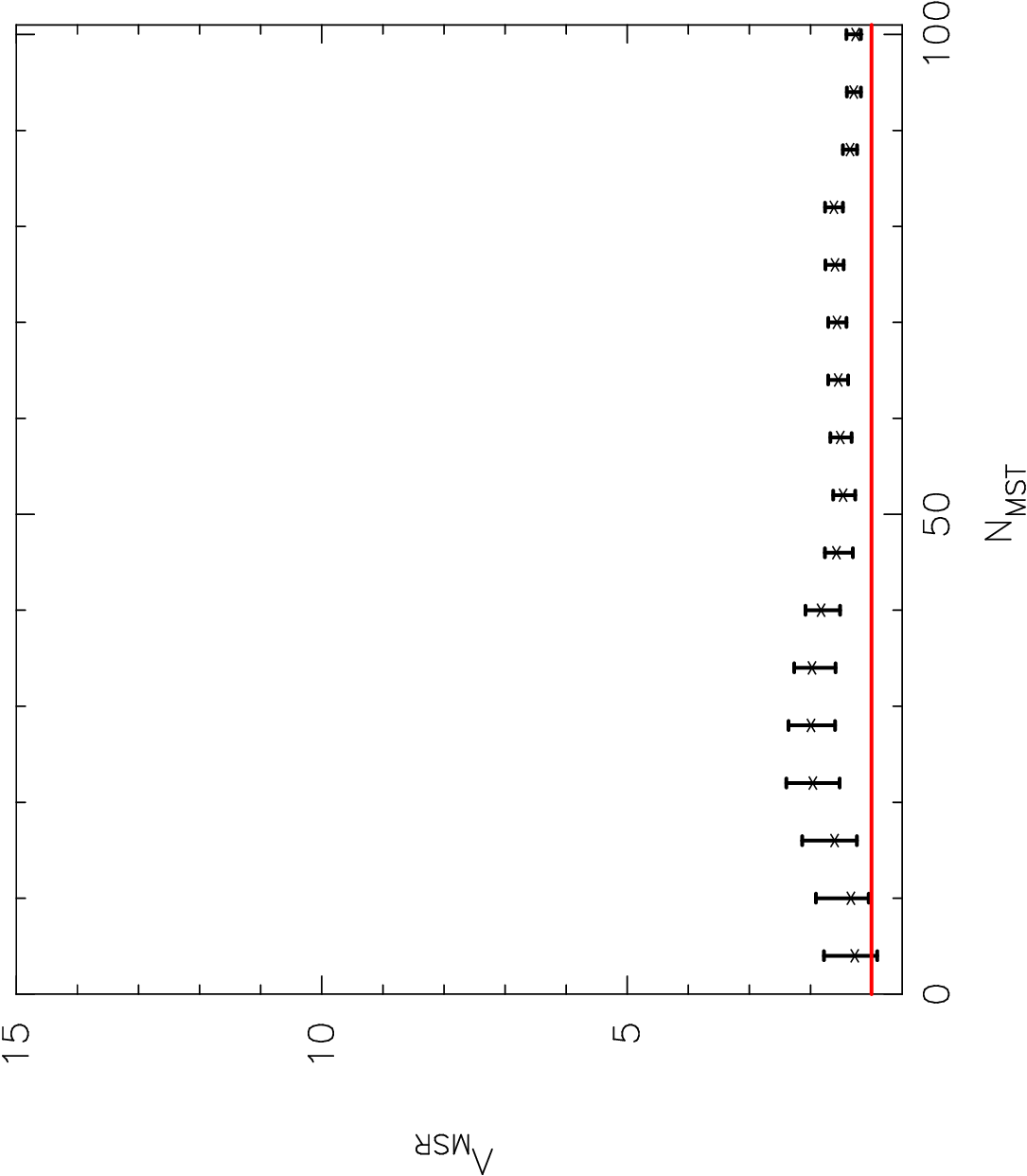}}}
\hspace*{-1.cm}
\subfigure[$\Sigma - m$, 0\,Myr]{\label{hot-g}\rotatebox{270}{\includegraphics[scale=0.26]{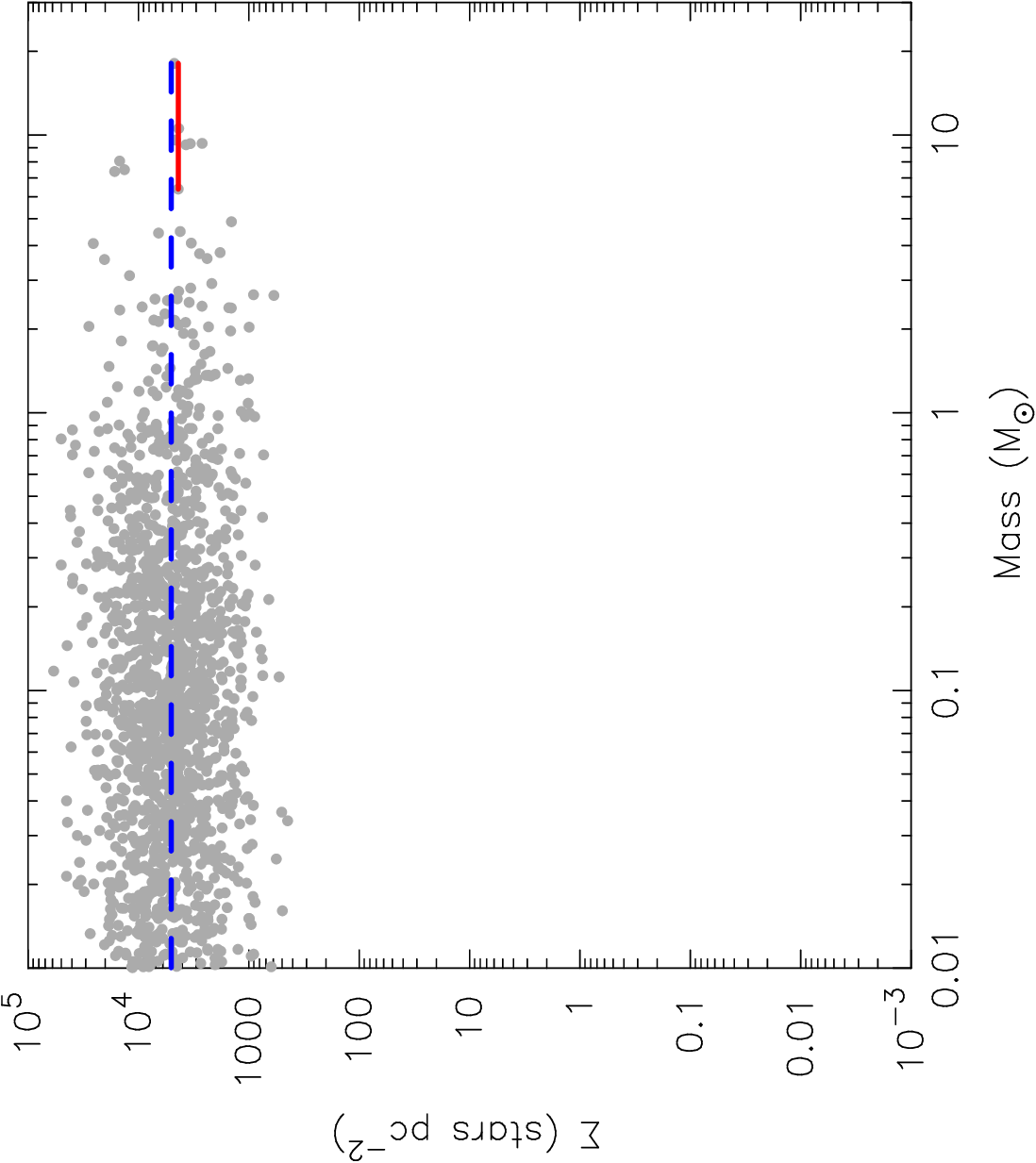}}}
\hspace*{0.3cm} 
\subfigure[$\Sigma - m$, 5\,Myr]{\label{hot-h}\rotatebox{270}{\includegraphics[scale=0.26]{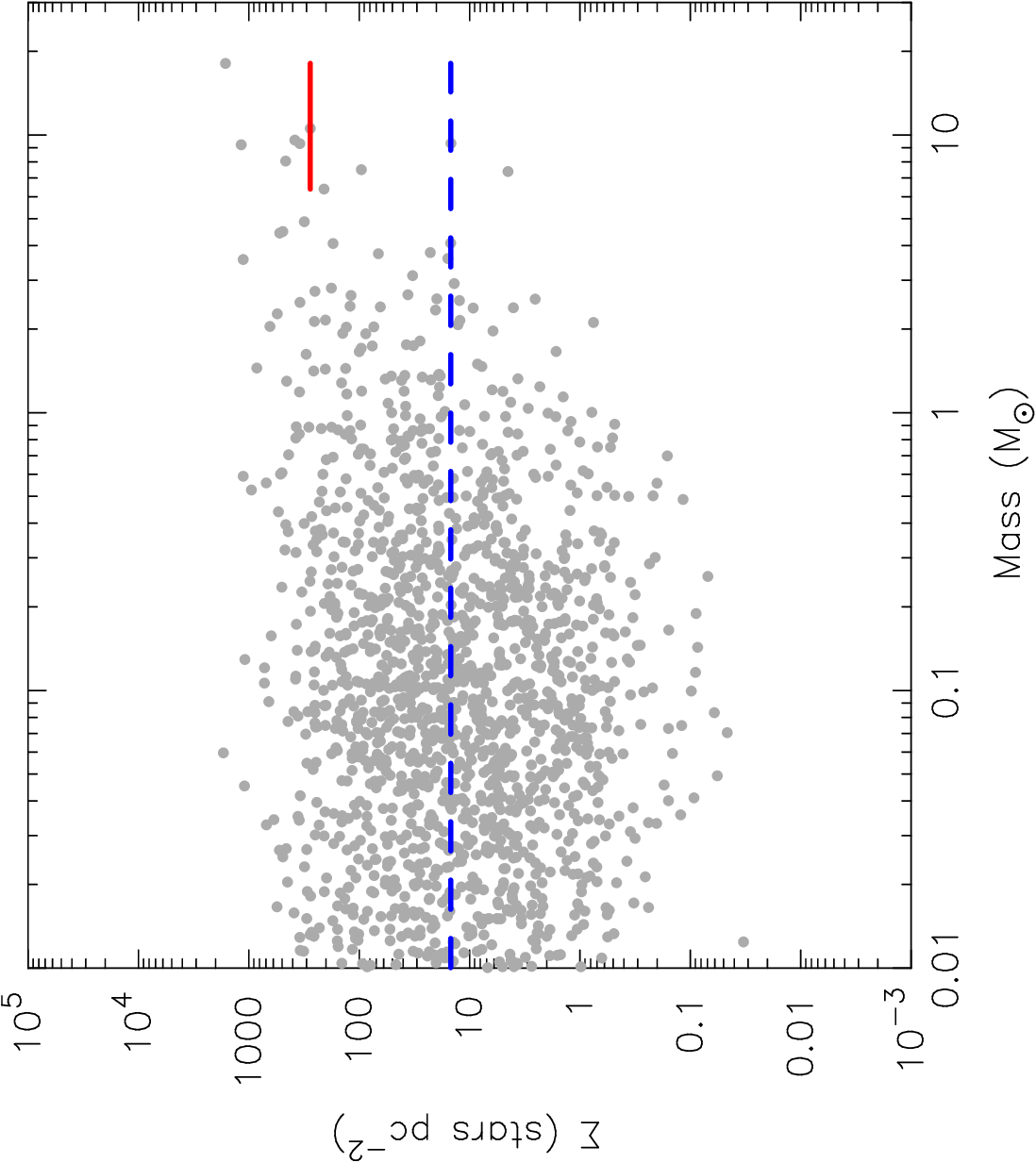}}}
\hspace*{0.3cm} 
\subfigure[$\Sigma - m$, 10\,Myr]{\label{hot-i}\rotatebox{270}{\includegraphics[scale=0.26]{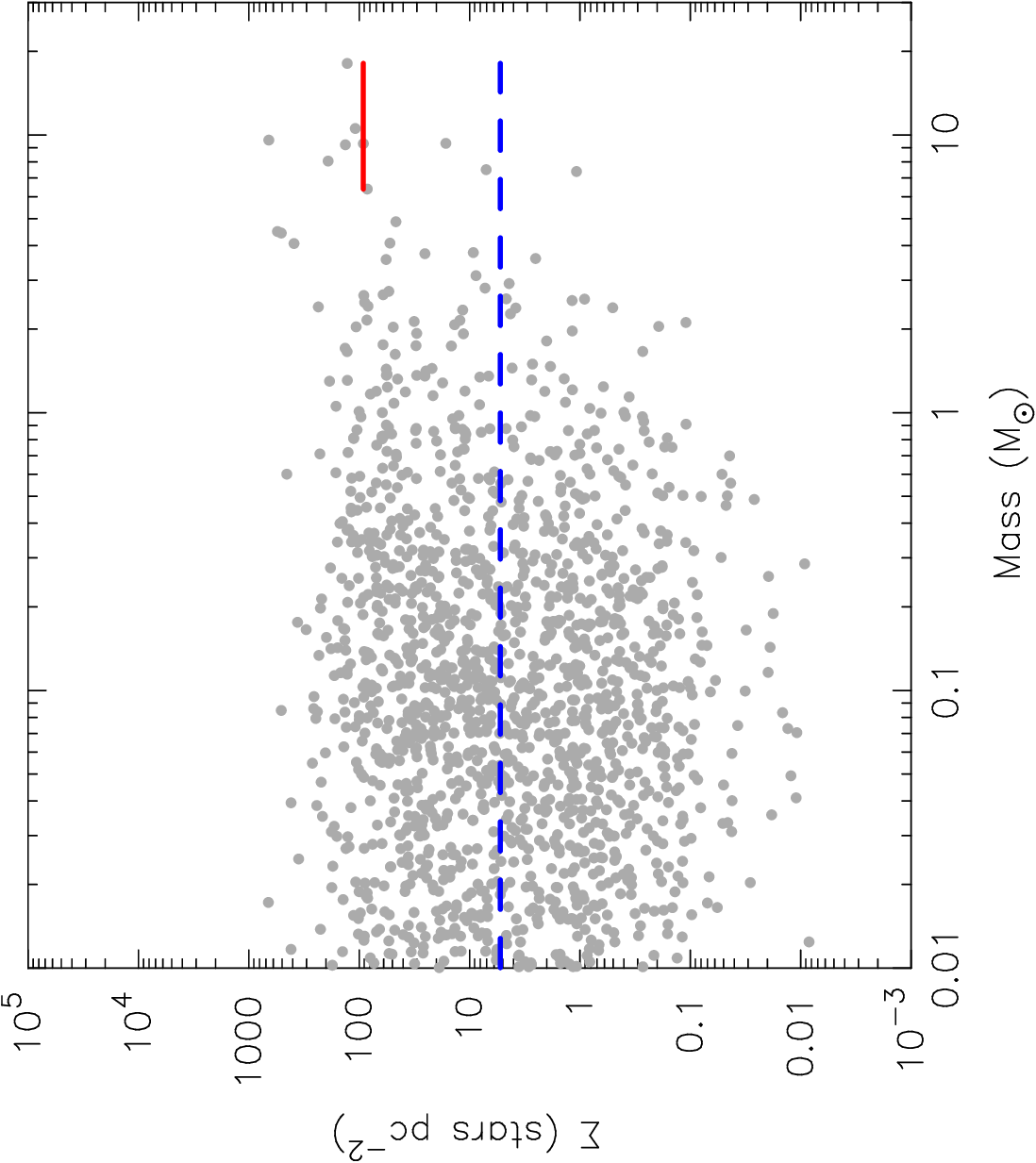}}}
\hspace*{-1.cm}
\subfigure[Evolution of $\Lambda_{\rm MSR}$]{\label{hot-j}\rotatebox{270}{\includegraphics[scale=0.26]{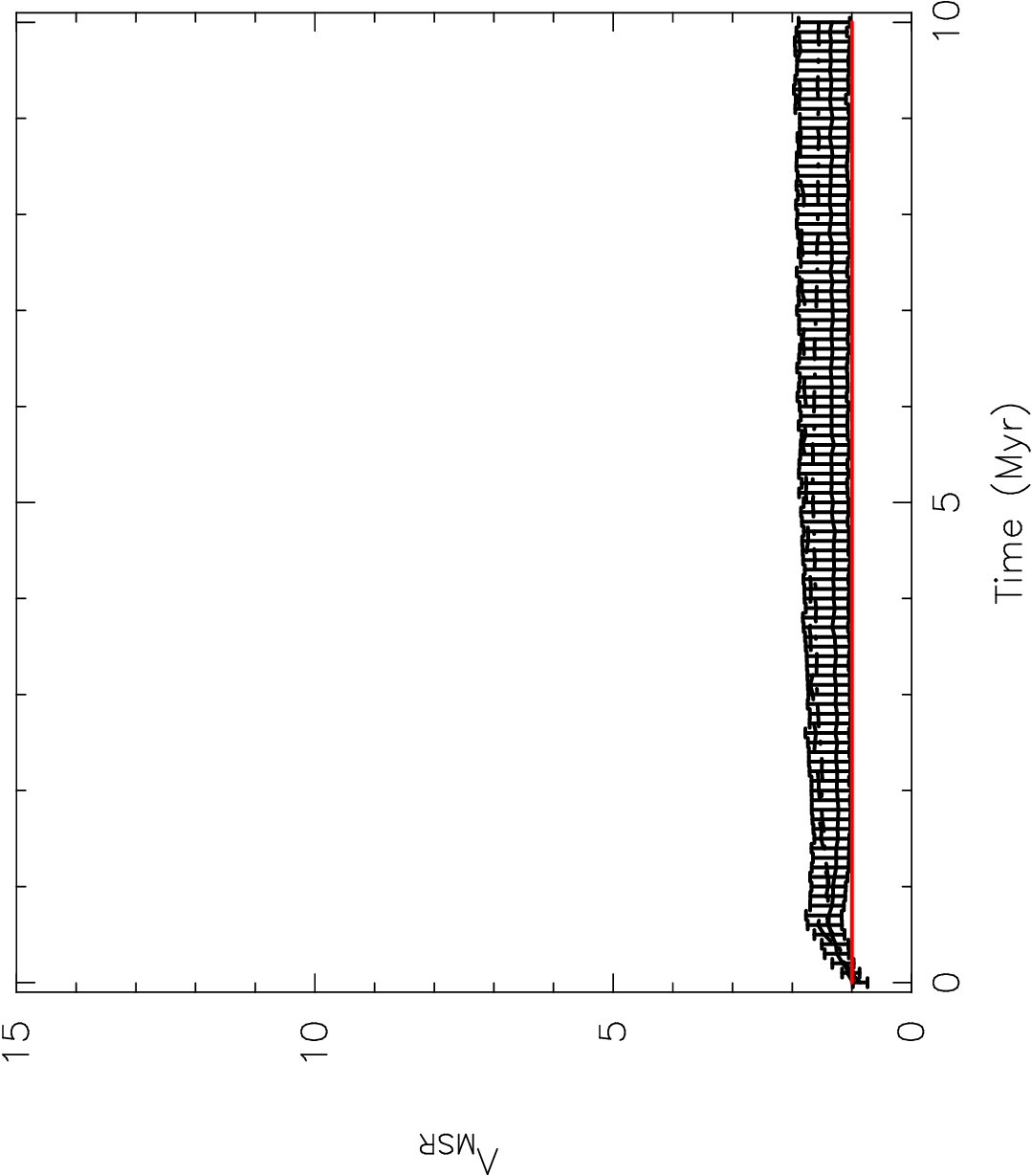}}}
\hspace*{0.3cm} 
\subfigure[Evolution of $\Sigma_{\rm LDR}$]{\label{hot-k}\rotatebox{270}{\includegraphics[scale=0.26]{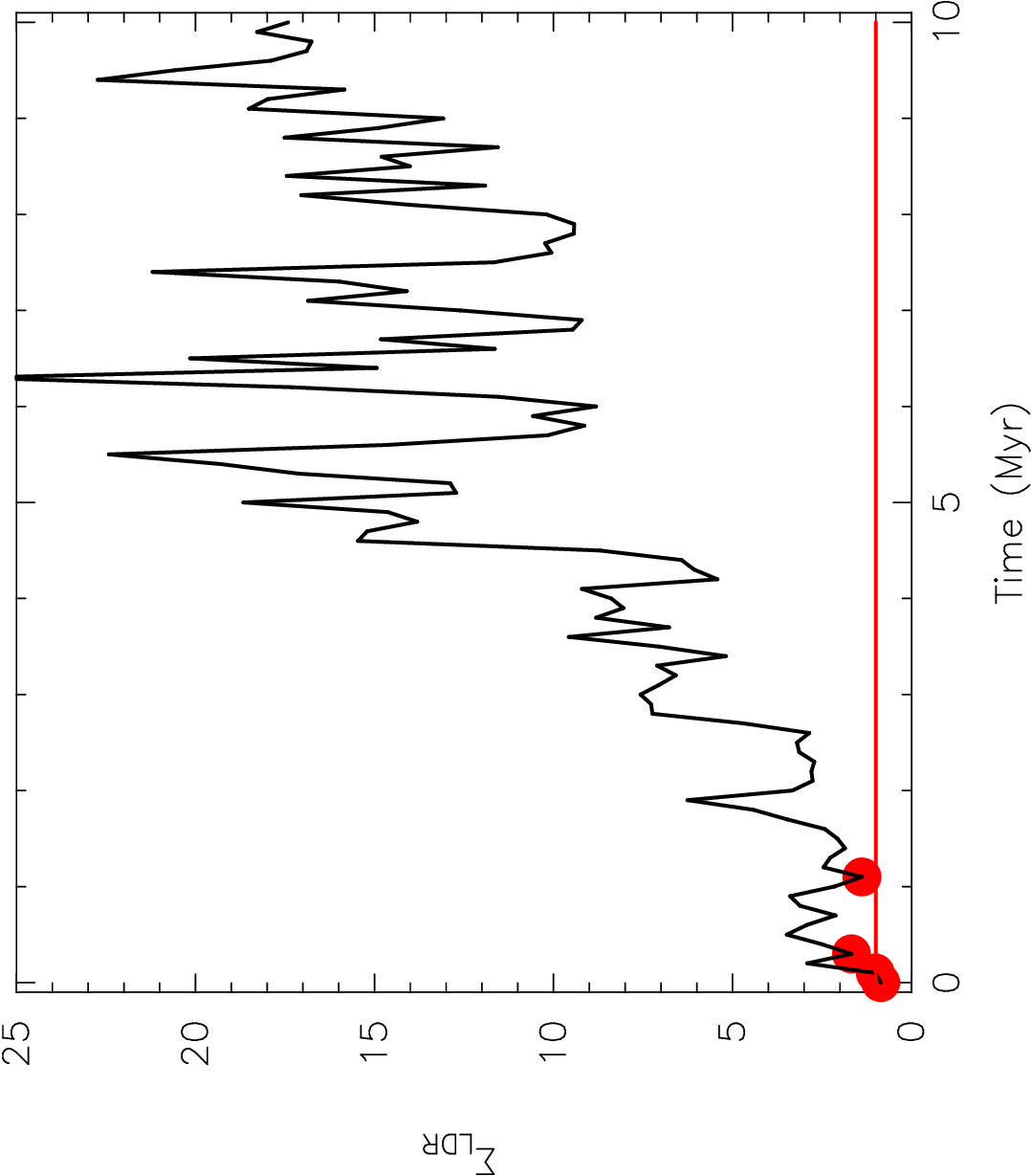}}}
\hspace*{0.3cm} 
\subfigure[Evolution of $\mathcal{Q}$--parameter]{\label{hot-l}\rotatebox{270}{\includegraphics[scale=0.26]{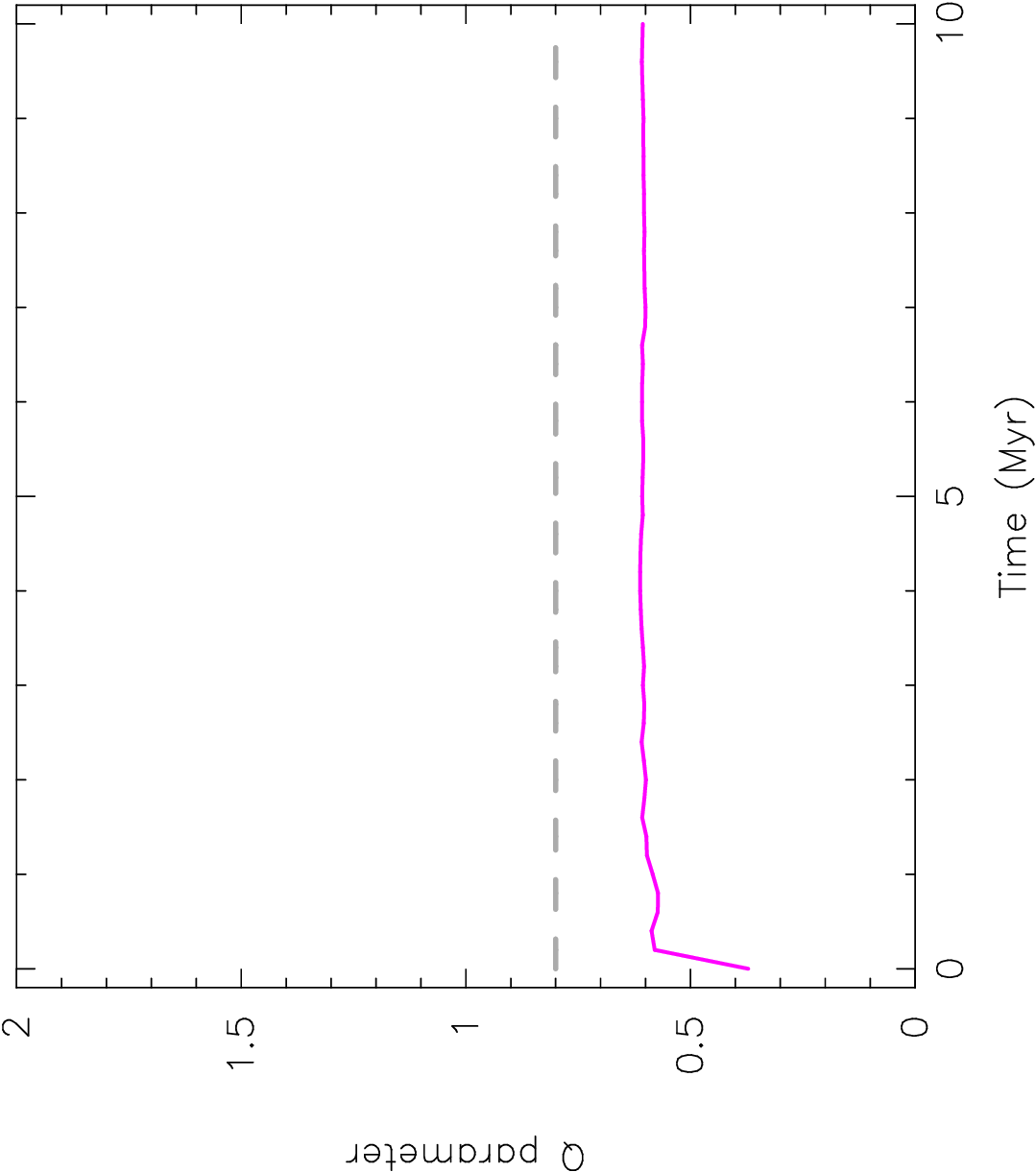}}}
\caption[bf]{Evolution of a supervirial ($\alpha_{\rm vir} = 1.5$), substructured star forming region ($D = 1.6$) with $N = 1500$ stars.  We show the morphology at 0, 5 and 10\,Myr (a -- c), the mass segregation ratio, $\Lambda_{\rm MSR}$ of the 
$N_{\rm MST}$ most massive stars stars at 0, 5 and 10\,Myr (d -- f), and the surface density as a function of stellar mass, $\Sigma - m$ (g -- i) at 0, 5 and 10\,Myr. We also show the evolution 
of $\Lambda_{\rm MSR}$ in panel (j) -- the solid, dashed and dot-dashed lines are the values of $\Lambda_{\rm MSR}$ for the 10, 20 and 50 most massive stars respectively; the ratio of the surface density of the 10 most massive stars to the median surface density of all stars in the region ($\Sigma_{\rm LDR} = \tilde{\Sigma}_{10}/\tilde{\Sigma}_{\rm all}$ -- panel k) and the evolution of the $\mathcal{Q}$--parameter  (panel l), with time.  }
\label{hot_indiv}
  \end{center}
\end{figure*}

In Fig.~\ref{hot_indiv} we show the evolution of a `typical' 
$N = 1500$ stars, supervirial ($\alpha_{\rm vir} = 1.5$), substructured ($D = 1.6$) star
forming region. As in Fig.~\ref{cool_indiv}, in panels (a) -- (c) we show the
morphology at 0, 5 and 10\,Myr respectively.  The most massive stars
(shown by the red triangles) are initially randomly placed in the
substructured fractal. The global motion of the region causes it to
expand and after 5\,Myr two distinct subclusters have formed,
separated by a distance of $\sim$20\,pc.  However, there are also
stars in between, and the region has evolved into an association-like
complex. The association expands further until the simulation end-time
at 10\,Myr. 

We measure $\Lambda_{\rm MSR}$ for this region; in panel (d) the mass
segregation ratio as a function of the number of stars in the minimum
spanning tree, $N_{\rm MST}$ is shown at 0\,Myr (i.e.\,\,before any
dynamical  evolution has occured). Unlike the subvirial, collapsing
fractal shown in Fig.~\ref{cool_indiv}, this supervirial, expanding
fractal does not show any evidence of dynamical mass segregation at
5~or~10\,Myr (panels (e) and (f), respectively).  This is unsurprising
as the massive stars have no opportunity to become concentrated
together. 

Note that this is not simply due to the inclusion of `outliers' in the determination of $\Lambda_{\rm MSR}$; 
varying the number of stars in the MST does not change the result.

In panel (g) we show the
local surface density of every star in the region as a function of the
star's mass (the grey points). The median stellar surface density for
the entire region   is indicated by the blue dashed line in
Fig.~\ref{hot-g}. Initially, the median surface density is
$\tilde{\Sigma}_{\rm all} = 5052$\,stars\,pc$^{-2}$, whereas the 10
most massive stars have a median surface density of
$\tilde{\Sigma}_{10}  = 4356$\,stars\,pc$^{-2}$ as shown by the solid
red line. This difference is not significant --  a two dimensional KS
test returns a p-value of 0.58 that the two subsets share the same
parent distribution. However, the region massive stars in the region subsequently 
attain much higher local densities than the average star in the region. After 5\,Myr
(panel h) the most massive stars have a median surface density of
$\tilde{\Sigma}_{10}  = 277$\,stars\,pc$^{-2}$  compared to the region
median value of $\tilde{\Sigma}_{\rm all} = 15$\,stars
\,pc$^{-2}$. Now the KS test between these two subsets returns a
p-value of $< 10^{-3}$, indicating a significant difference. After 10\,Myr the 10 most massive stars again have a
higher surface density $\tilde{\Sigma}_{10} = 92$\,stars\,pc$^{-2}$,
whereas the median surface density for the whole association is
$\tilde{\Sigma}_{\rm all} = 5$\,stars\,pc$^{-2}$.

The evolution of $\Lambda_{\rm MSR}$ for the duration of the
simulation is shown in Fig.~\ref{hot-j}. The value of $\Lambda_{\rm
  MSR}$ for the 10 most massive stars is shown by the solid line (with
error bars), and $\Lambda_{\rm MSR}$ for the 20 and 50 most massive
stars is shown by the dashed and dot-dashed lines,  respectively. We
plot $\Lambda_{\rm MSR} = 1$, i.e.\,\,no mass segregation, shown by
the horizontal red line. It is apparent that the region does not show
any mass segregation according to the definition of $\Lambda_{\rm
  MSR}$. 
 
Conversely, the strong difference in the density distribution of the massive stars from the $\Sigma -m$
measure at 5\,Myr (Fig.~\ref{hot-h})  and 10\,Myr (Fig.~\ref{hot-i})
is apparent throughout the lifetime of the region. In Fig.~\ref{hot-k}
we show the evolution of the surface density of the 10 most  massive
stars divided by the surface density for the whole region
($\Sigma_{\rm LDR} = \tilde{\Sigma}_{10}/\tilde{\Sigma}_{\rm all}$)
as a function of time. We show $\Sigma_{\rm LDR} = 1$, i.e.\,\,no difference in local density as a function of mass, by the horizontal red line. We plot a filled red circle
at times when the difference  between the most massive stars and every
star in the region is not significant -- which occurs at 0, 0.25 and
1.2\,Myr. However, after this time the  $\Sigma_{\rm LDR}$ ratio rises
steadily during the remainder of the simulation.

The very different behaviours of $\Lambda_{\rm MSR}$ and $\Sigma_{\rm
  LDR}$ are because they trace different physical processes.  $\Lambda_{\rm MSR}$ is tracing the relative closeness of
the massive stars {\em to each other}.  As the region expands the
massive stars approximately keep their initial relative distributions
as there is no way that they can know about each other, hence
$\Lambda_{\rm MSR}$ remains low.  But $\Sigma_{\rm LDR}$ measures the
local density of stars around each {\em individual} massive star.  The
increase in $\Sigma_{\rm LDR}$ tells us that whilst the massive stars
know nothing about each other, they do act as a potential well for
nearby low-mass stars to fall into, increasing the local density around
them.   $\Sigma_{\rm LDR}$ continues to increase as the massive stars
build larger and larger `retinues' of low-mass stars.

We plot the evolution of structure in this model (as measured by the
$\mathcal{Q}$--parameter) in Fig.~\ref{hot-l}. The boundary between
substructured and centrally concentrated morphologies ($\mathcal{Q} =
0.8$) is shown by the grey dashed line. The $\mathcal{Q}$--parameter
rises rapidly from  its original value of $\mathcal{Q} = 0.4$, but
then remains at $\mathcal{Q} \simeq 0.6$ for the remainder of the
simulation, indicating that the region is still substructured (as is
evident in the morphology at 5 and 10\,Myr (panels (b) and (c),
respectively).  In the initial rise in $\mathcal{Q}$ is due to the dispersion of some
local substructure, and mergers between some nearby regions of
substructure.  


\subsection{Evolution of all regions}

The models presented in Figs.~\ref{cool_indiv}~and~\ref{hot_indiv}
were chosen as `typical' examples from each suite of 20 initially {\em
  statistically} identical simulations. However, the dynamical evolution of clusters
can be highly stochastic; \citet{Allison10}  showed that statistically 
identical initial conditions can result in very different evolution. 
With this in mind, it is essential to consider ensembles of 
simulations in order to examine the spread in evolution and outcomes. 

\subsubsection{Evolution of the $\mathcal{Q}$-parameter}

\begin{figure*}
  \begin{center}
\setlength{\subfigcapskip}{10pt}
\vspace*{-0.3cm}
\hspace*{-1.cm}\subfigure[$\alpha_{\rm vir} = 0.3, D = 1.6$]{\label{Qpar-a}\rotatebox{270}{\includegraphics[scale=0.27]{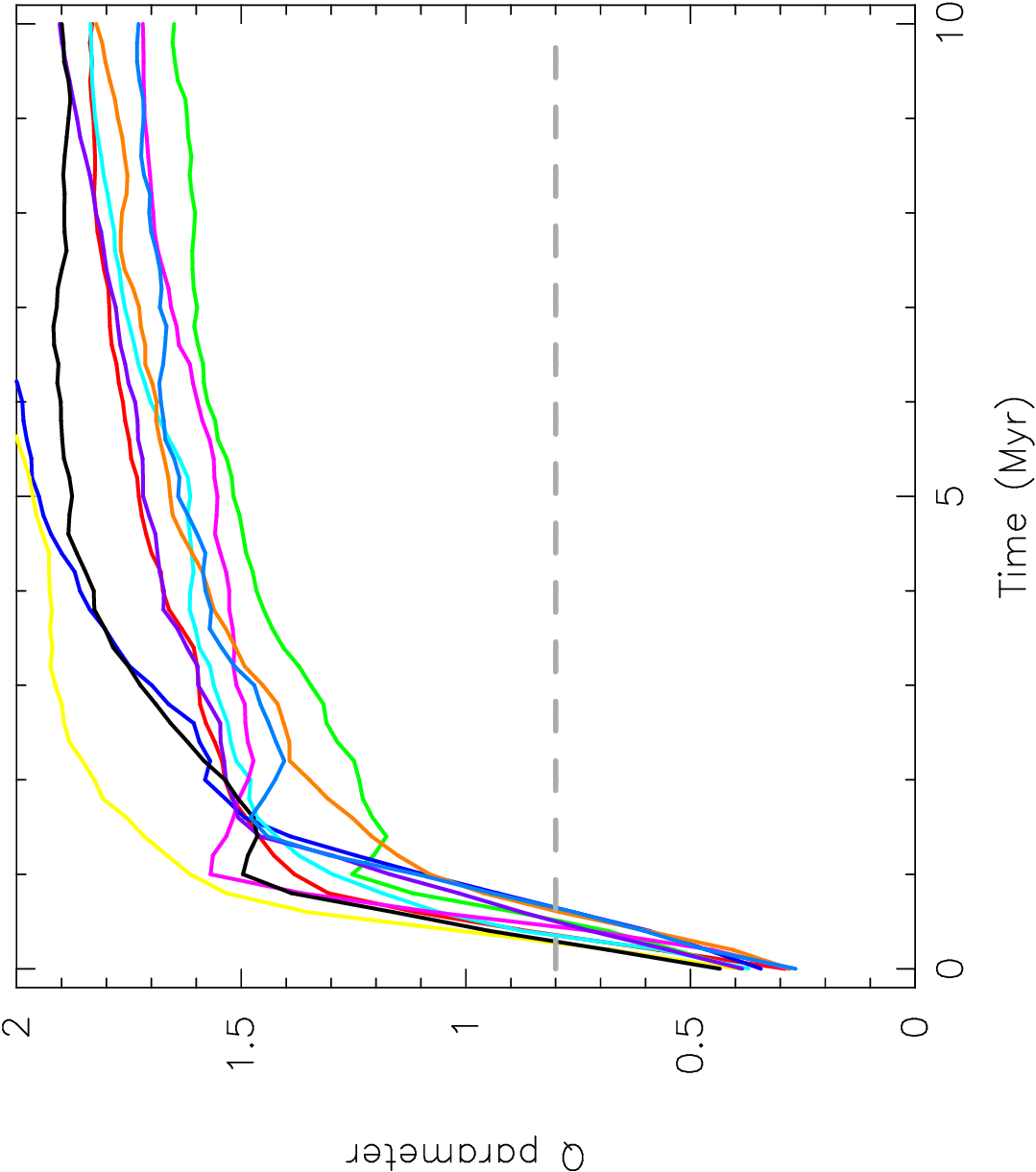}}}
\hspace*{0.3cm} 
\subfigure[$\alpha_{\rm vir} = 0.3, D = 2.0$]{\label{Qpar-b}\rotatebox{270}{\includegraphics[scale=0.27]{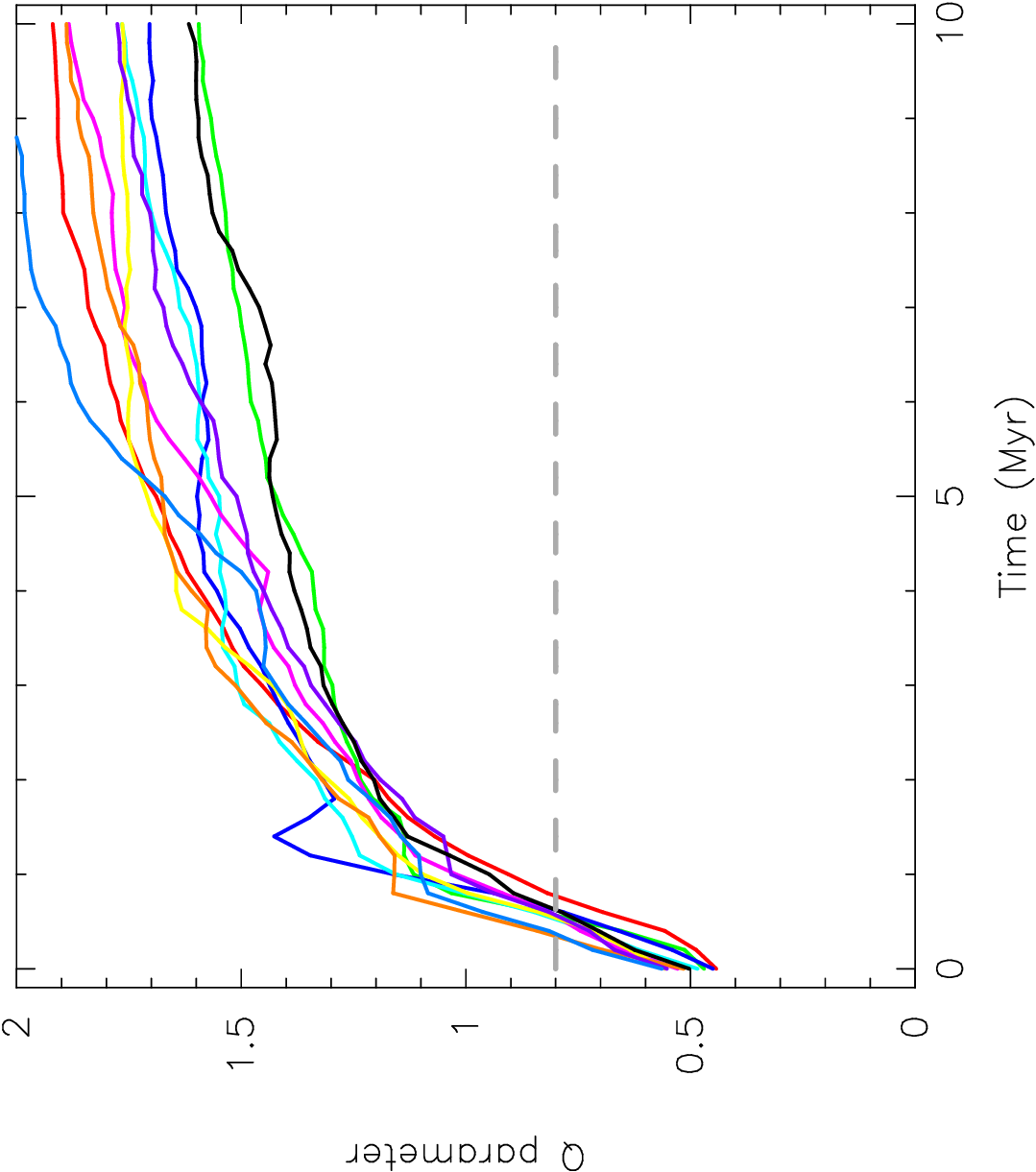}}}
\hspace*{0.3cm} 
\subfigure[$\alpha_{\rm vir} = 0.3, D = 3.0$]{\label{Qpar-c}\rotatebox{270}{\includegraphics[scale=0.27]{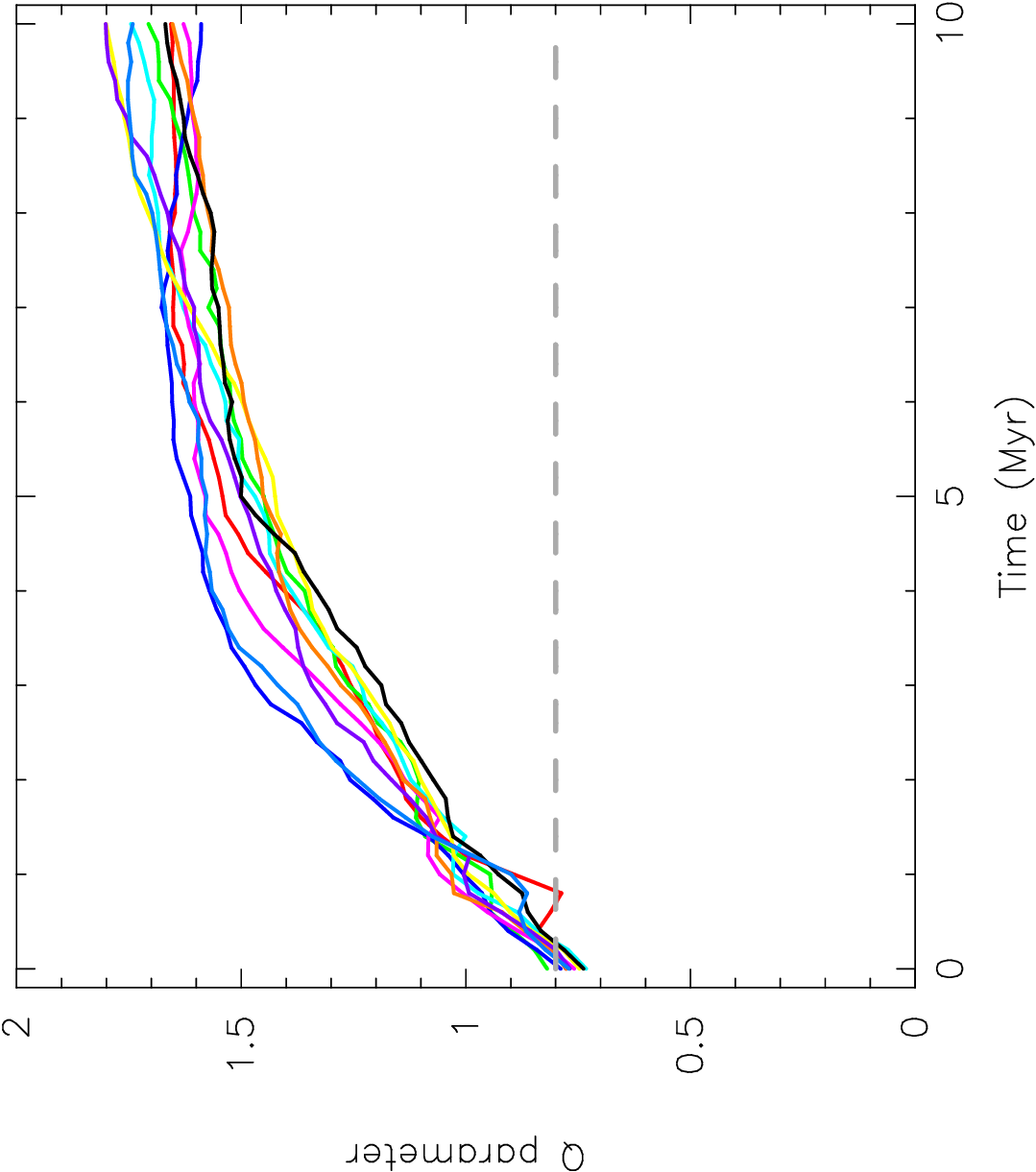}}}
\hspace*{-1.cm}
\subfigure[$\alpha_{\rm vir} = 0.5, D = 1.6$]{\label{Qpar-d}\rotatebox{270}{\includegraphics[scale=0.27]{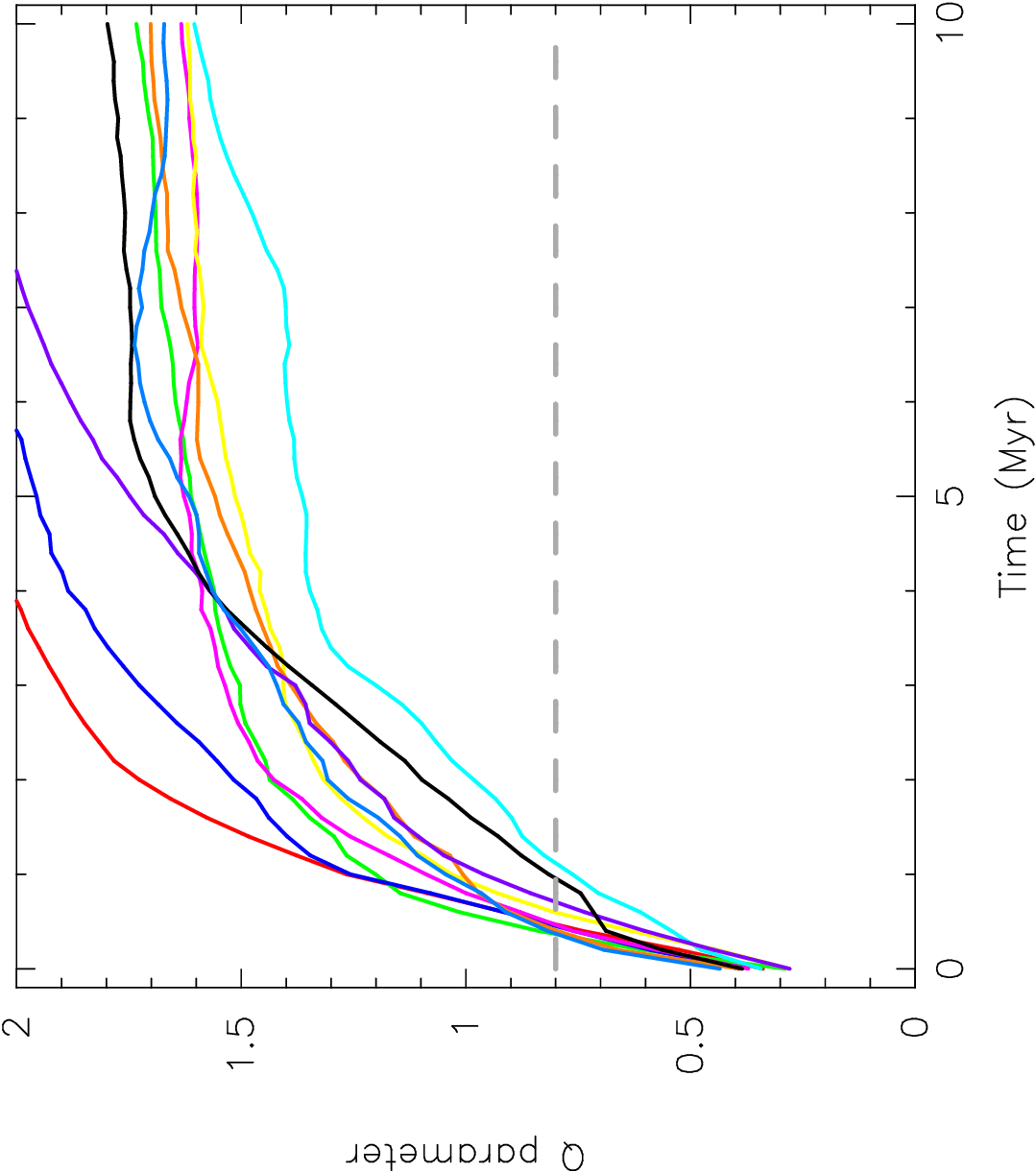}}}
\hspace*{0.3cm} 
\subfigure[$\alpha_{\rm vir} = 0.5, D = 2.0$]{\label{Qpar-e}\rotatebox{270}{\includegraphics[scale=0.27]{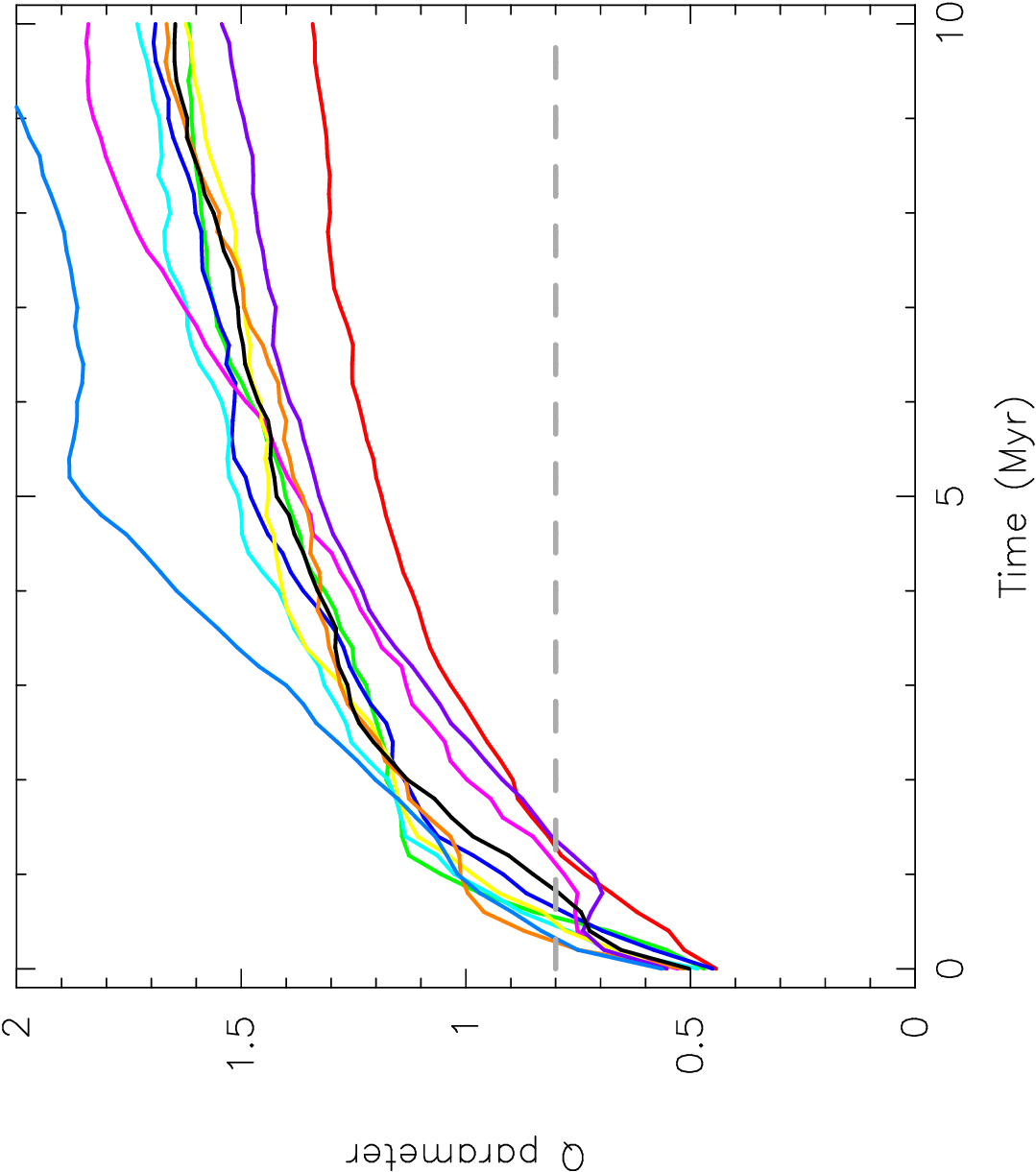}}}
\hspace*{0.3cm} 
\subfigure[$\alpha_{\rm vir} = 0.5, D = 3.0$]{\label{Qpar-f}\rotatebox{270}{\includegraphics[scale=0.27]{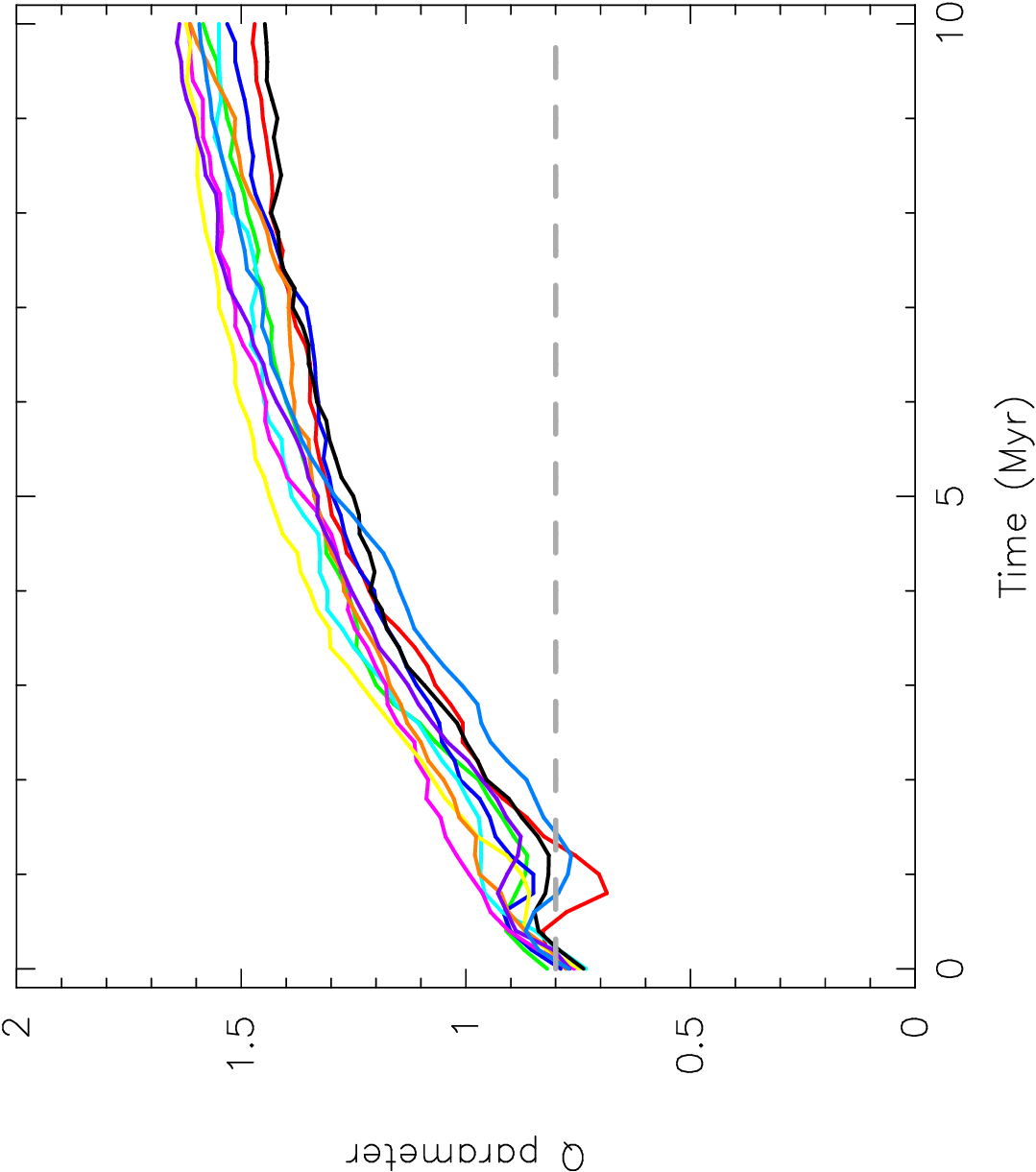}}}
\hspace*{-1.cm}
\subfigure[$\alpha_{\rm vir} = 1.5, D = 1.6$]{\label{Qpar-g}\rotatebox{270}{\includegraphics[scale=0.27]{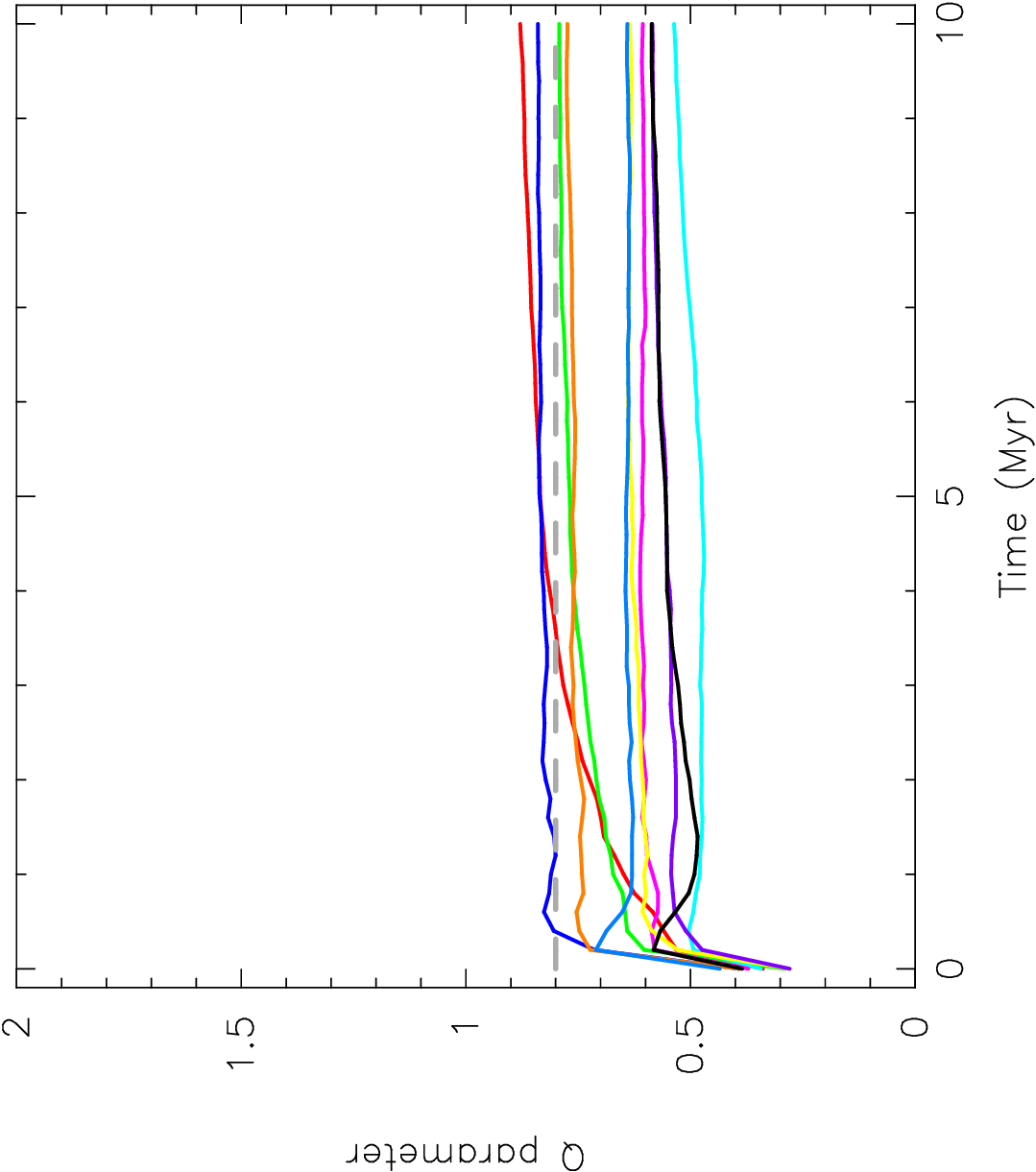}}}
\hspace*{0.3cm} 
\subfigure[$\alpha_{\rm vir} = 1.5, D = 2.0$]{\label{Qpar-h}\rotatebox{270}{\includegraphics[scale=0.27]{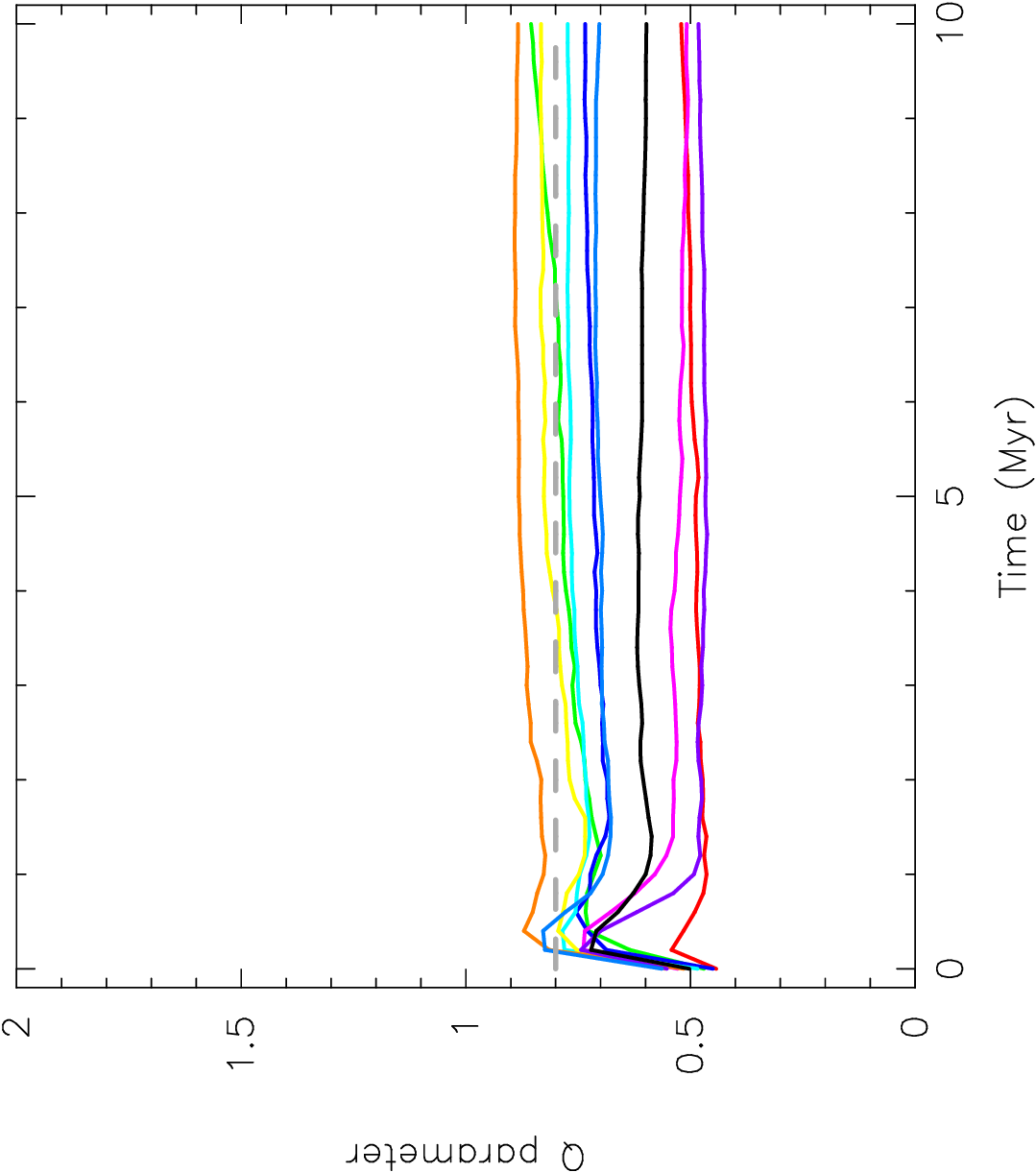}}}
\hspace*{0.3cm} 
\subfigure[$\alpha_{\rm vir} = 1.5, D = 3.0$]{\label{Qpar-i}\rotatebox{270}{\includegraphics[scale=0.27]{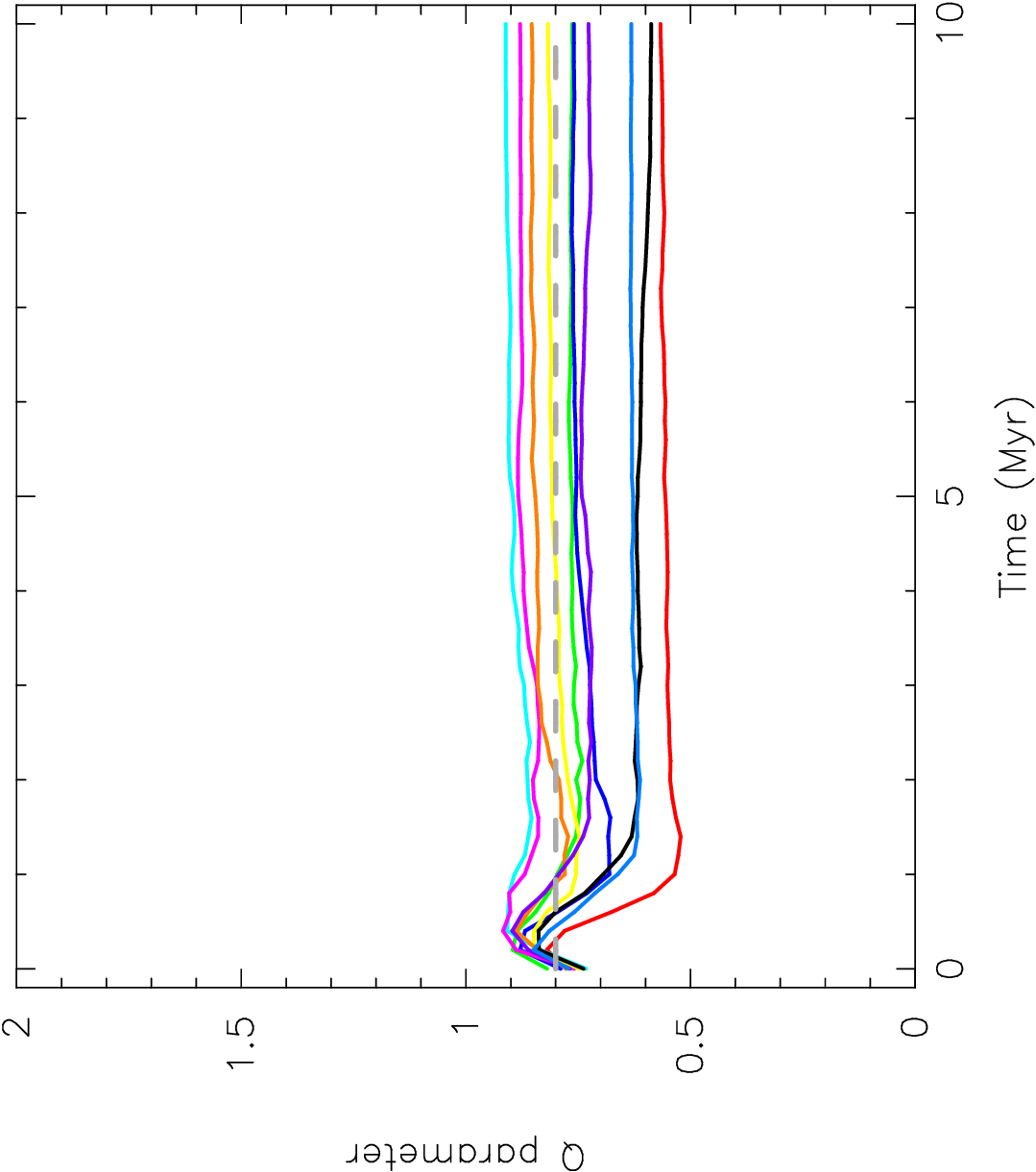}}}
\caption[bf]{Evolution of the $\mathcal{Q}$--parameter with time for the simulations with $N = 1500$ stars. For clarity, we only show the first 10 simulations in each suite (we consider the $\mathcal{Q}$--parameter from all 20 simulations at several specific times in 
Figs.~\ref{Q_MSR}~and~\ref{Q_Sig}).  }
\label{Qpar_all}
  \end{center}
\end{figure*}

We first consider the evolution of spatial structure, as measured by
the $\mathcal{Q}$--parameter, in each of our nine suites of
simulations ($N = 1500$ stars and $\alpha_{\rm vir} = 0.3, 0.5, 1.5$ paired with $D = 1.6, 2.0,
3.0$). In Fig.~\ref{Qpar_all} we show the evolution of the
$\mathcal{Q}$--parameter with time -- the boundary between
substructured and radially smooth clusters ($\mathcal{Q} = 0.8$) is
shown by the dashed grey line. For clarity, we only show the first 10
simulations (rather than all 20), but later we will include the
$\mathcal{Q}$--parameter from every simulation at specific times in
Figs.~\ref{Q_MSR}~and~\ref{Q_Sig}. 

If we compare the cool regions (Figs.~\ref{Qpar-a}--\ref{Qpar-c}) with
the tepid regions (Figs.~\ref{Qpar-d}--\ref{Qpar-f}), we see that
substructure is erased on similar timescales  (i.e.\,\,$\mathcal{Q}$
becomes greater than 0.8 within the first 1\,Myr).  This shows the
erasure of substructure on roughly a crossing time \citep[see e.g.][]{Goodwin04a}.

On average, the
cool regions (top row panels (a)--(c)) become more centrally
concentrated (i.e.\,\,they reach higher
values of $\mathcal{Q}$) than  the tepid regions (middle row, panels 
(d)--(f)).  Similarly, regions with more initial substructure
($D=1.6$ in panels (a) and (d) in the first column) become more centrally
concentrated than initially fairly smooth regions ($D=3.0$ in panels
(c) and (f) in the third column).  This is due to the lower energy of
the cool regions, and greater potential energy stored in substructure
in the clumpy regions allowing a deeper collapse \citep[e.g.][]{Allison09b,Allison10}. 

However, the spread
between the 10 simulations with identical initial conditions is so
large that it becomes difficult to distinguish between different
initial fractal dimensions and virial ratios for any {\em individual}
region using the $\mathcal{Q}$--parameter alone.  All sets of initial
conditions can produce systems with $\mathcal{Q} = 1.5$ after 10~Myr.
But it is worth noting that values of $\mathcal{Q}$ near 2 are only
formed from initially substructured initial conditions.

In contrast, all supervirial regions ($\alpha_{\rm vir} = 1.5$ on the
bottom row, panels (g)--(i)) keep a low ($\leq 0.8$) value of the
$\mathcal{Q}$--parameter.  In the case of initially substructured
regions (panels (g) and (h)) this is because they expand and so are
unable to erase their substructure \citep[see][]{Goodwin04a}.

Interestingly, several of the initally smooth ($D = 3.0$, panel (i))
regions develop substructure as they expand ($\mathcal{Q}$ falls).
This is due to the way the initial conditions are implimented.  Whilst
the $D = 3.0$ systems are roughly a uniform density sphere they do contain
velocity substructure inherited from their `parents' (see Section~\ref{method}).  However, unless
one believes that fairly uniform density regions would form with
uncorrelated and well mixed velocities, this may well not be
particularly unphysical.

\subsubsection{Evolution of $\Lambda_{\rm MSR}$}

\begin{figure*}
  \begin{center}
\setlength{\subfigcapskip}{10pt}
\vspace*{-0.3cm}
\hspace*{-1.cm}\subfigure[$\alpha_{\rm vir} = 0.3, D = 1.6$]{\label{msr_med-a}\rotatebox{270}{\includegraphics[scale=0.27]{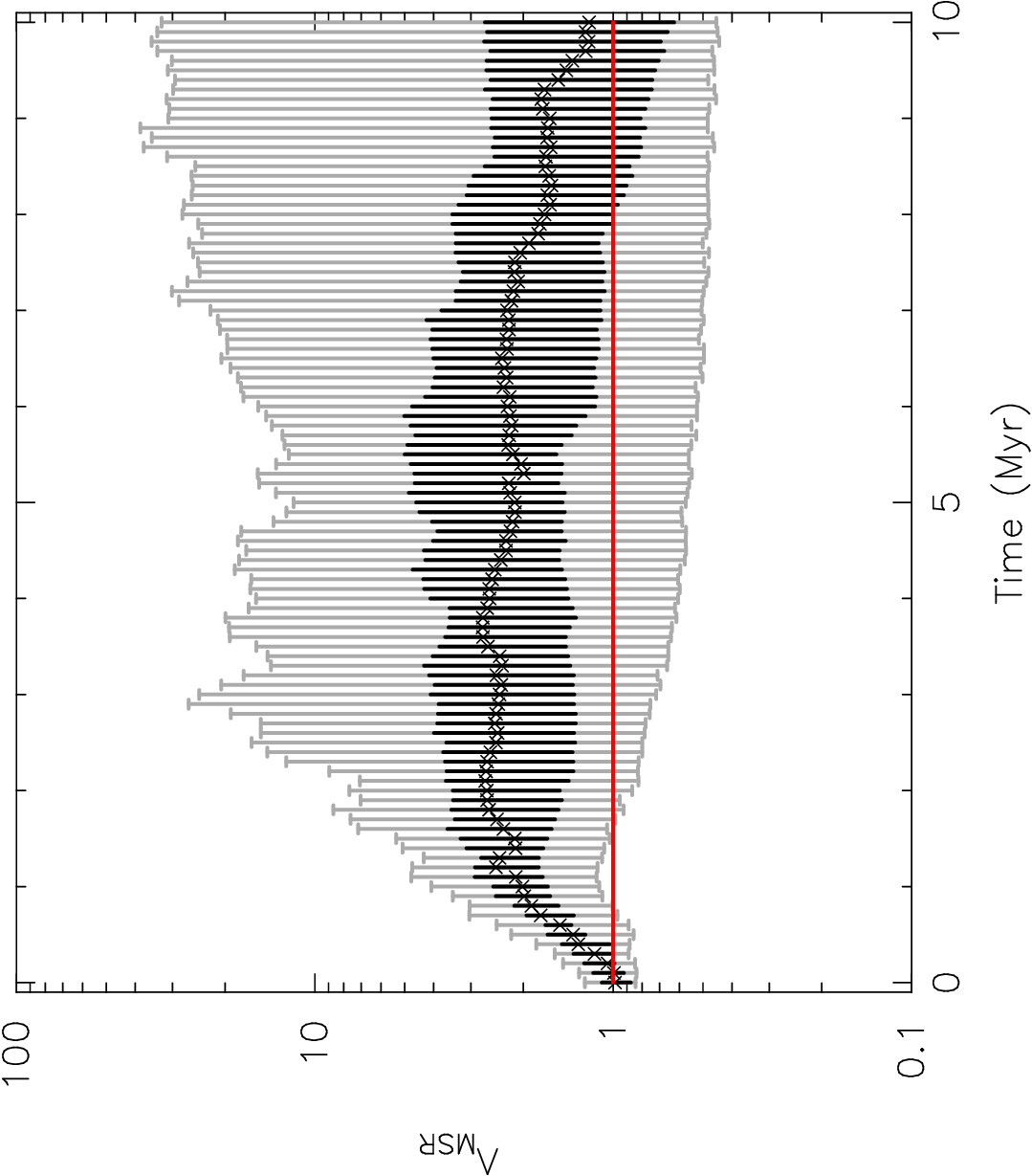}}}
\hspace*{0.3cm} 
\subfigure[$\alpha_{\rm vir} = 0.3, D = 2.0$]{\label{msr_med-b}\rotatebox{270}{\includegraphics[scale=0.27]{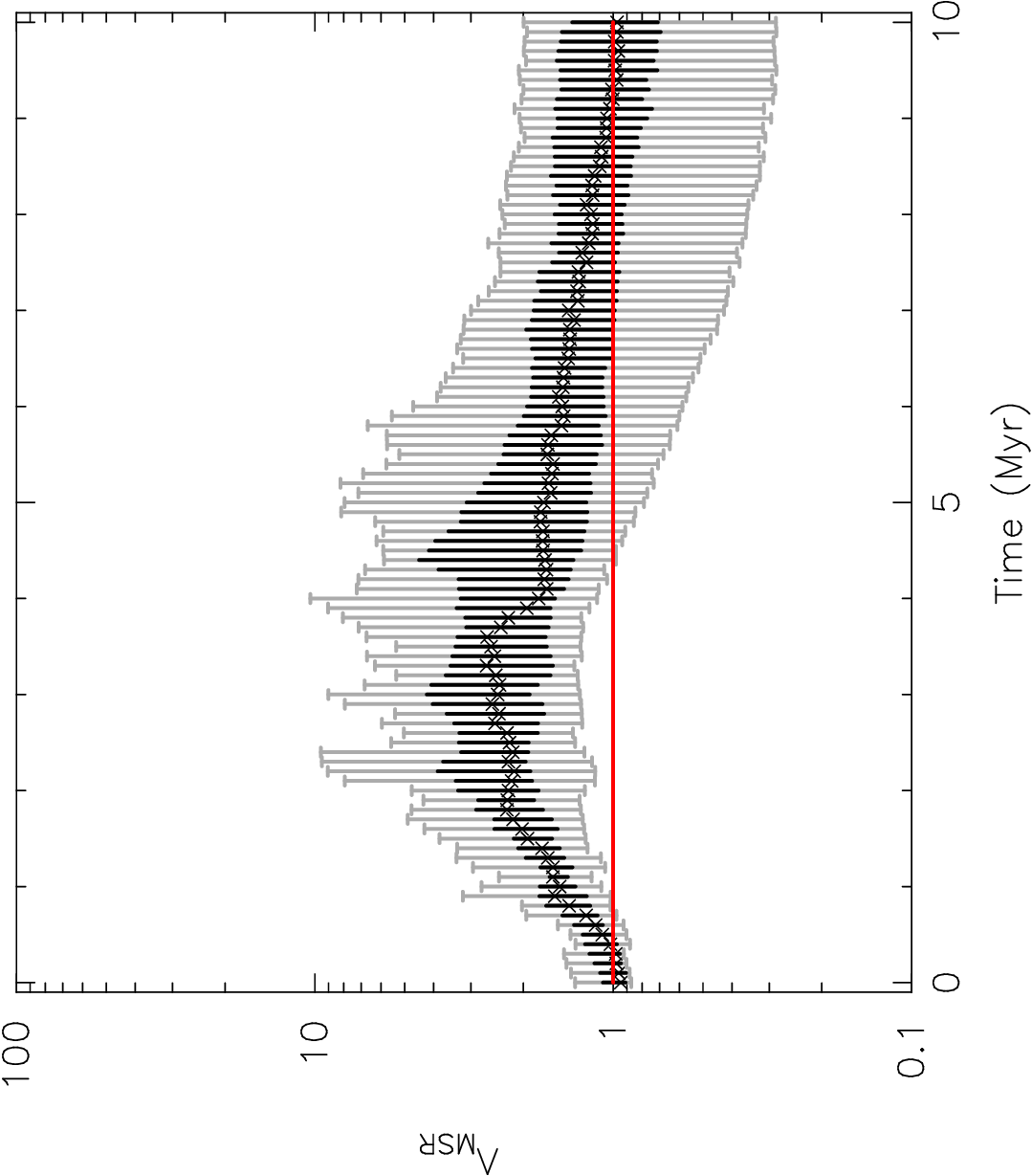}}}
\hspace*{0.3cm} 
\subfigure[$\alpha_{\rm vir} = 0.3, D = 3.0$]{\label{msr_med-c}\rotatebox{270}{\includegraphics[scale=0.27]{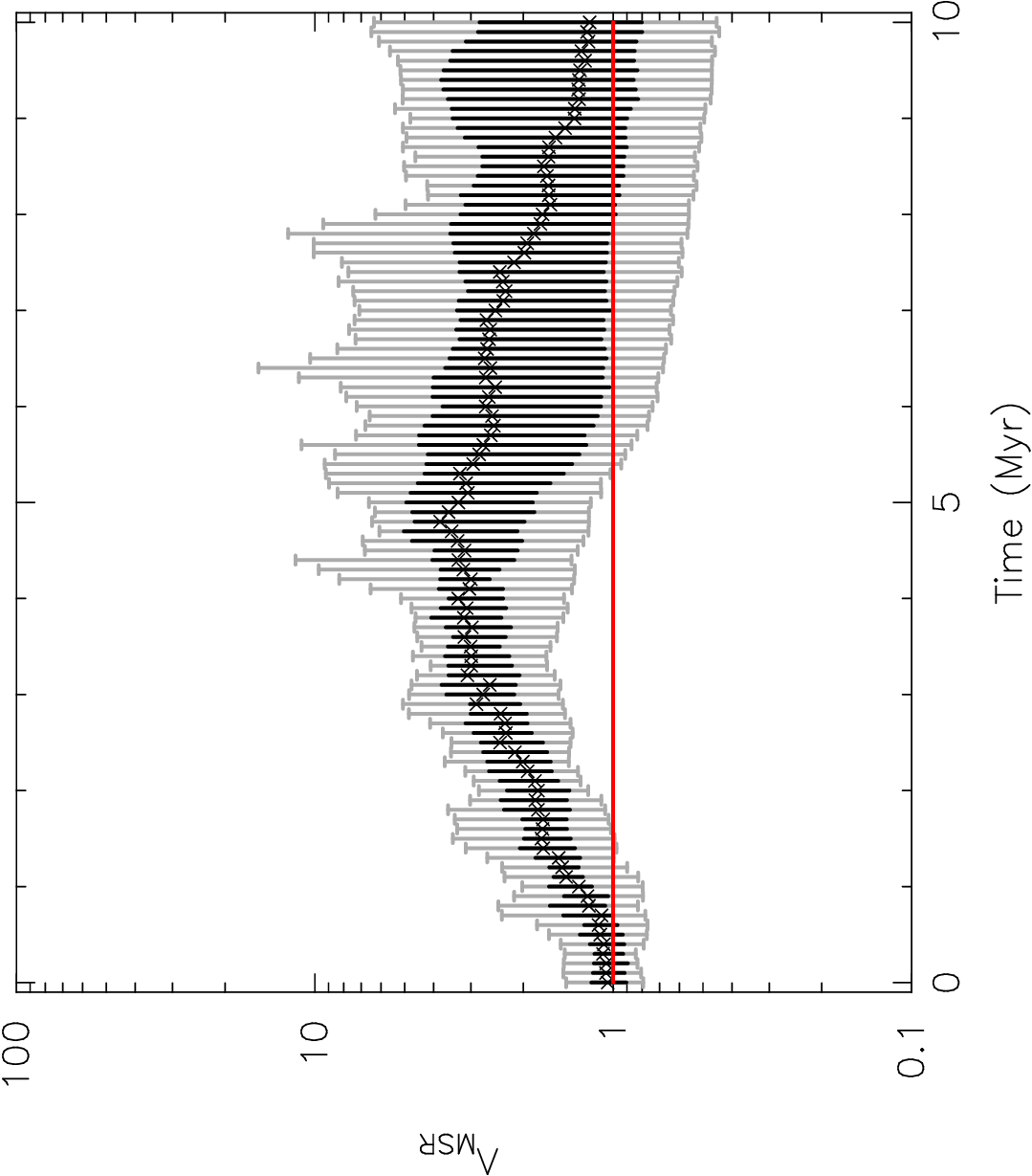}}}
\hspace*{-1.cm}
\subfigure[$\alpha_{\rm vir} = 0.5, D = 1.6$]{\label{msr_med-d}\rotatebox{270}{\includegraphics[scale=0.27]{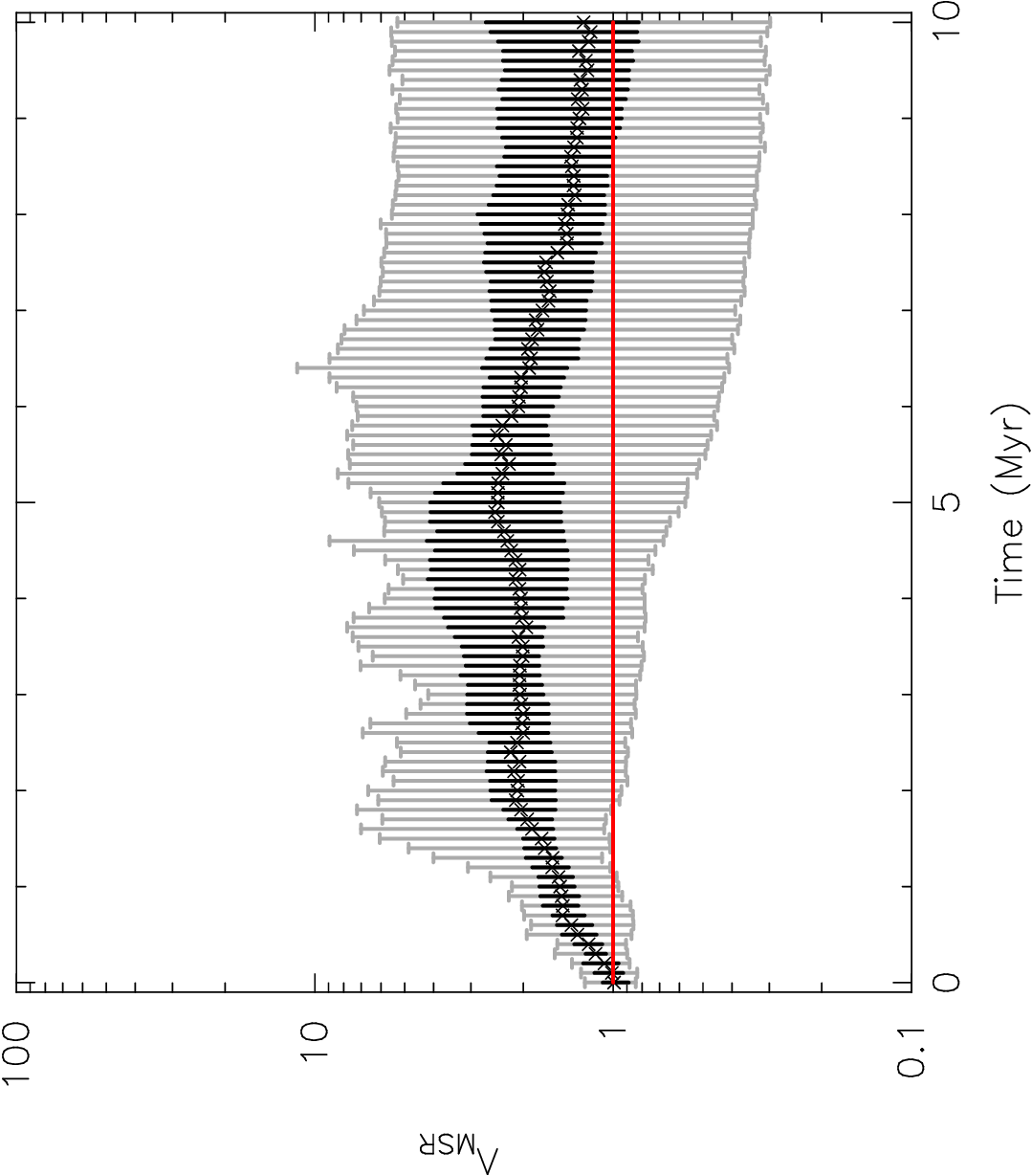}}}
\hspace*{0.3cm} 
\subfigure[$\alpha_{\rm vir} = 0.5, D = 2.0$]{\label{msr_med-e}\rotatebox{270}{\includegraphics[scale=0.27]{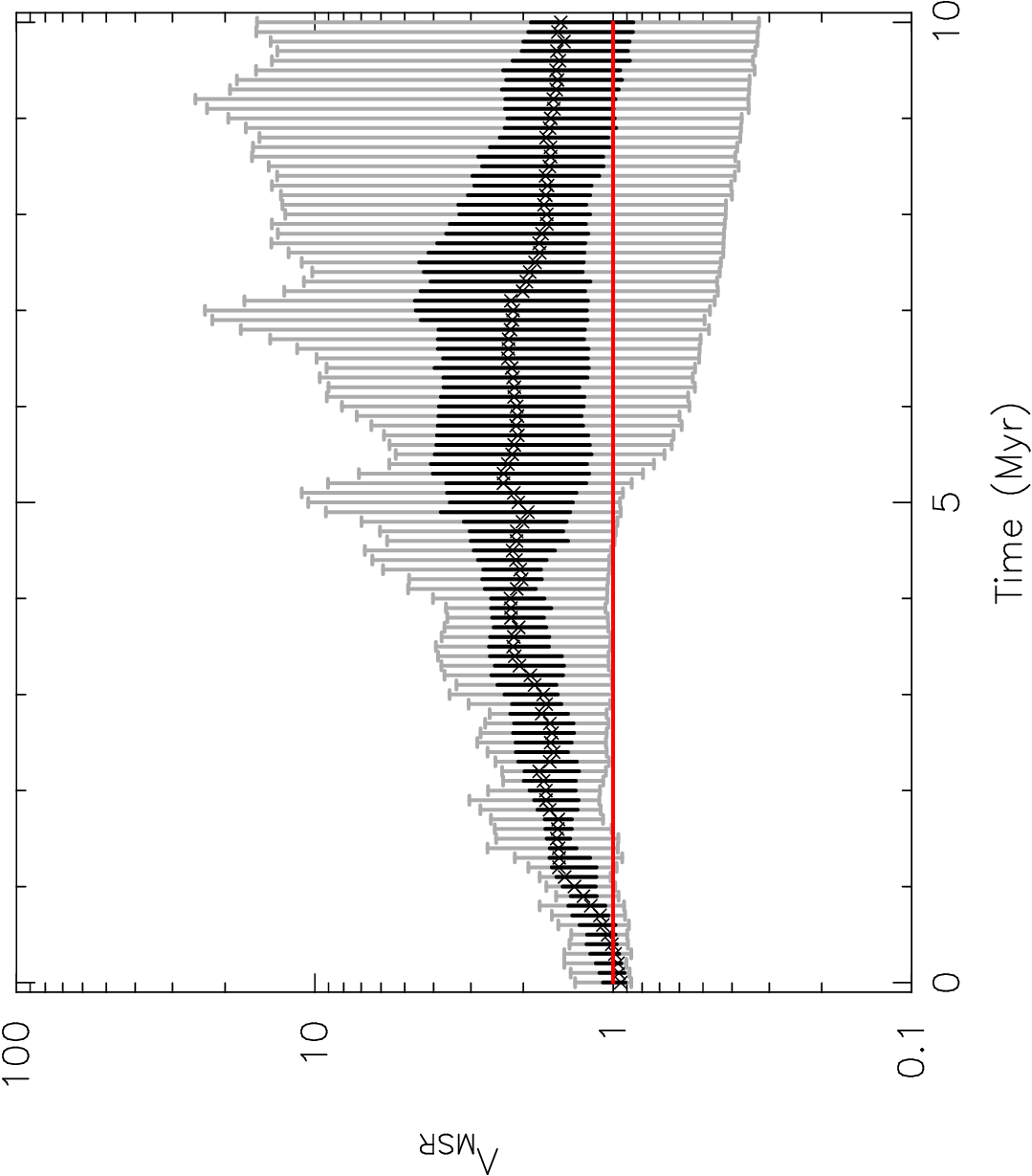}}}
\hspace*{0.3cm} 
\subfigure[$\alpha_{\rm vir} = 0.5, D = 3.0$]{\label{msr_med-f}\rotatebox{270}{\includegraphics[scale=0.27]{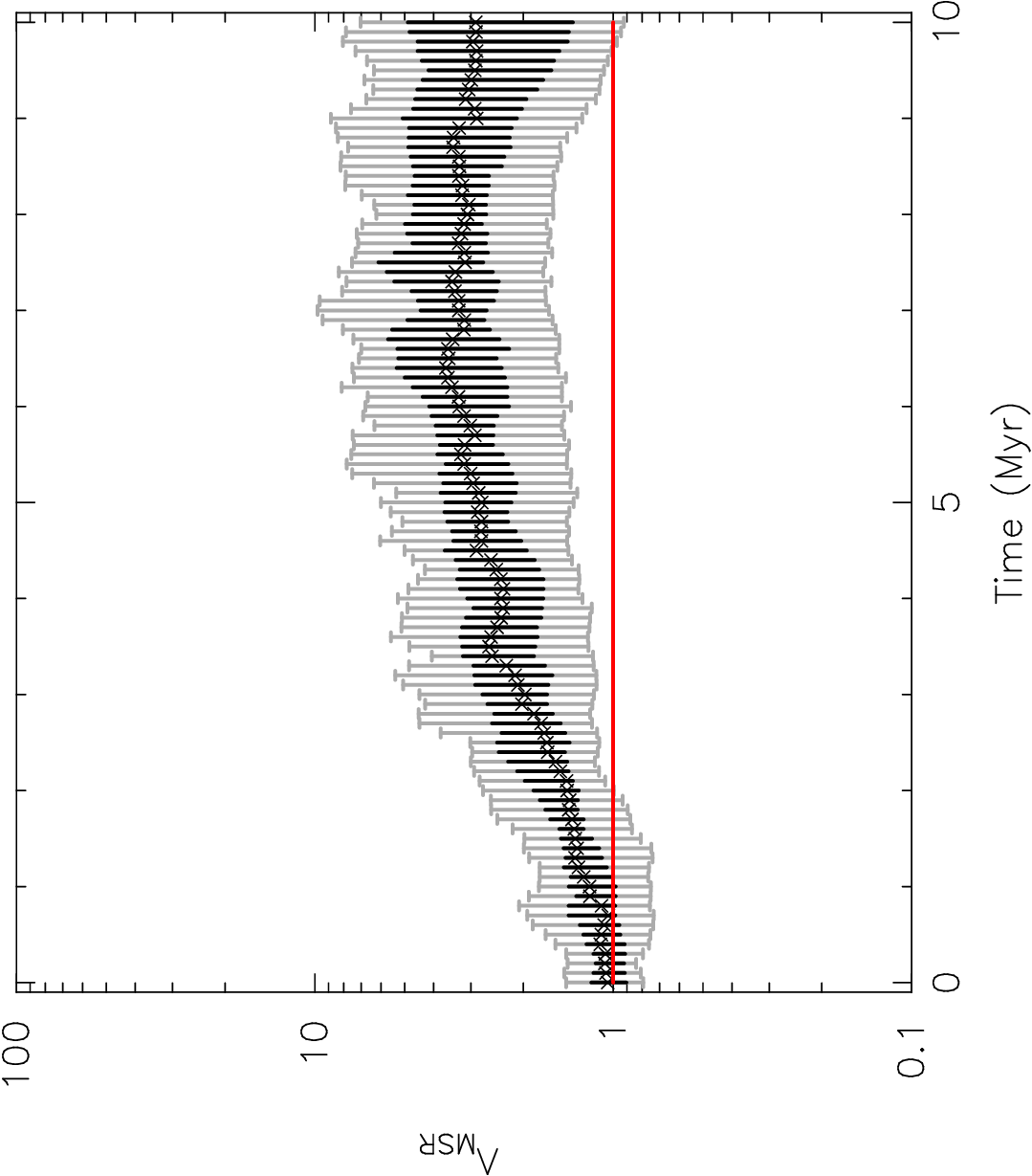}}}
\hspace*{-1.cm}
\subfigure[$\alpha_{\rm vir} = 1.5, D = 1.6$]{\label{msr_med-g}\rotatebox{270}{\includegraphics[scale=0.27]{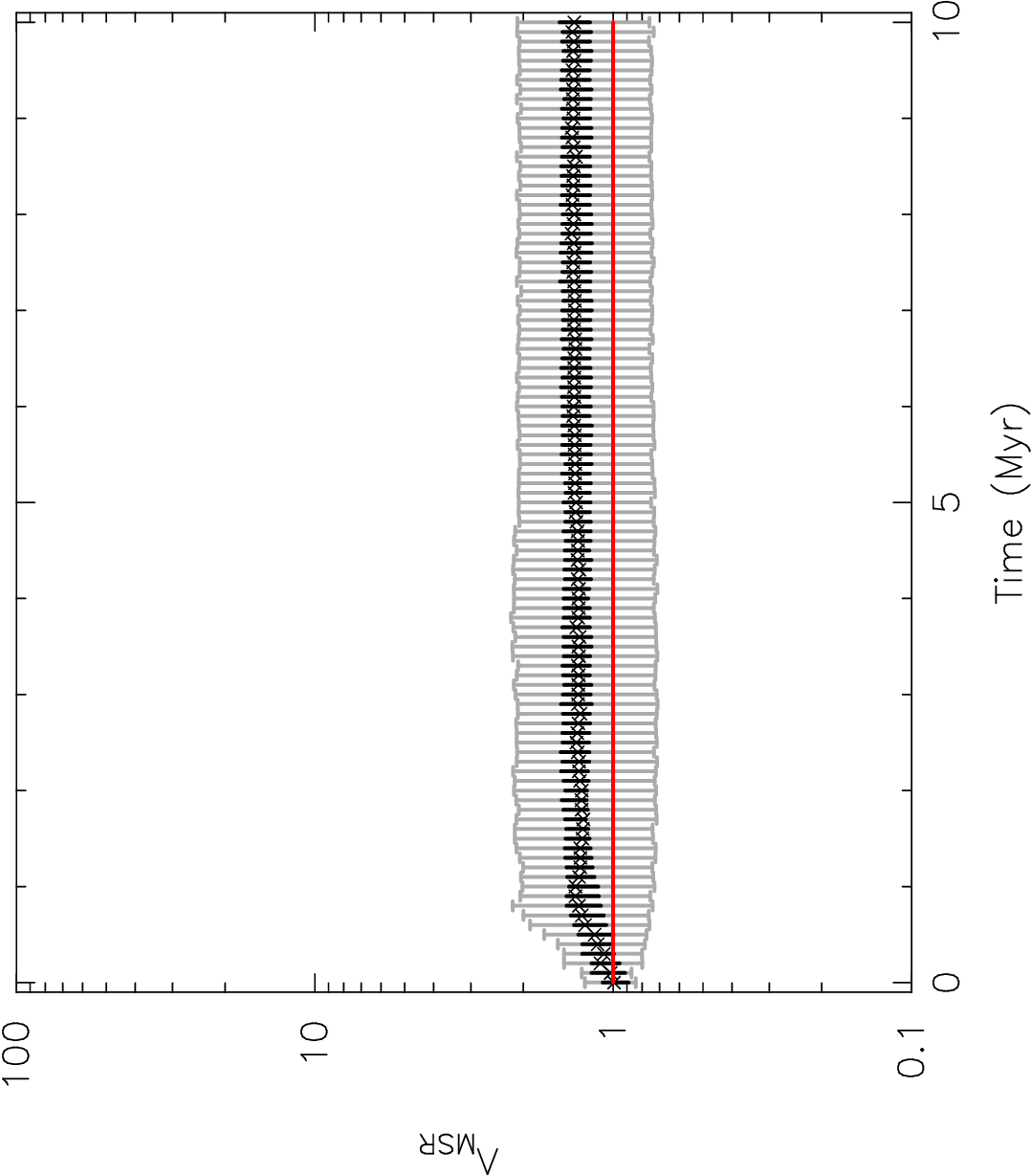}}}
\hspace*{0.3cm} 
\subfigure[$\alpha_{\rm vir} = 1.5, D = 2.0$]{\label{msr_med-h}\rotatebox{270}{\includegraphics[scale=0.27]{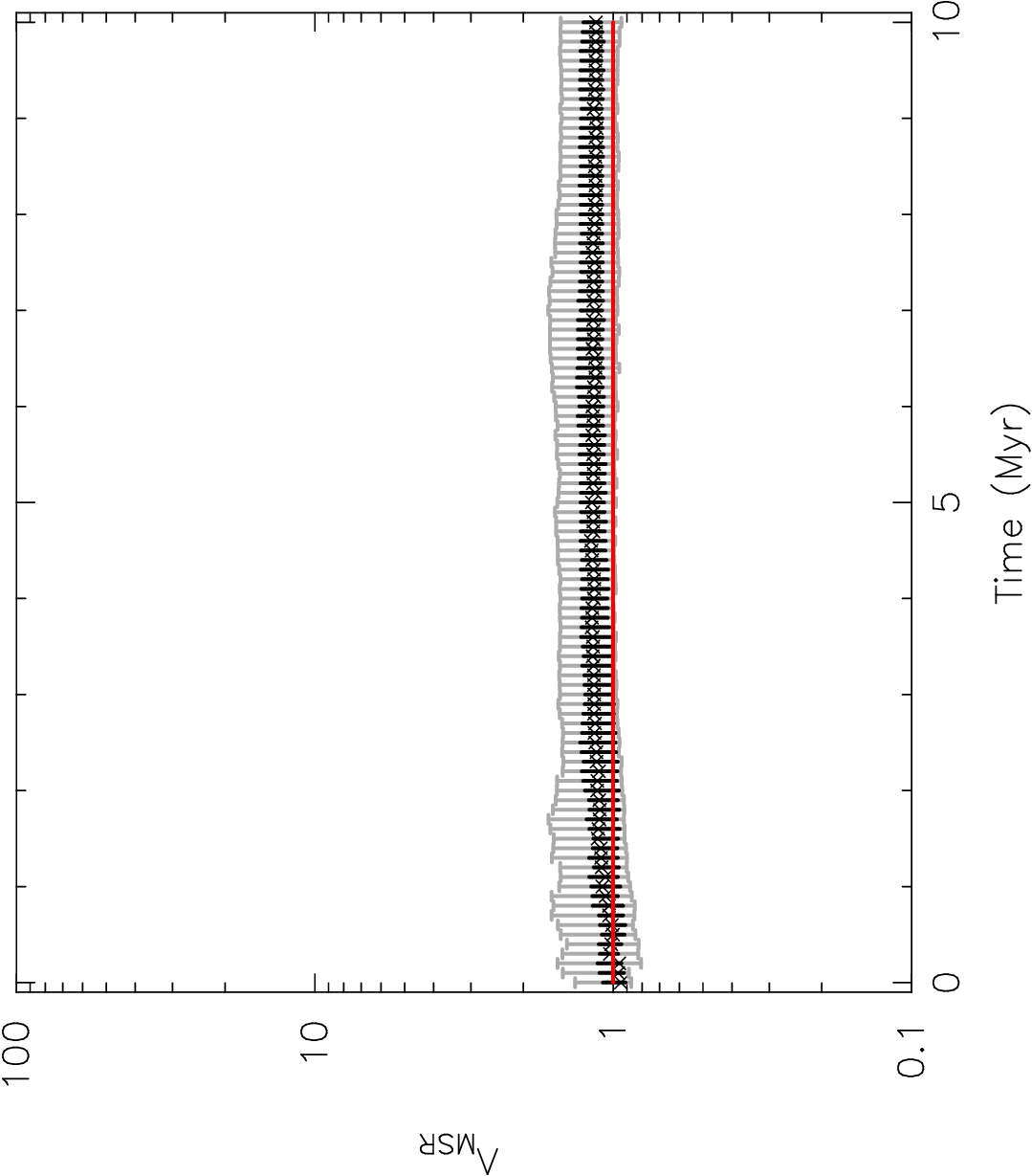}}}
\hspace*{0.3cm} 
\subfigure[$\alpha_{\rm vir} = 1.5, D = 3.0$]{\label{msr_med-i}\rotatebox{270}{\includegraphics[scale=0.27]{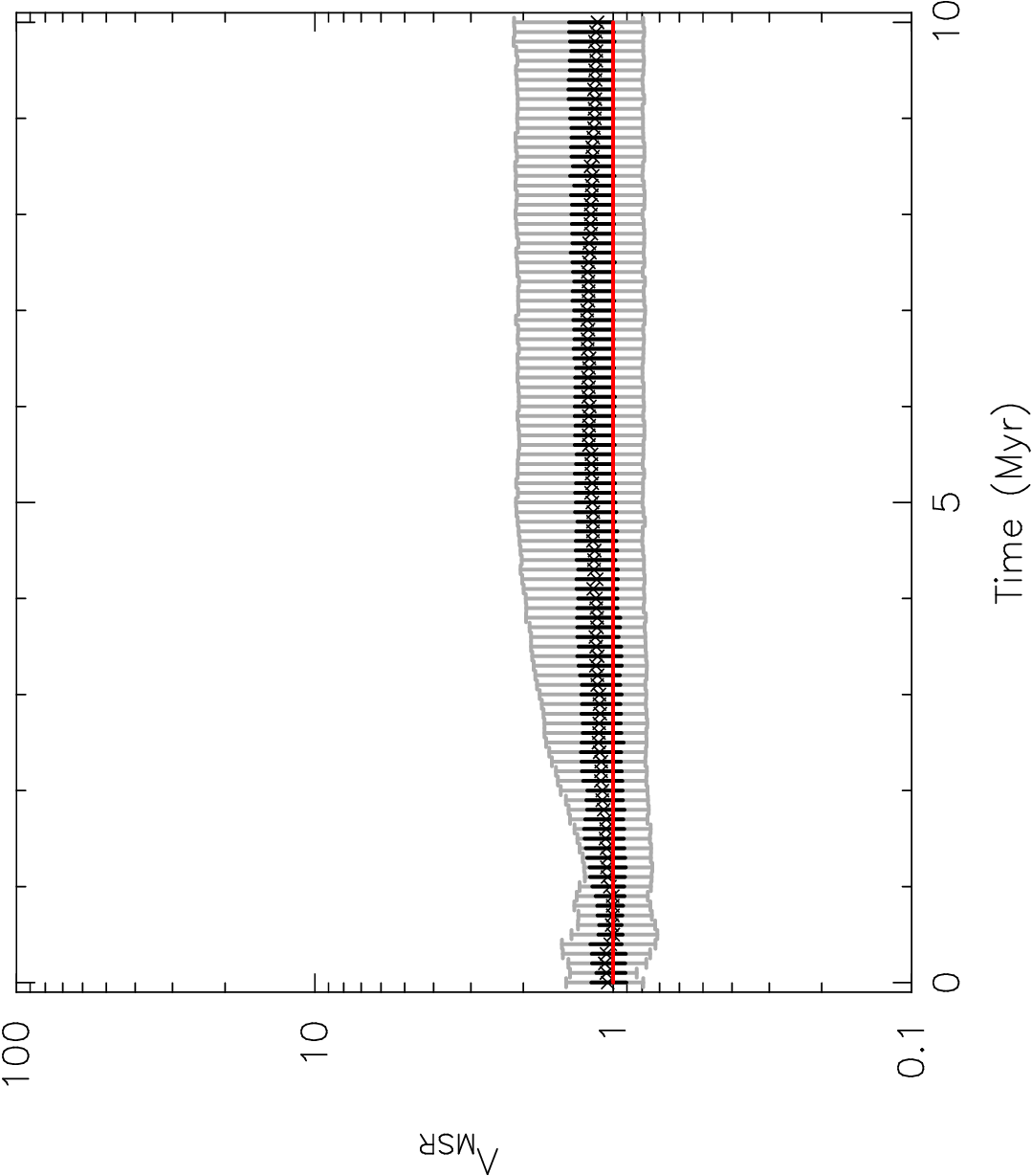}}}
\caption[bf]{Evolution of $\Lambda_{\rm MSR}$ with time for simulations with $N = 1500$. Each panel shows the median $\Lambda_{\rm MSR}$ value of 20 simulations with identical initial conditions (the crosses) and the darker error bars indicate 
25 and 75 percentile values. The entire range of possible values from the 20 sets of initial conditions is shown by the lighter error bars. }
\label{msr_med}
  \end{center}
\end{figure*}

In Fig.~\ref{msr_med} we show the evolution of the $\Lambda_{\rm MSR}$
(always for the tenth most massive star) mass segregation ratio for 
all nine ensembles of initial conditions in our
parameter space (again, $N = 1500$ stars and $\alpha_{\rm vir} = 0.3, 0.5, 1.5$ paired with 
$D = 1.6, 2.0, 3.0$). 

Unlike in Fig.~\ref{Qpar_all} above, the results are far too variable
to plot each individual simulation in an ensemble without producing a
completely unreadable figure.  Instead, at each time we plot the 
median value of $\Lambda_{\rm MSR}$ in each ensemble by the black
cross.  The range covered by half of the simulations (i.e.\,\,between the
25 and 75 percentiles) are shown by the black bars
(these are {\em not} error bars in the
conventional sense).  The the whole range covered at any time is
indicated by the grey bars.  

It should be noted that the evolution of any single system is not a
simple passage through this range.  Systems with high-$\Lambda_{\rm
  MSR}$ at 1~Myr may have either high- or low-$\Lambda_{\rm MSR}$ by
10~Myr and vice-versa (see Allison et al. 2010 for a fuller
investigation of the evolution of $\Lambda_{\rm MSR}$).

Unsurprisingly, the bottom row of Fig.~\ref{msr_med} for hot regions
shows no sign of mass segregation in $\Lambda_{\rm MSR}$ no matter
what the initial level of substructure.  $\Lambda_{\rm MSR}$ starts at
roughly unity and remains there as the regions expand giving the
massive stars no chance to concentrate in any one place (values of
about 2 are achievable if by chance 2 or 3 massive stars are
initially close to one another).

Cool and tepid regions show a wide variety in the evolution of
$\Lambda_{\rm MSR}$.  Because of the initial conditions every region
starts at $\Lambda_{\rm MSR} = 1$.  As found by \citet{Allison10} 
the evolution of these regions can be very stochastic.  Generally,
regions will relax and collapse (more violently the lower both $D$ and
$\alpha_{\rm vir}$ are).  This leads to an increase in $\Lambda_{\rm
  MSR}$ over the first few Myr as the massive stars come closer
together and are able to dynamically mass segregate (see all panels
(a)--(f)).  But once dynamical mass segregation has occured they can
evolve in many ways.  Especially in low-$D$ and/or low-$\alpha_{\rm
  vir}$ regions hierachical systems of massive stars can form which
may violently decay (resulting in a rapid decline in $\Lambda_{\rm
  MSR}$ as seen above in Fig.~\ref{cool_indiv} for our `typical' cool system), or survive
(resulting in $\Lambda_{\rm MSR}$ remaining high).  Massive binaries
can form that completely disrupt some regions.  In some cases, the
decay of the massive multiple system ejects high-mass stars resulting
in a $\Lambda_{\rm MSR} < 1$ (inverse mass segregation).

\subsubsection{Evolution of $\Sigma_{\rm LDR}$}

\begin{figure*}
  \begin{center}
\setlength{\subfigcapskip}{10pt}
\vspace*{-0.3cm}
\hspace*{-1.cm}\subfigure[$\alpha_{\rm vir} = 0.3, D = 1.6$]{\label{sigm_med-a}\rotatebox{270}{\includegraphics[scale=0.27]{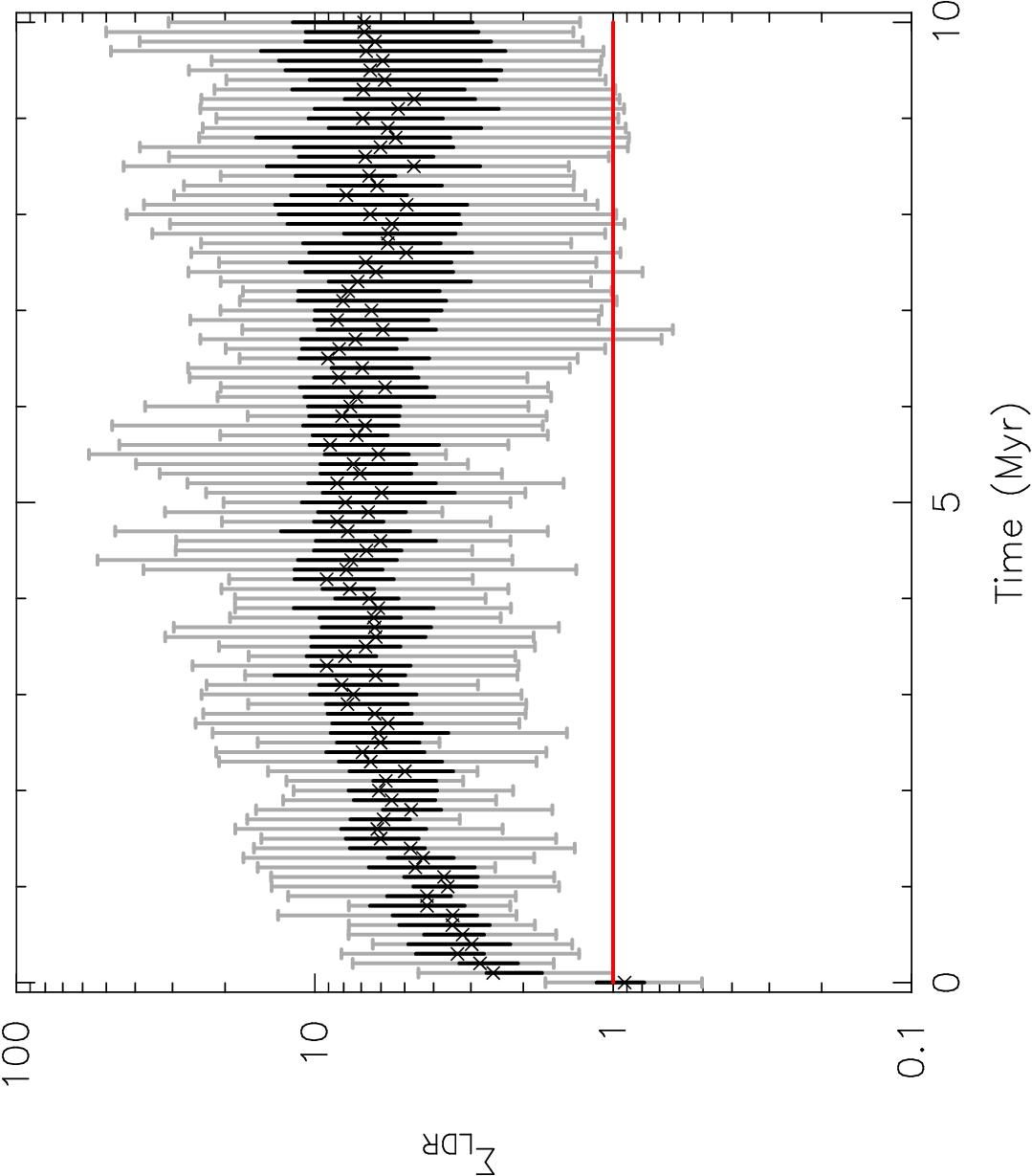}}}
\hspace*{0.3cm} 
\subfigure[$\alpha_{\rm vir} = 0.3, D = 2.0$]{\label{sigm_med-b}\rotatebox{270}{\includegraphics[scale=0.27]{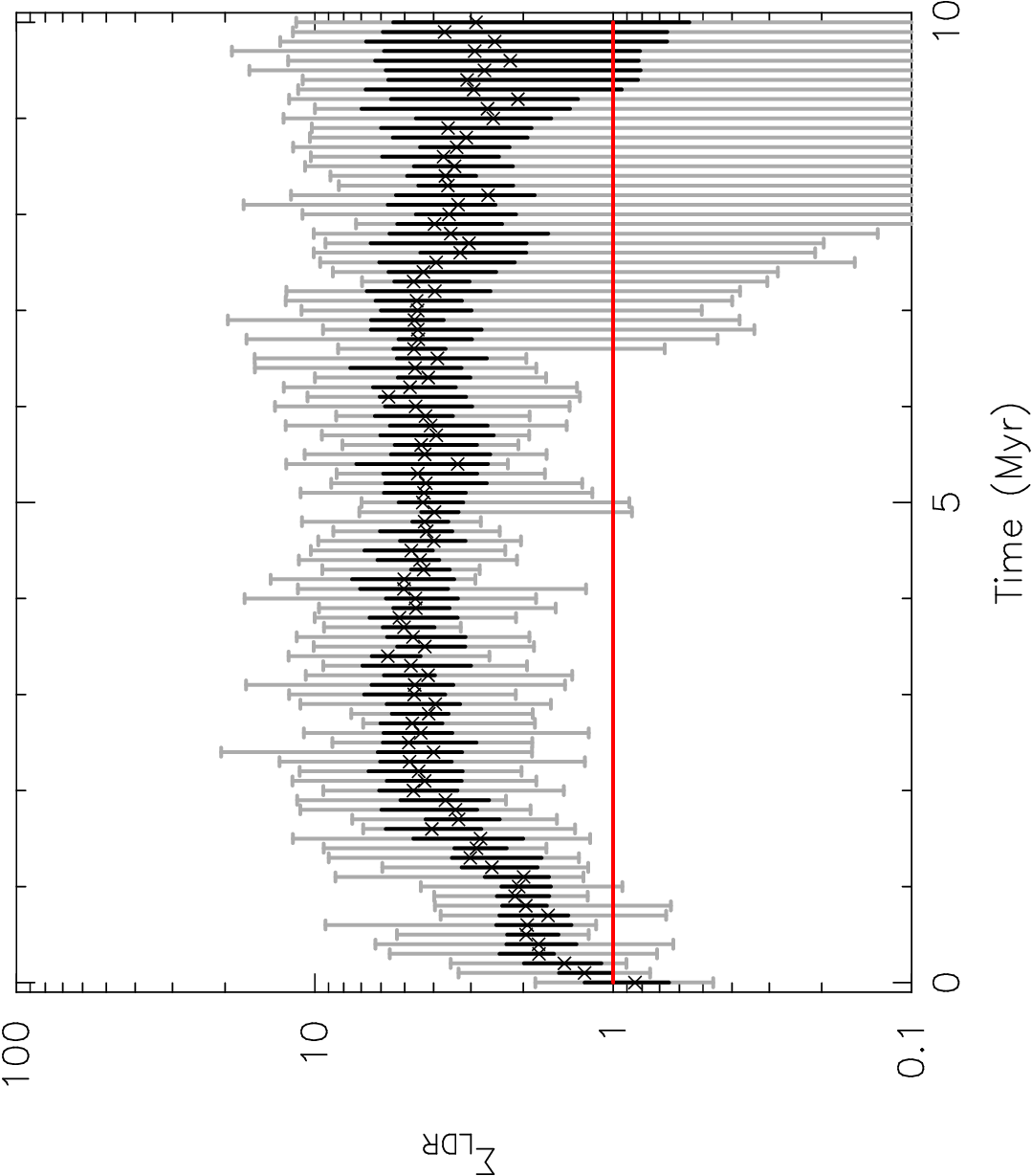}}}
\hspace*{0.3cm} 
\subfigure[$\alpha_{\rm vir} = 0.3, D = 3.0$]{\label{sigm_med-c}\rotatebox{270}{\includegraphics[scale=0.27]{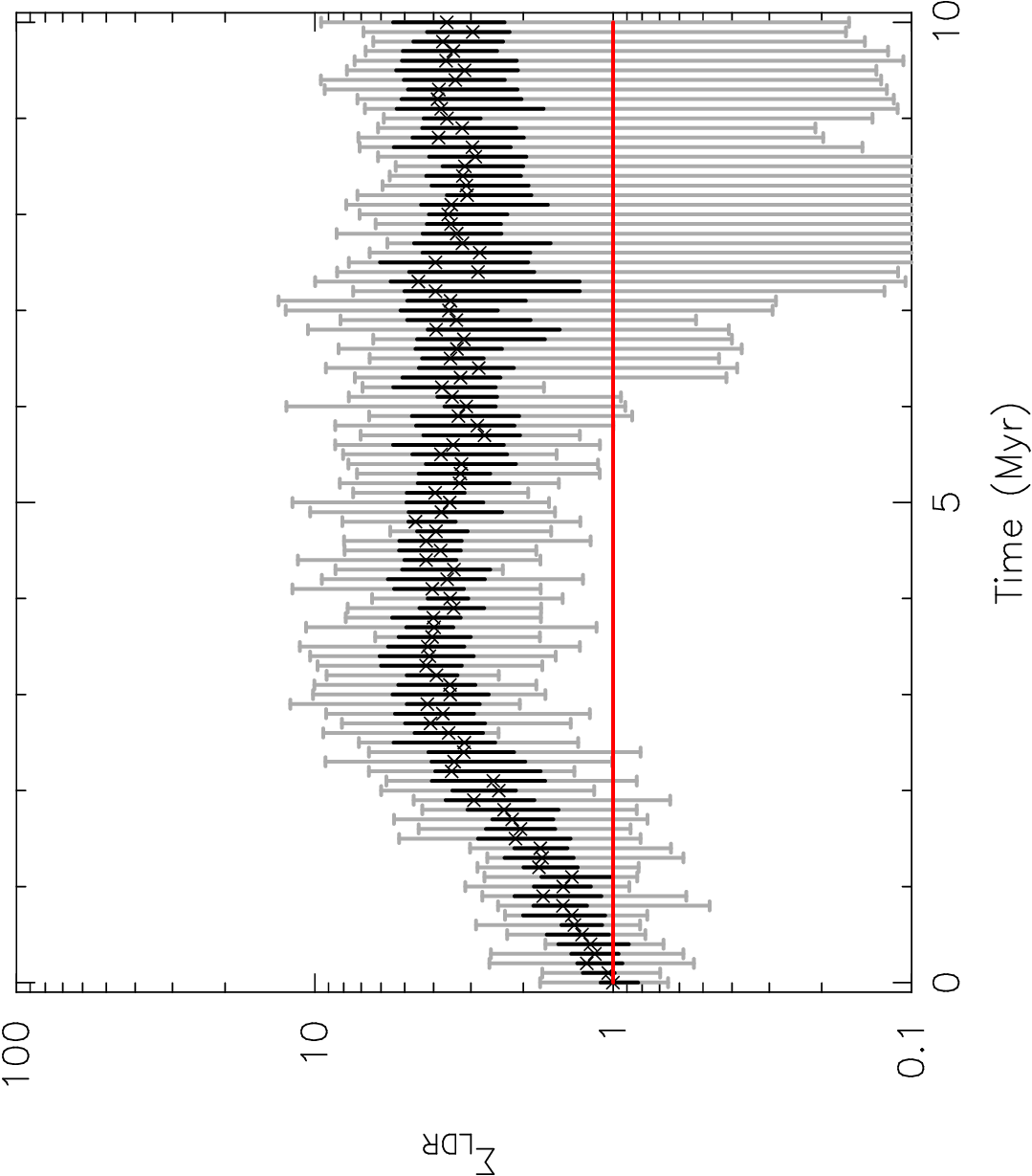}}}
\hspace*{-1.cm}
\subfigure[$\alpha_{\rm vir} = 0.5, D = 1.6$]{\label{sigm_med-d}\rotatebox{270}{\includegraphics[scale=0.27]{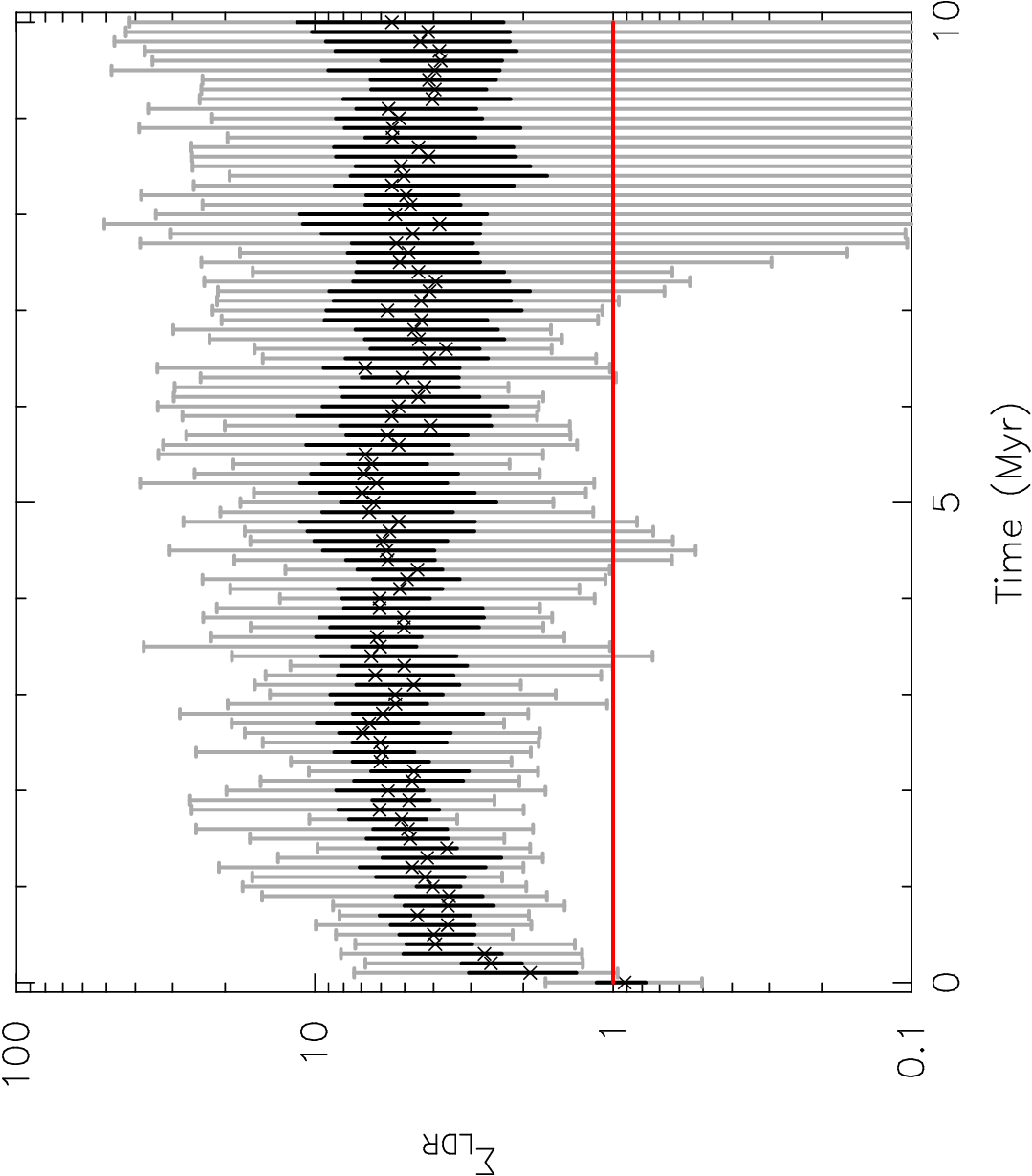}}}
\hspace*{0.3cm} 
\subfigure[$\alpha_{\rm vir} = 0.5, D = 2.0$]{\label{sigm_med-e}\rotatebox{270}{\includegraphics[scale=0.27]{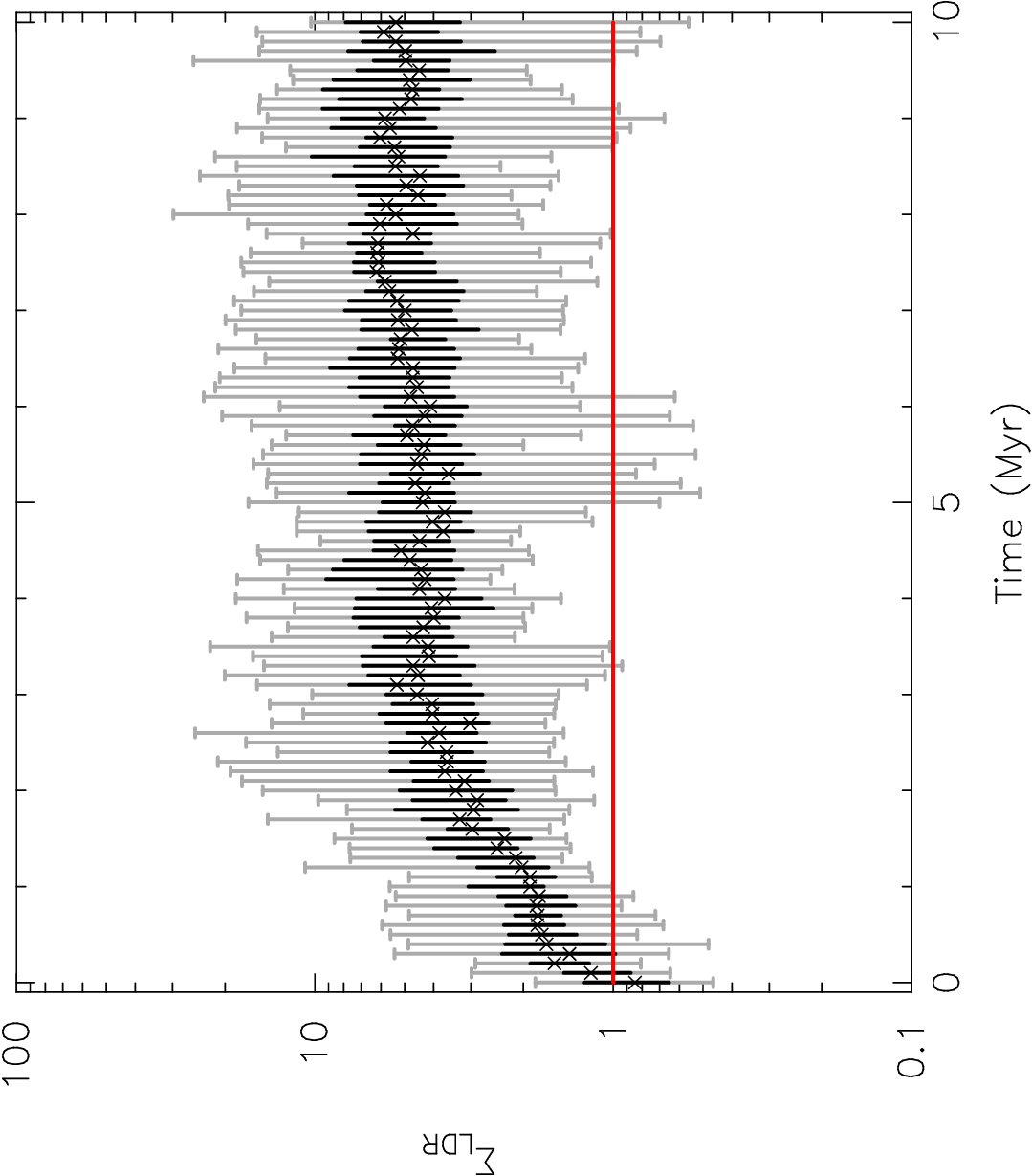}}}
\hspace*{0.3cm} 
\subfigure[$\alpha_{\rm vir} = 0.5, D = 3.0$]{\label{sigm_med-f}\rotatebox{270}{\includegraphics[scale=0.27]{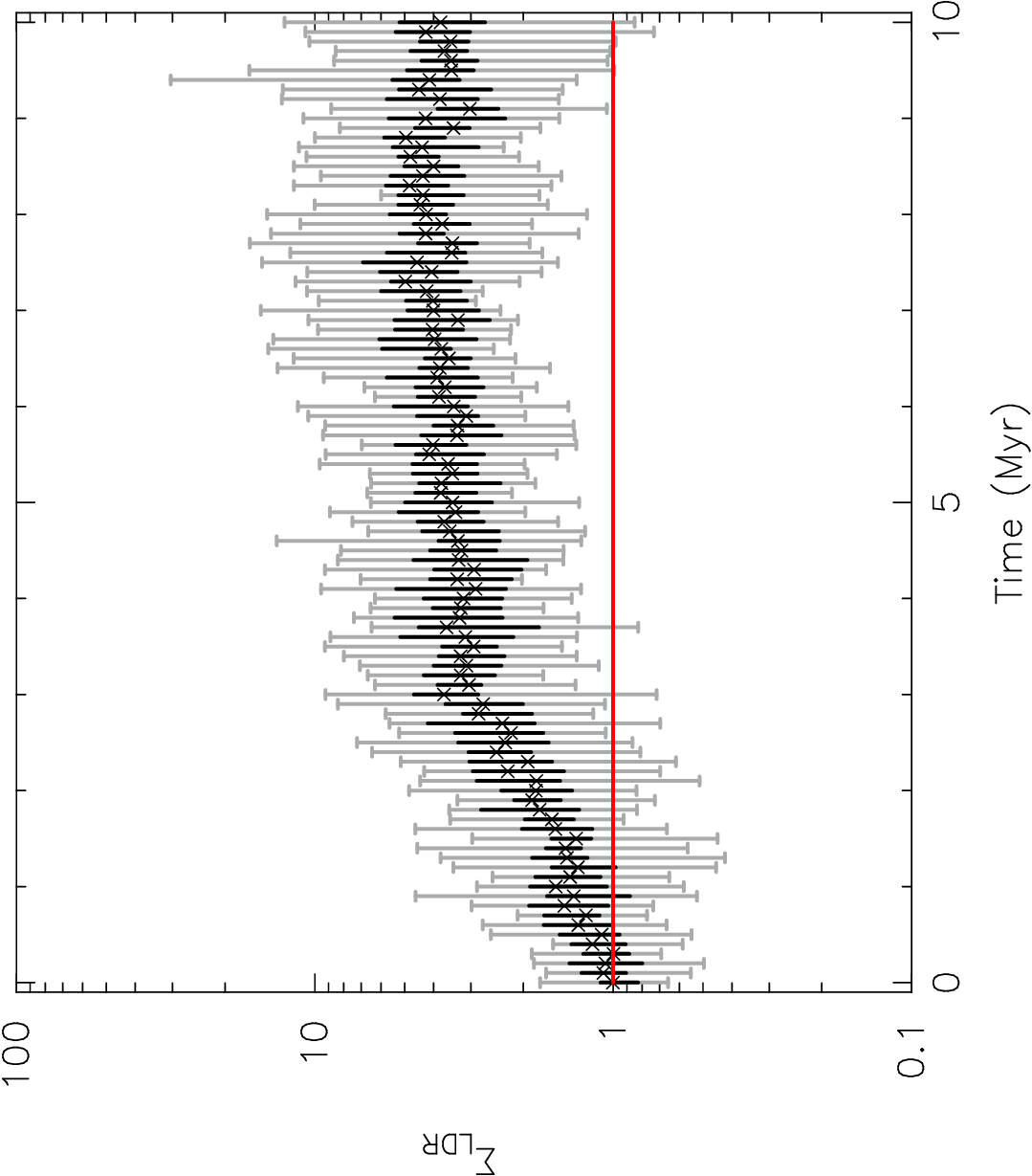}}}
\hspace*{-1.cm}
\subfigure[$\alpha_{\rm vir} = 1.5, D = 1.6$]{\label{sigm_med-g}\rotatebox{270}{\includegraphics[scale=0.27]{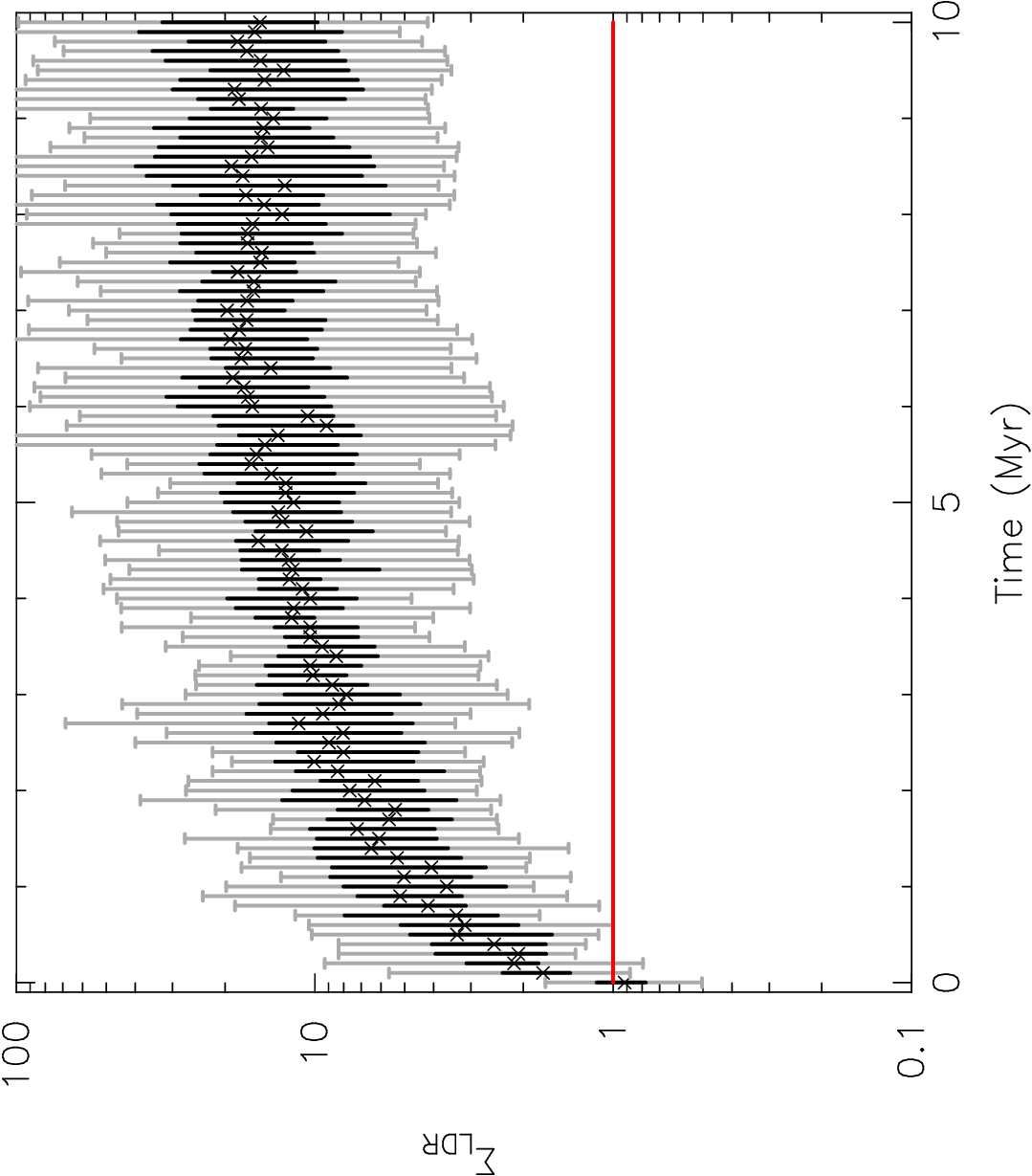}}}
\hspace*{0.3cm} 
\subfigure[$\alpha_{\rm vir} = 1.5, D = 2.0$]{\label{sigm_med-h}\rotatebox{270}{\includegraphics[scale=0.27]{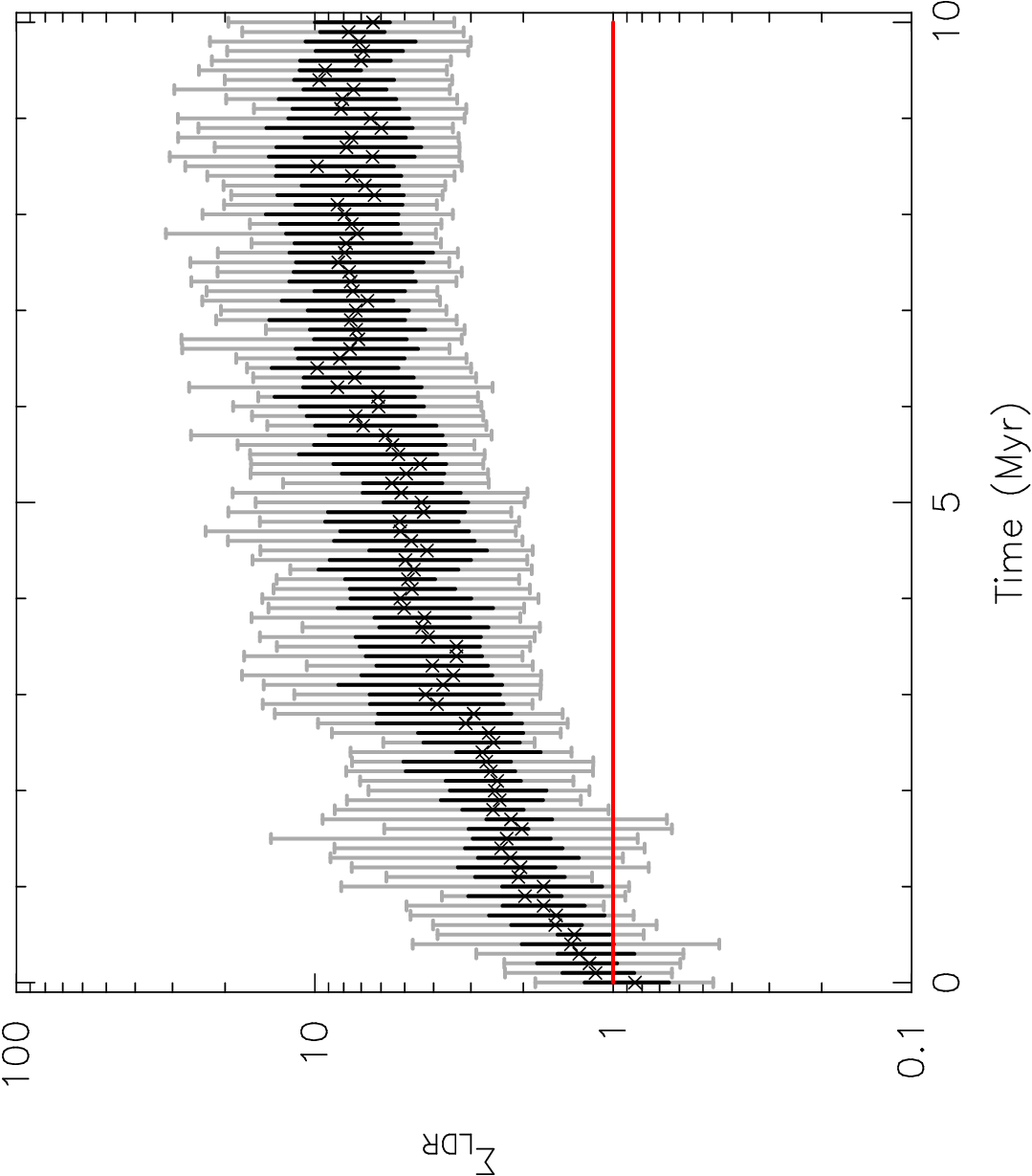}}}
\hspace*{0.3cm} 
\subfigure[$\alpha_{\rm vir} = 1.5, D = 3.0$]{\label{sigm_med-i}\rotatebox{270}{\includegraphics[scale=0.27]{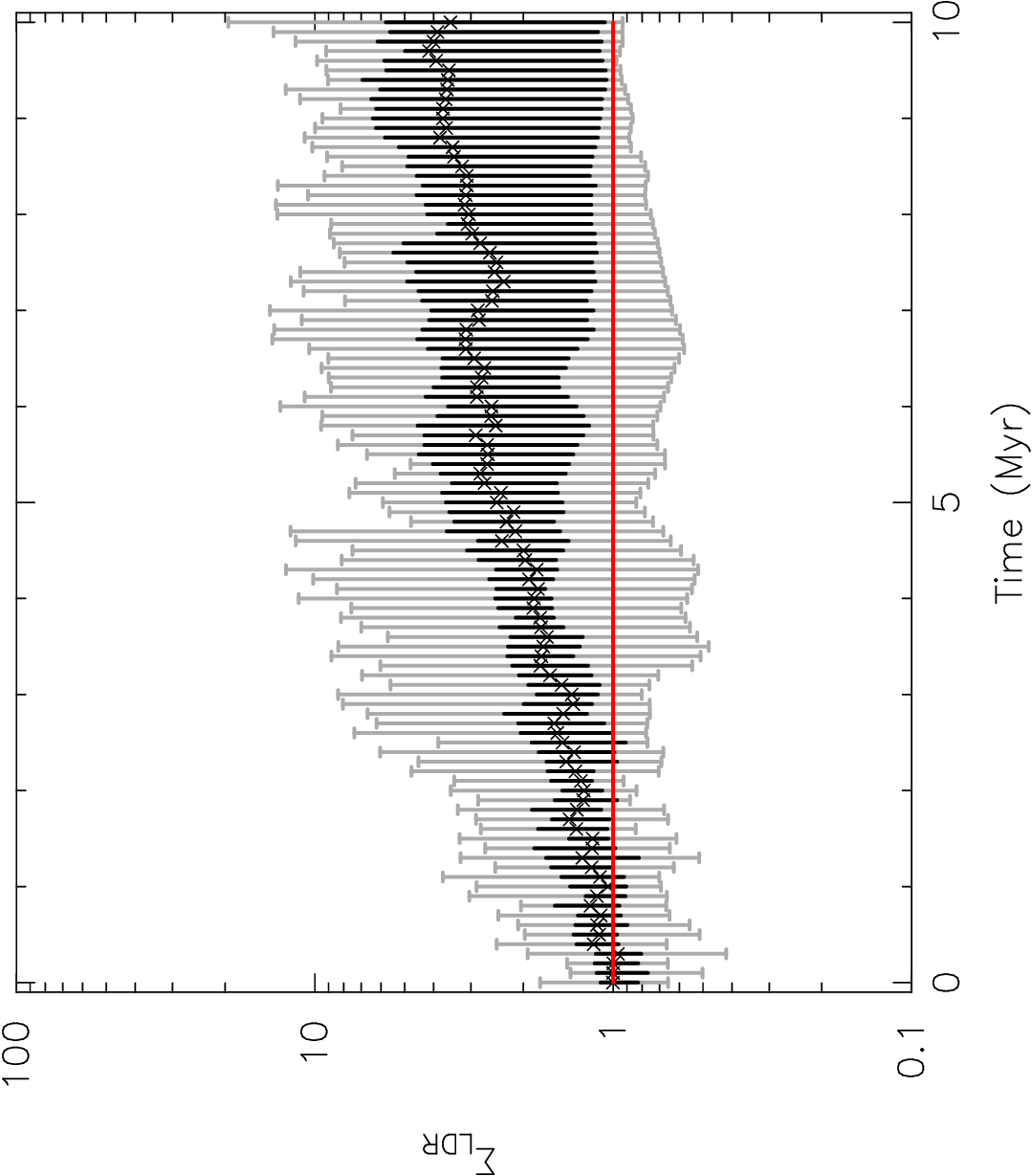}}}
\caption[bf]{Evolution of $\Sigma_{\rm LDR}$ with time for simulations with $N = 1500$ stars. Each panel shows the median $\Sigma_{\rm LDR}$ value from 20 simulations with identical initial conditions (the crosses) and the darker error bars indicate 
25 and 75 percentile values. The entire range of possible values from the 20 sets of initial conditions is shown by the lighter error bars. }
\label{sigm_med}
  \end{center}
\end{figure*}

In Fig.~\ref{sigm_med} we show the evolution of the $\Sigma_{\rm LDR}$
ratio for all nine sets of initial conditions (again, $N = 1500$ stars and $\alpha_{\rm vir} =
0.3, 0.5, 1.5$ for all of $D = 1.6, 2.0, 3.0$). As in 
Fig.~\ref{msr_med} we plot the 
median  $\Sigma_{\rm LDR}$ value for 20 simulations by the black cross
(all values are for the 10 most massive stars). We show the 25 and 75
percentiles with the black bars, and indicate the
extrema  by the grey bars.  Again note that the `error bars' capture
the range covered at each time, and that the evolution of each
particular simulation's $\Sigma_{\rm LDR}$ can be very complicated.

We show $\Sigma_{\rm LDR} = 1$
by the solid red horizontal line. Usually, if
$\Sigma_{\rm LDR} > 2$ the most massive stars have (statistically)
signficantly higher densities than the average stars in the
region. Indeed, if we recall  Figs.~\ref{cool-k}~and~\ref{hot-k}, we
see that the red circles (indicating an insignificant $\Sigma_{\rm
  LDR}$ ratio) are only present for $\Sigma_{\rm LDR} < 2$.

In contrast to $\Lambda_{\rm MSR}$, in {\em all} ensembles $\Sigma_{\rm
  LDR}$ tends to increase with time.  Indeed, the median $\Sigma_{\rm
  LDR}$ by 10~Myr for almost all sets of initial conditions is very
similar, with values of  $\Sigma_{\rm
  LDR } \sim 5-10$.  This means that the most
massive stars almost always find themselves at significantly higher local surface
densities than an average star, even in a hot, expanding region.

The time taken for $\Sigma_{\rm LDR}$ to increase depends on the level
of substructure.  In the first column with $D=1.6$ (panels (a), (d), and
(g)) $\Sigma_{\rm LDR}$ rises to significant levels almost
immediately.  But in the third column with $D=3.0$ (panels (c), (f),
and (i)) $\Sigma_{\rm LDR}$ takes a few~Myr to reach significant levels.

The reason for this is that the massive stars act as a local potential
well which can trap low-mass stars.  $\Sigma_{\rm LDR}$ measures the
size of the retinue of low (or high) mass stars collected by a massive
star.  If initial substructure is present, the high-mass star is
likely to find itself with a ready-made retinue to attract, hence
$\Sigma_{\rm LDR}$ rises very quickly.  But when the stellar
distribution is smooth it takes some time for the massive stars to
collect a significant retinue of low-mass stars and so $\Sigma_{\rm
  LDR}$ rises more slowly.  Indeed, the most significant rise 
in $\Sigma_{\rm LDR}$ is seen in the
hottest and most substructured regions (panel (g)) where local
substructure is bound and collected by the massive stars 
within a globally unbound region.

In a few cases, as seen in the large spread in extremes of
$\Sigma_{\rm LDR}$ in panels (b) and (d) especially, the decay of
higher-order massive star multiples can eject massive stars without a
retinue causing $\Sigma_{\rm LDR}$ to fall below unity (i.e.\,\,the
massive stars have very low local densities).  \citet{Allison11} showed that higher-order Trapezium-like massive star systems
are more likely to form at moderate $D$ or $\alpha_{\rm vir}$ which is
why panel (a) with the most extreme $D$ and $\alpha_{\rm vir}$ does
not show this decay and consequent $\Sigma_{\rm LDR} < 1$ values as
it is much less likely to contain a relatively long-lived
Trapezium-like system.

\subsubsection{Structure versus mass segregation and local density}
\label{structms}

\begin{figure*}
  \begin{center}
\setlength{\subfigcapskip}{10pt}
\vspace*{-0.3cm}
\hspace*{-1.cm}\subfigure[$\alpha_{\rm vir} = 0.3, D = 1.6$]{\label{Q_MSR-a}\rotatebox{270}{\includegraphics[scale=0.28]{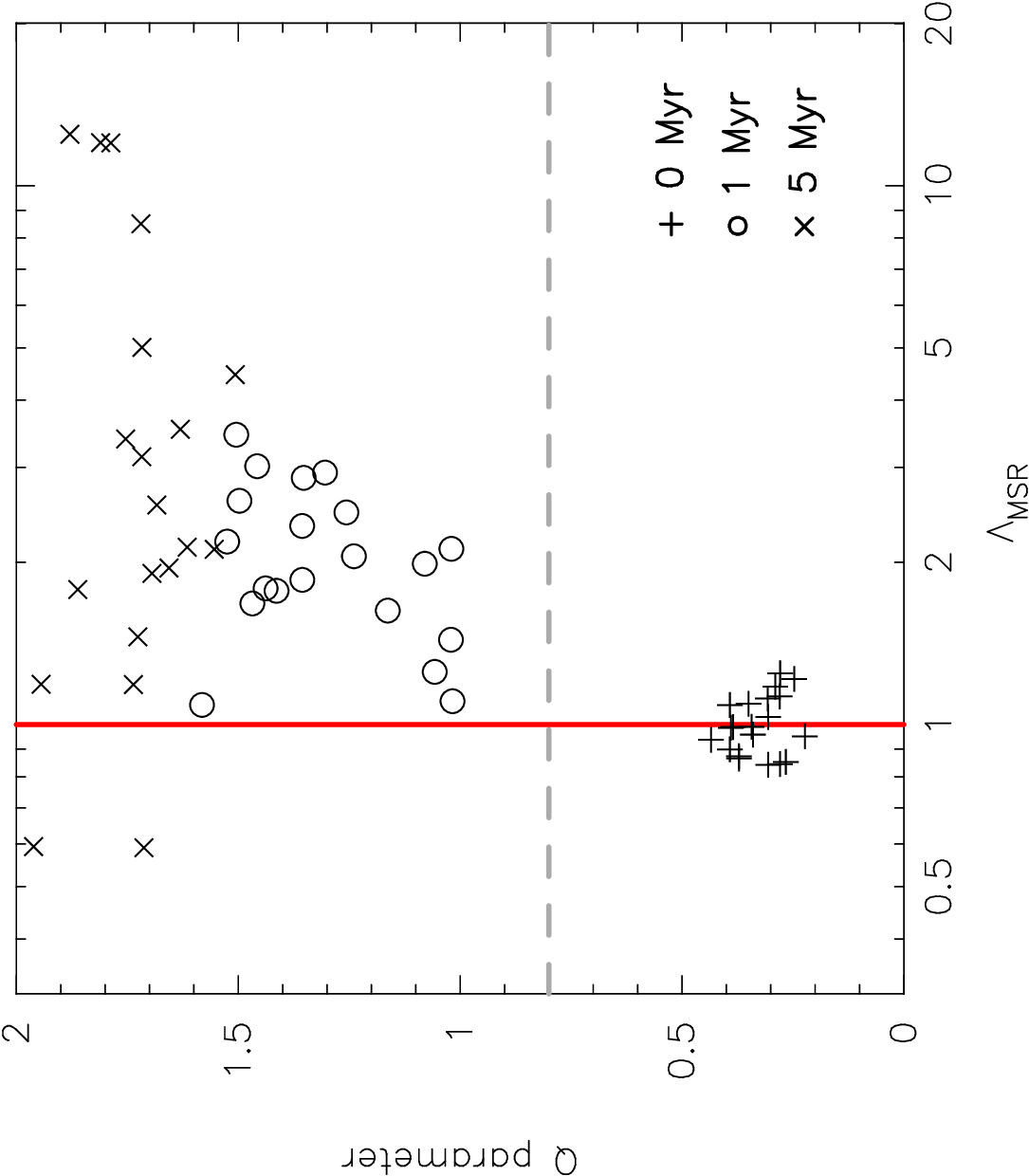}}}
\hspace*{0.1cm} 
\subfigure[$\alpha_{\rm vir} = 0.3, D = 2.0$]{\label{Q_MSR-b}\rotatebox{270}{\includegraphics[scale=0.28]{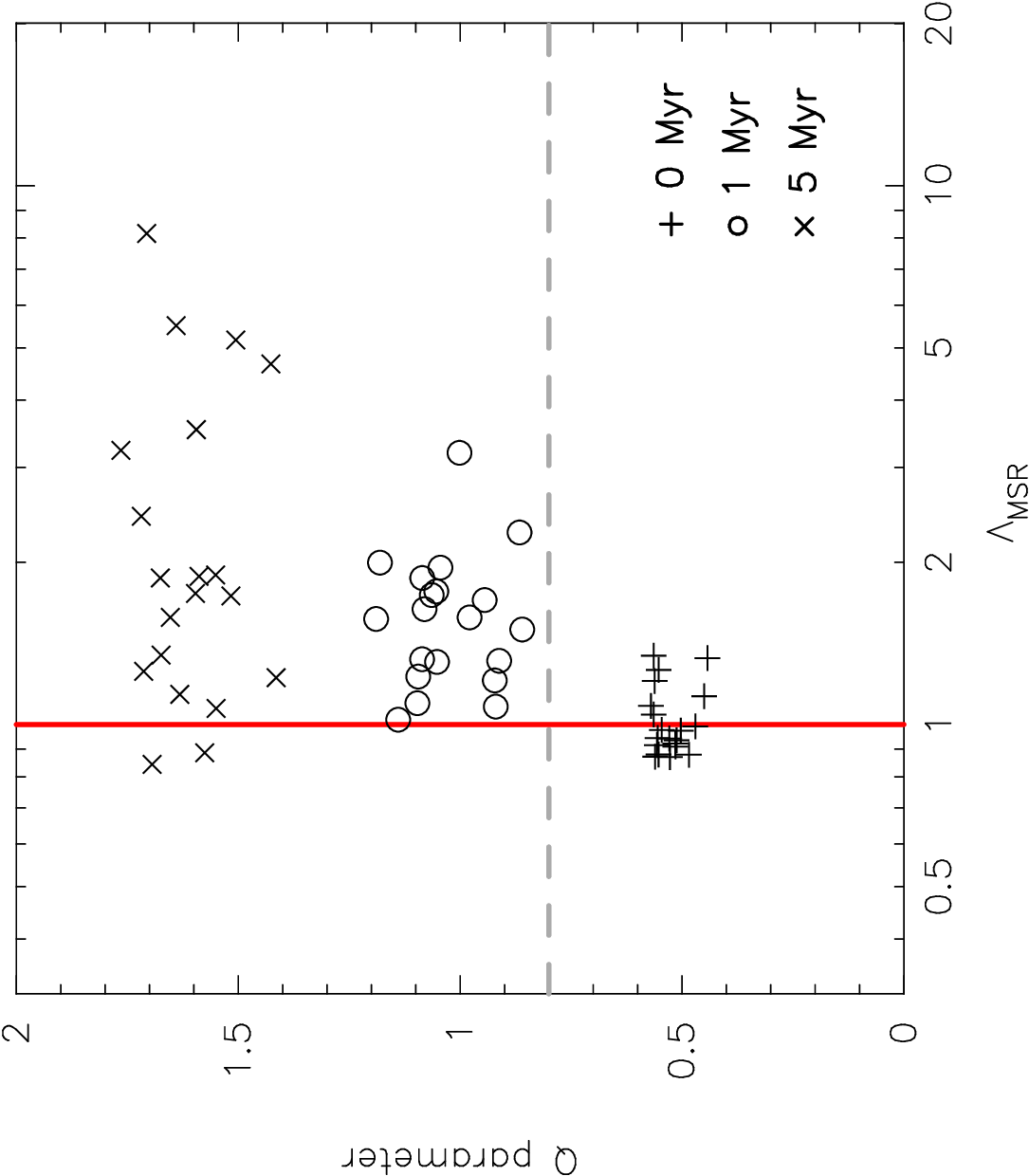}}}
\hspace*{0.1cm} 
\subfigure[$\alpha_{\rm vir} = 0.3, D = 3.0$]{\label{Q_MSR-c}\rotatebox{270}{\includegraphics[scale=0.28]{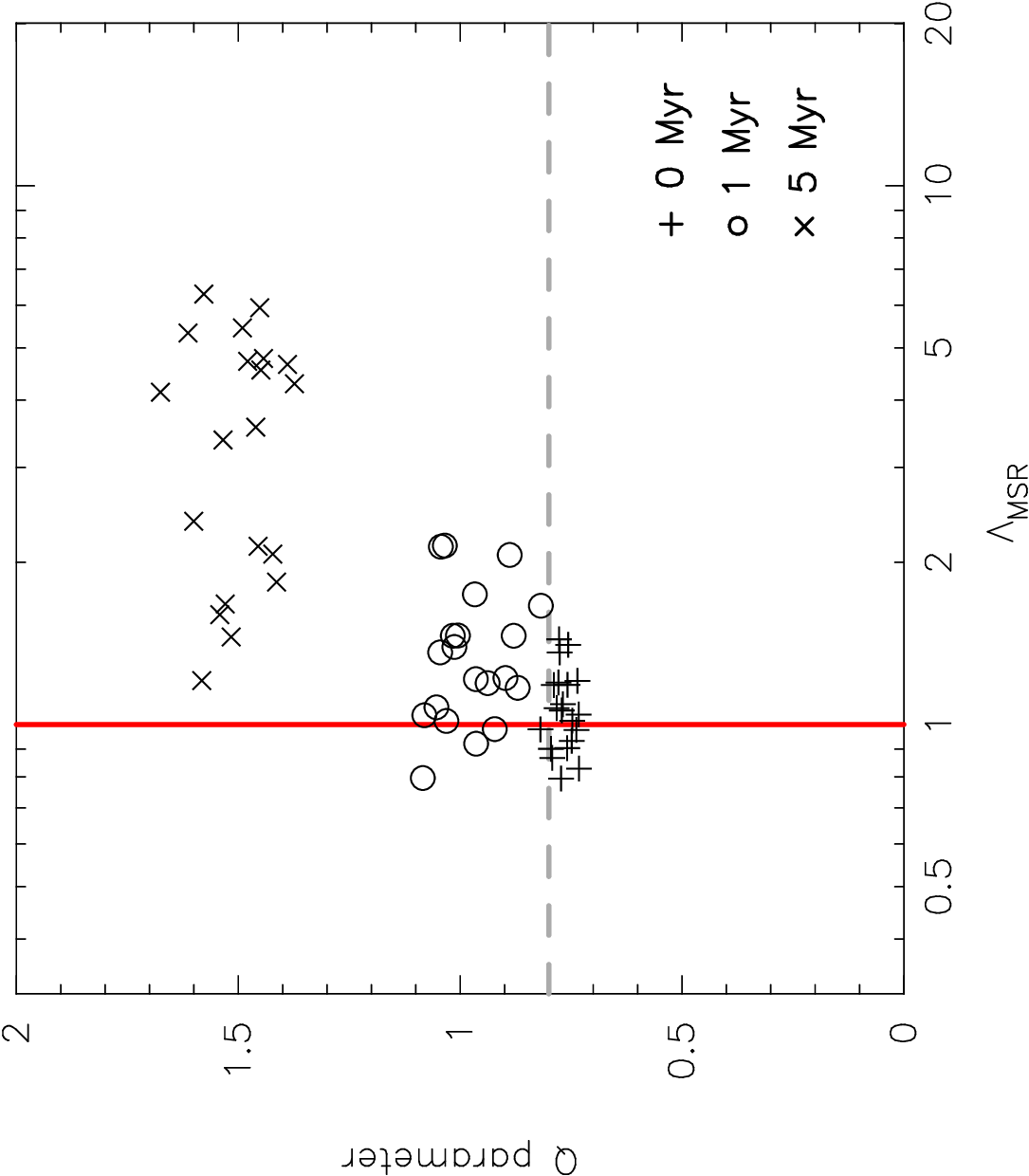}}}
\hspace*{-1.cm}
\subfigure[$\alpha_{\rm vir} = 0.5, D = 1.6$]{\label{Q_MSR-d}\rotatebox{270}{\includegraphics[scale=0.28]{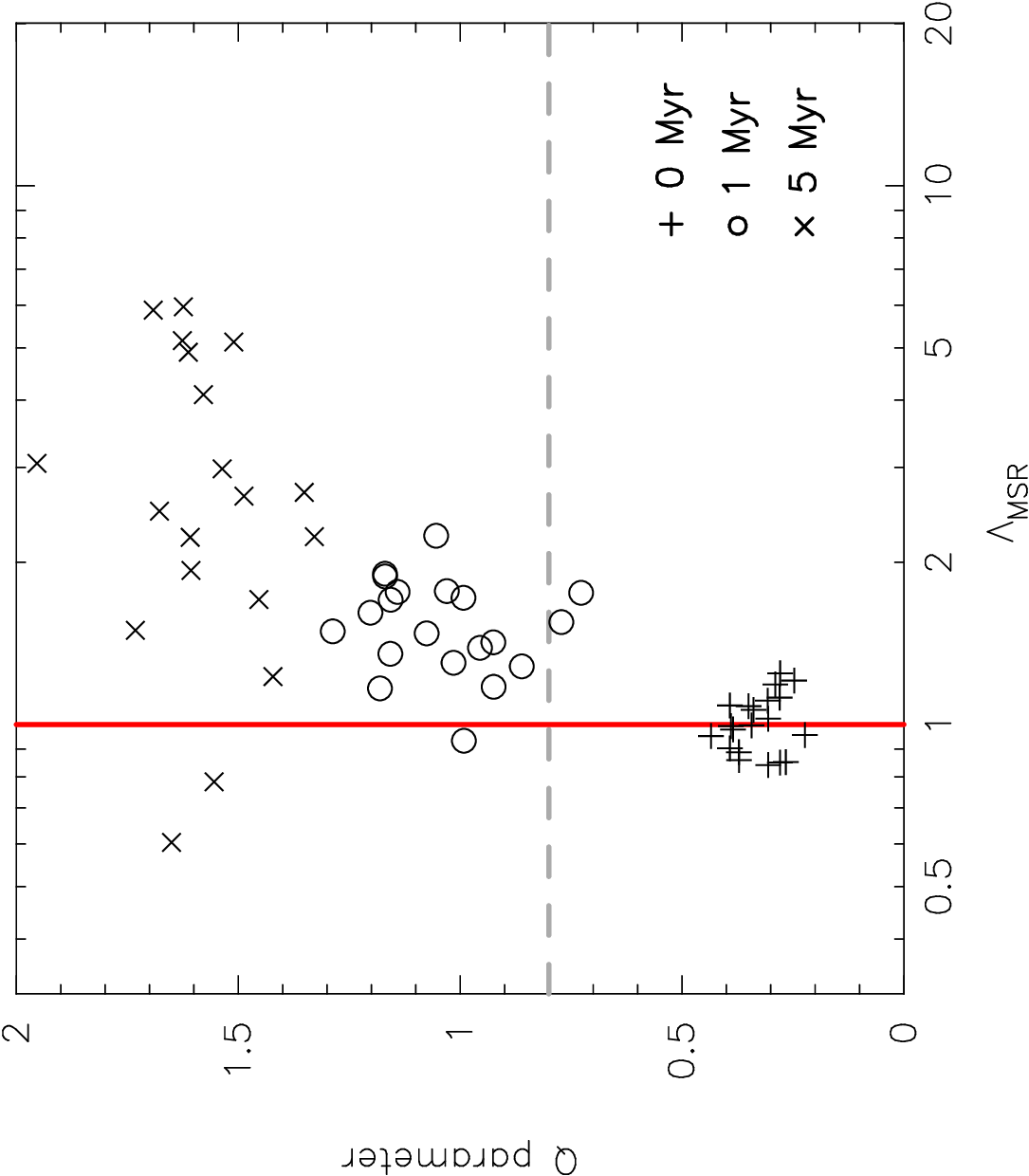}}}
\hspace*{0.1cm} 
\subfigure[$\alpha_{\rm vir} = 0.5, D = 2.0$]{\label{Q_MSR-e}\rotatebox{270}{\includegraphics[scale=0.28]{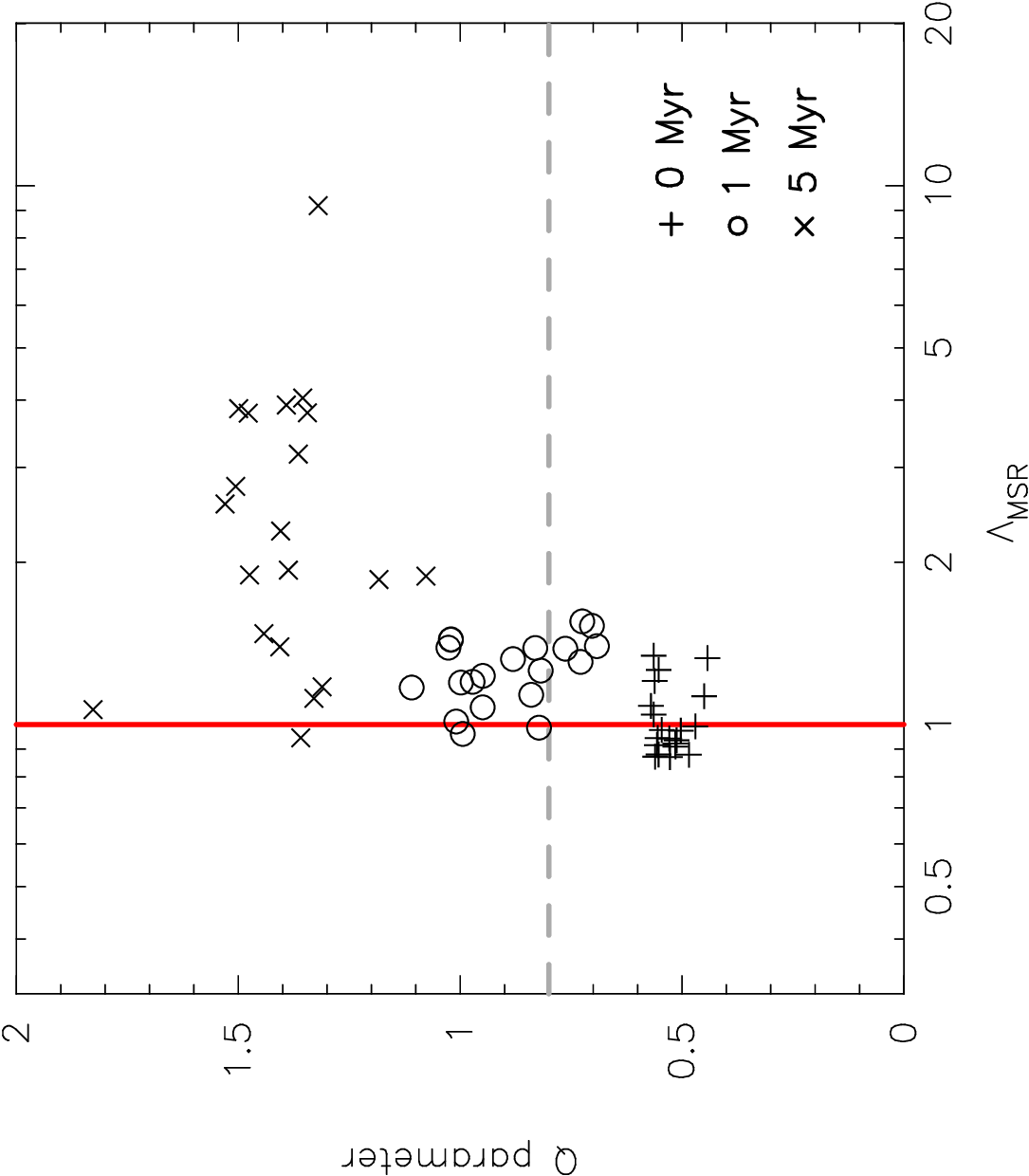}}}
\hspace*{0.1cm} 
\subfigure[$\alpha_{\rm vir} = 0.5, D = 3.0$]{\label{Q_MSR-f}\rotatebox{270}{\includegraphics[scale=0.28]{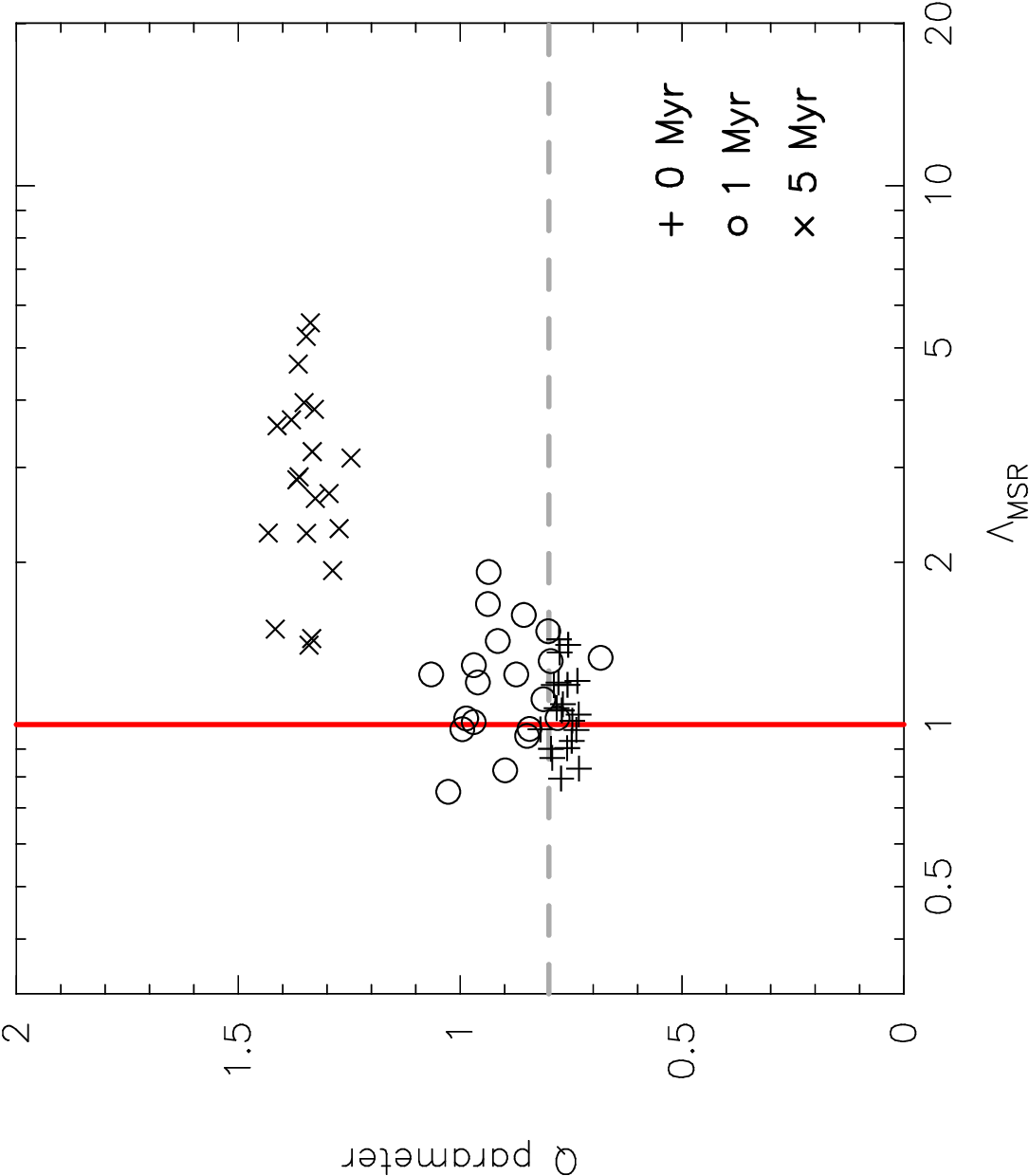}}}
\hspace*{-1.cm}
\subfigure[$\alpha_{\rm vir} = 1.5, D = 1.6$]{\label{Q_MSR-g}\rotatebox{270}{\includegraphics[scale=0.28]{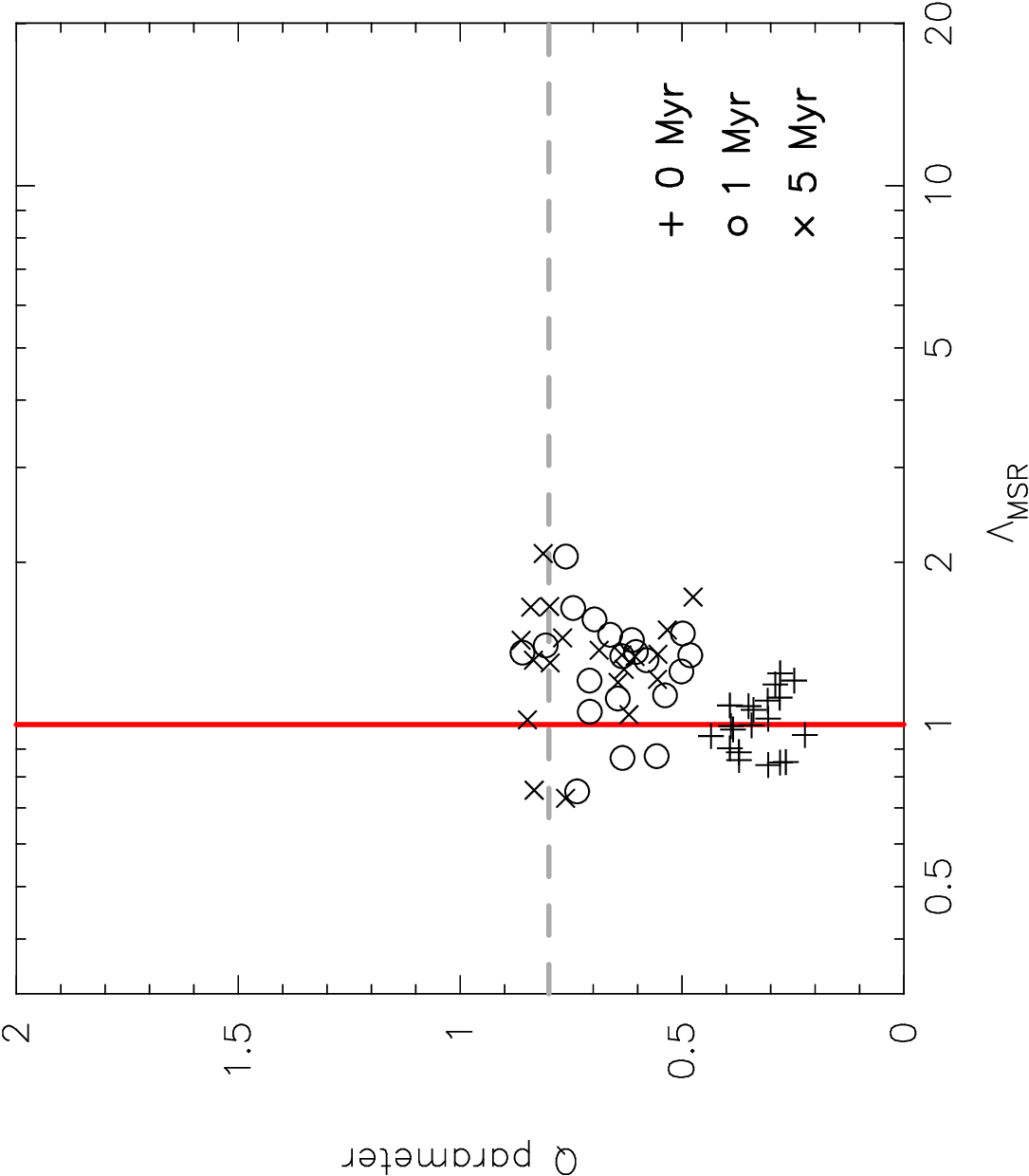}}}
\hspace*{0.1cm} 
\subfigure[$\alpha_{\rm vir} = 1.5, D = 2.0$]{\label{Q_MSR-h}\rotatebox{270}{\includegraphics[scale=0.28]{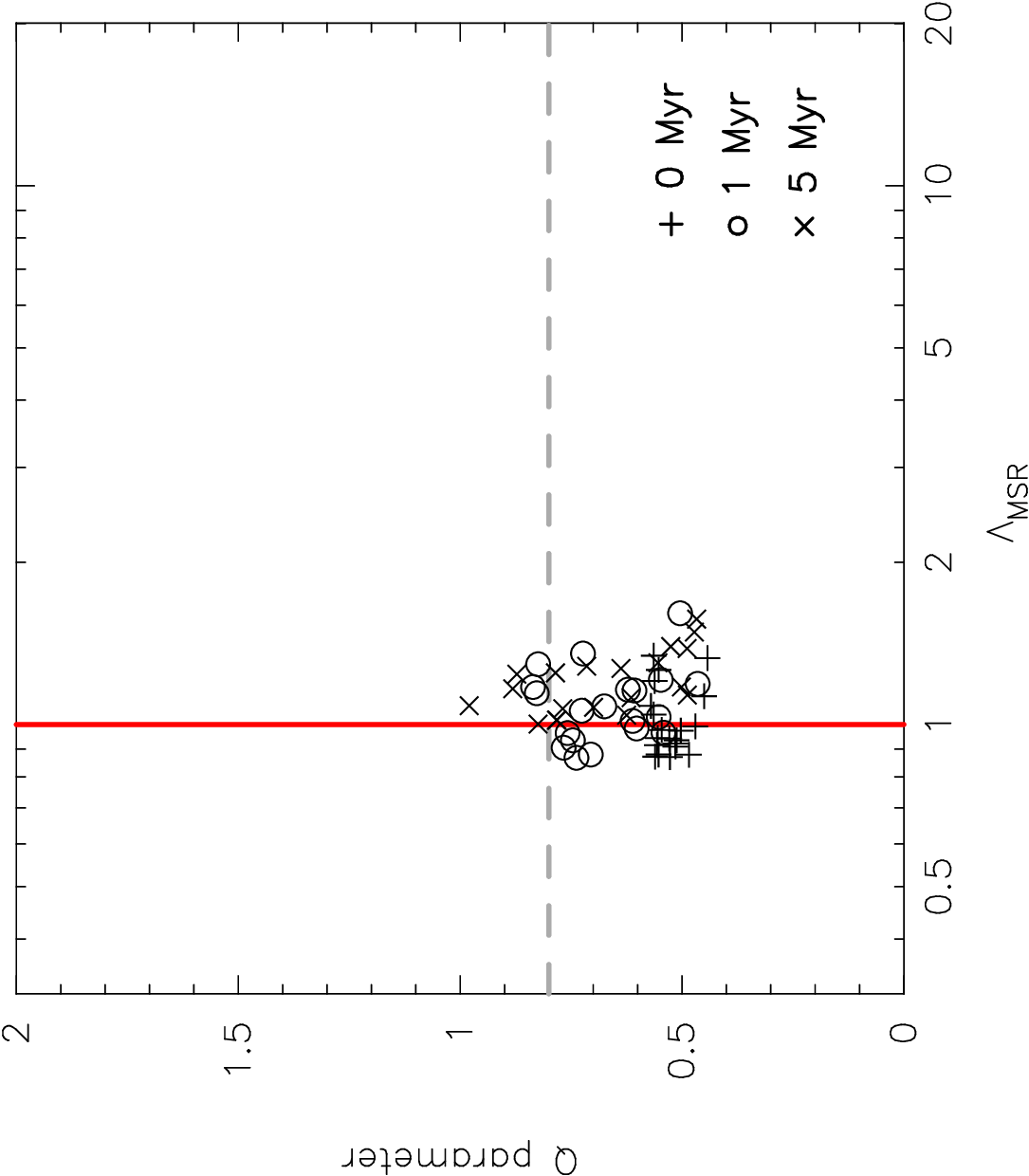}}}
\hspace*{0.1cm} 
\subfigure[$\alpha_{\rm vir} = 1.5, D = 3.0$]{\label{Q_MSR-i}\rotatebox{270}{\includegraphics[scale=0.28]{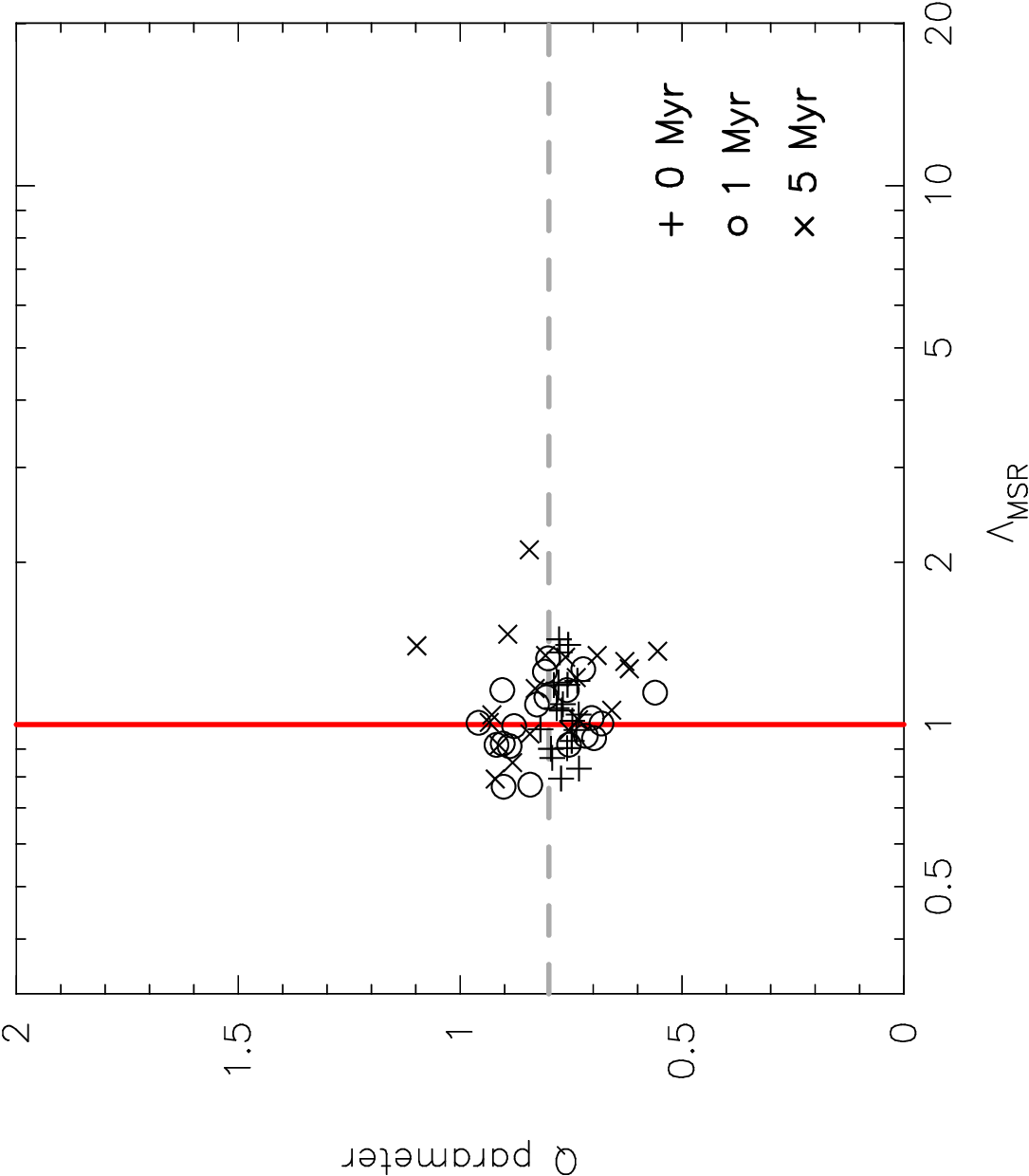}}}
\caption[bf]{The evolution of $\mathcal{Q}$--parameter versus $\Lambda_{\rm MSR}$ with time for the simulations with $N = 1500$ stars. For each simulation we plot the $\mathcal{Q}$--parameter and 
$\Lambda_{\rm MSR}$ at 0\,Myr (the plus signs), 1\,Myr (the open circles) and 5\,Myr (the crosses). The boundary between substructured associations and radially smooth clusters ($\mathcal{Q} = 0.8$) is indicated by the horizontal grey dashed line, 
and $\Lambda_{\rm MSR} =1$ is shown by the vertical red line.  }
\label{Q_MSR}
  \end{center}
\end{figure*}

\begin{figure*}
  \begin{center}
\setlength{\subfigcapskip}{10pt}
\vspace*{-0.3cm}
\hspace*{-1.cm}\subfigure[$\alpha_{\rm vir} = 0.3, D = 1.6$]{\label{Q_Sig-a}\rotatebox{270}{\includegraphics[scale=0.28]{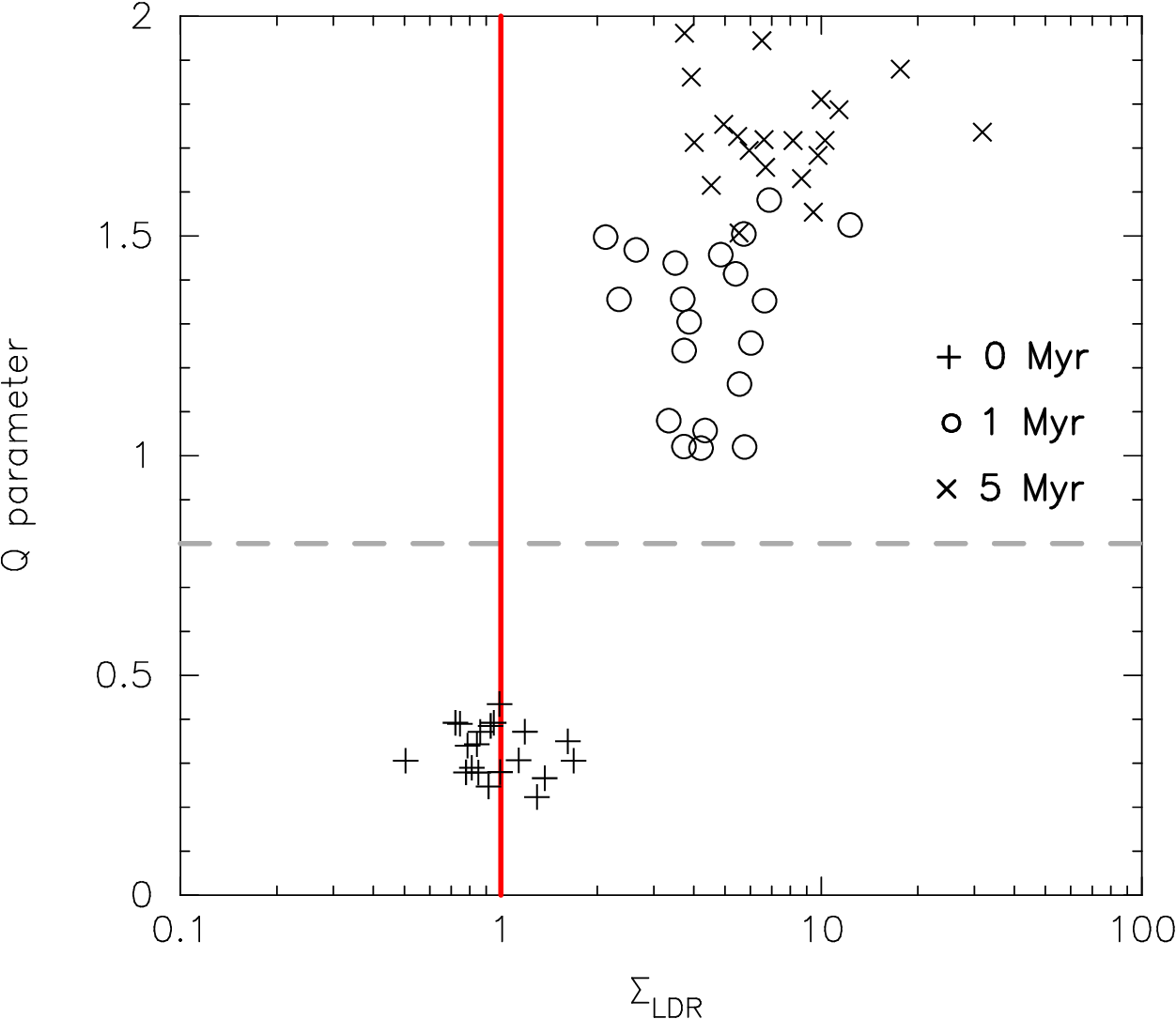}}}
\hspace*{0.1cm} 
\subfigure[$\alpha_{\rm vir} = 0.3, D = 2.0$]{\label{Q_Sig-b}\rotatebox{270}{\includegraphics[scale=0.28]{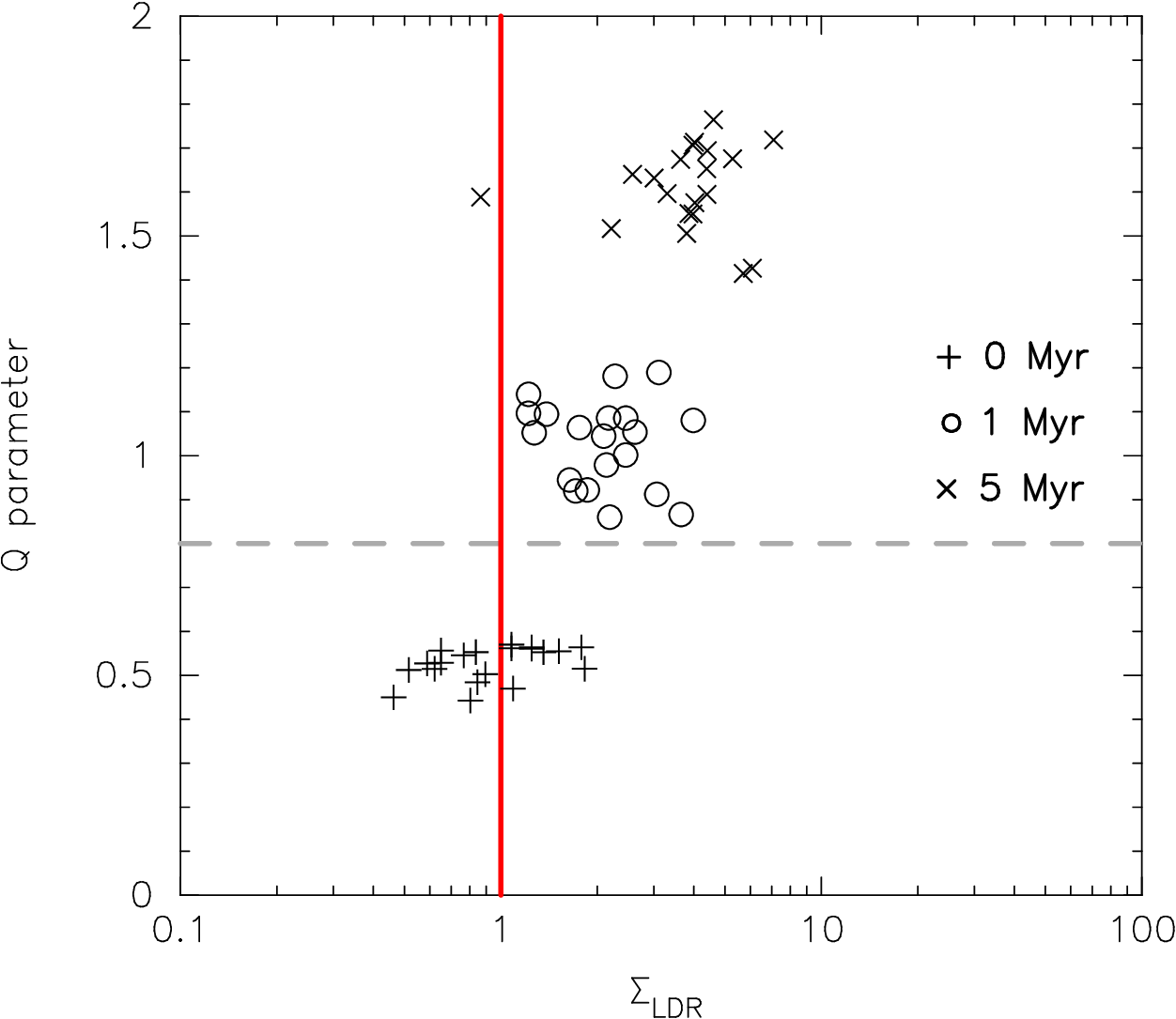}}}
\hspace*{0.1cm} 
\subfigure[$\alpha_{\rm vir} = 0.3, D = 3.0$]{\label{Q_Sig-c}\rotatebox{270}{\includegraphics[scale=0.28]{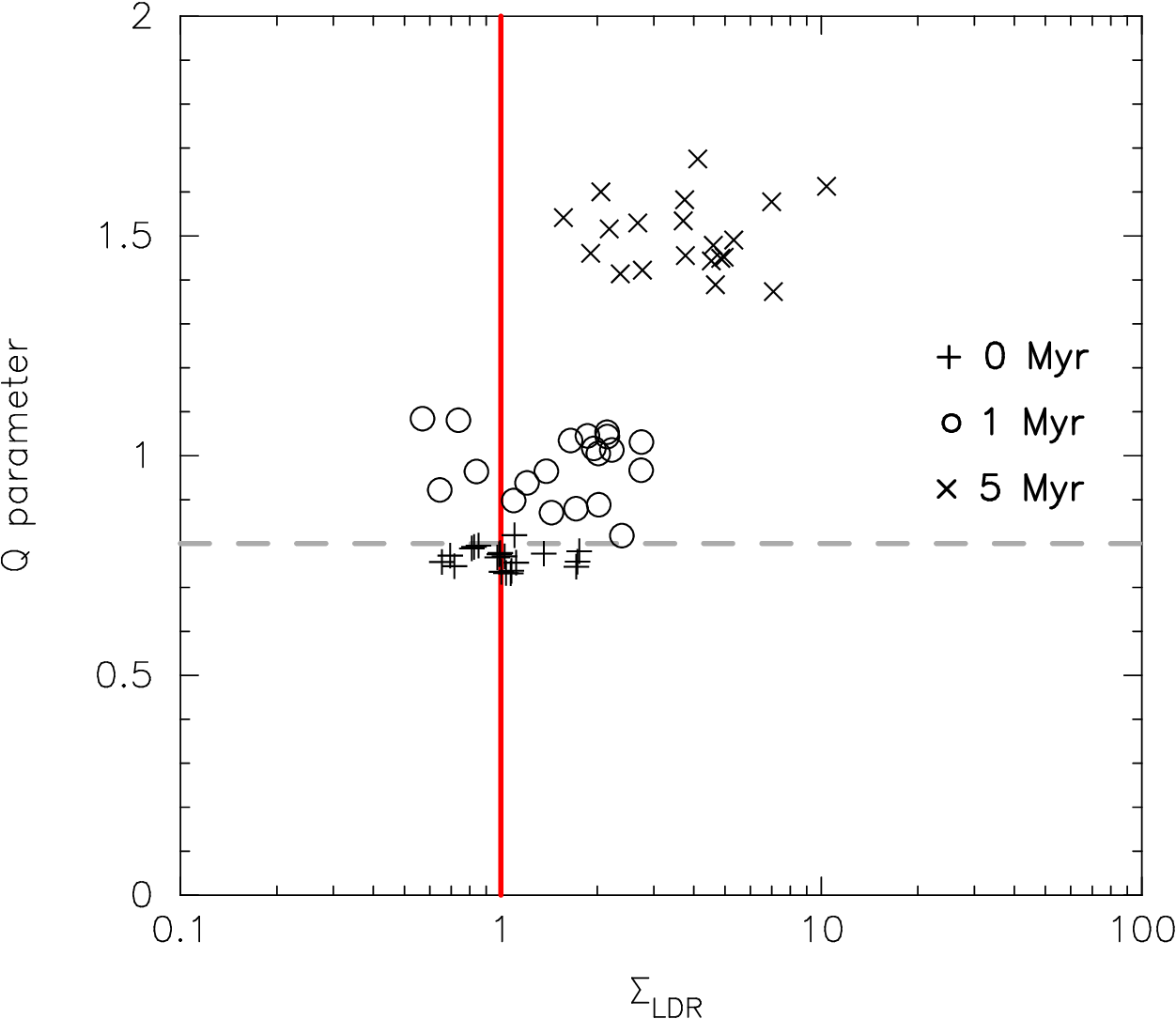}}}
\hspace*{-1.cm}
\subfigure[$\alpha_{\rm vir} = 0.5, D = 1.6$]{\label{Q_Sig-d}\rotatebox{270}{\includegraphics[scale=0.28]{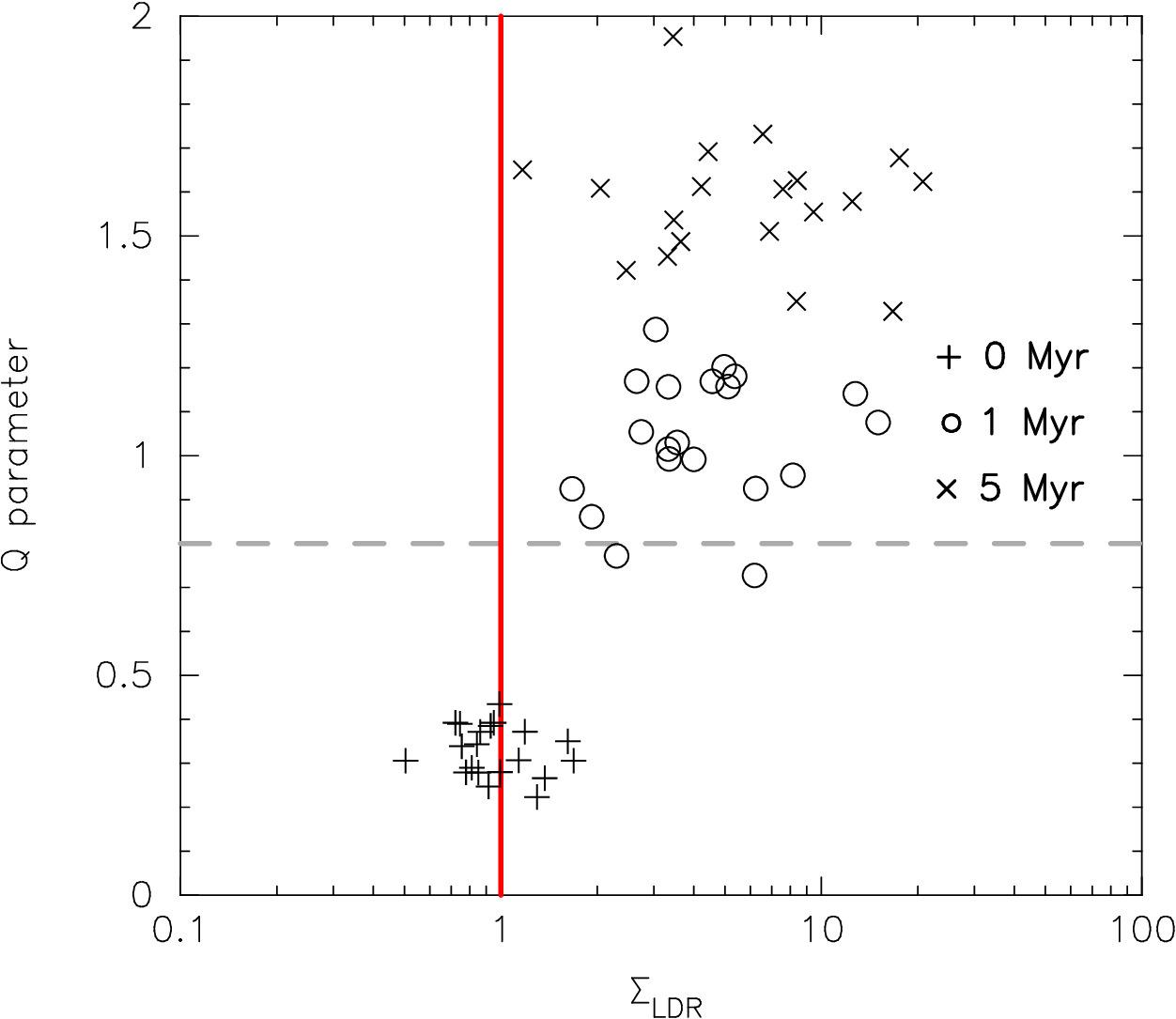}}}
\hspace*{0.1cm} 
\subfigure[$\alpha_{\rm vir} = 0.5, D = 2.0$]{\label{Q_Sig-e}\rotatebox{270}{\includegraphics[scale=0.28]{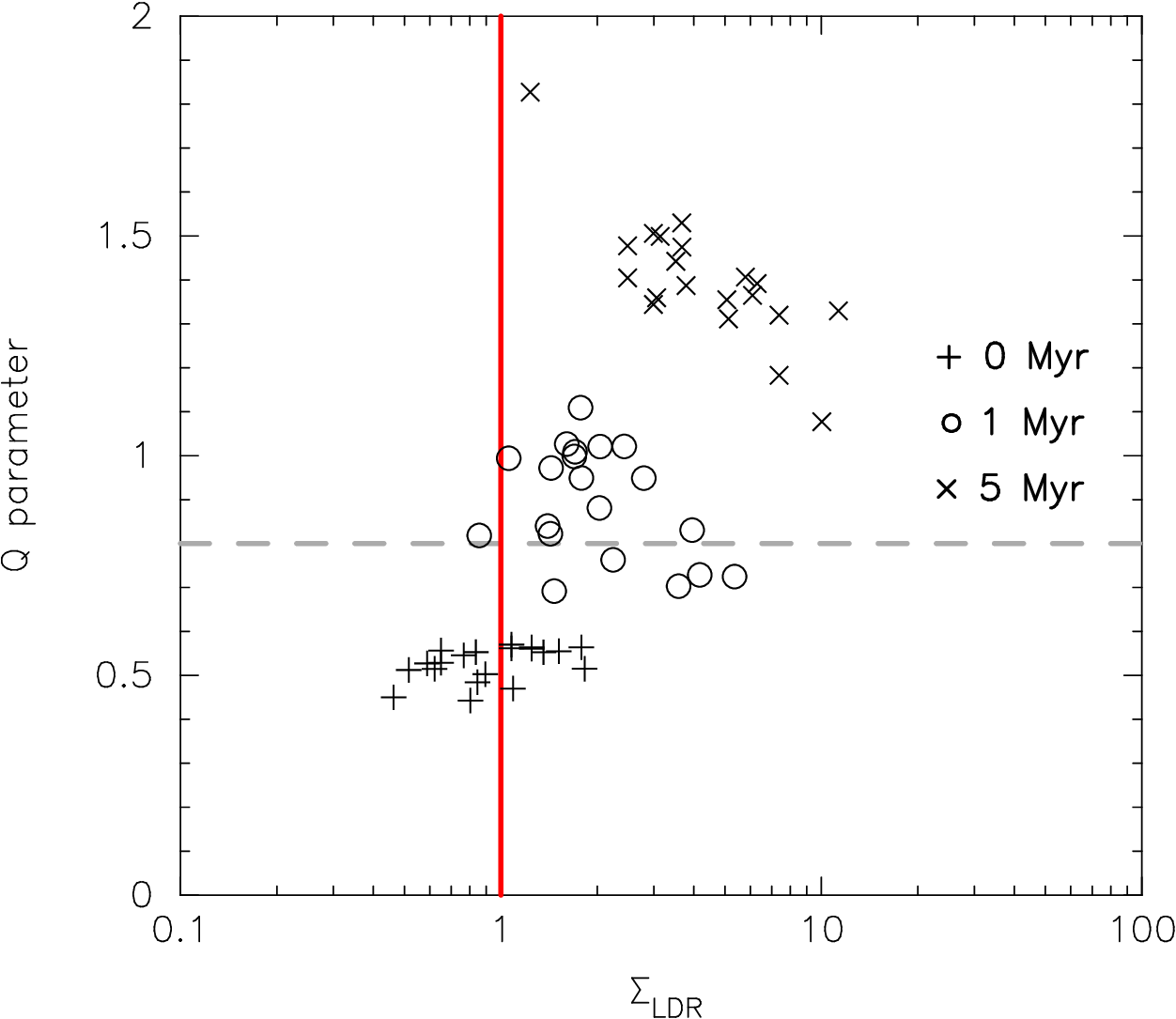}}}
\hspace*{0.1cm} 
\subfigure[$\alpha_{\rm vir} = 0.5, D = 3.0$]{\label{Q_Sig-f}\rotatebox{270}{\includegraphics[scale=0.28]{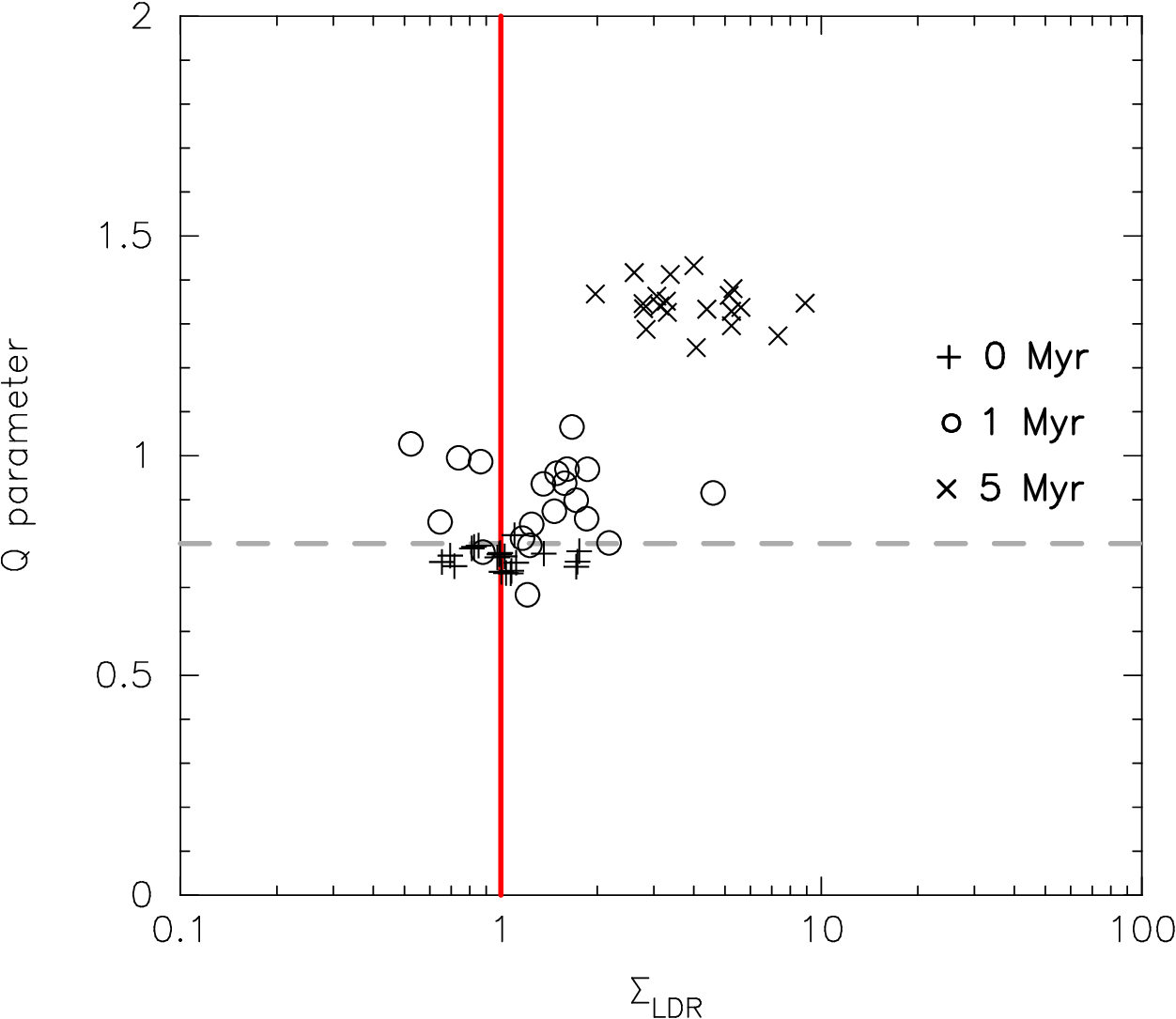}}}
\hspace*{-1.cm}
\subfigure[$\alpha_{\rm vir} = 1.5, D = 1.6$]{\label{Q_Sig-g}\rotatebox{270}{\includegraphics[scale=0.28]{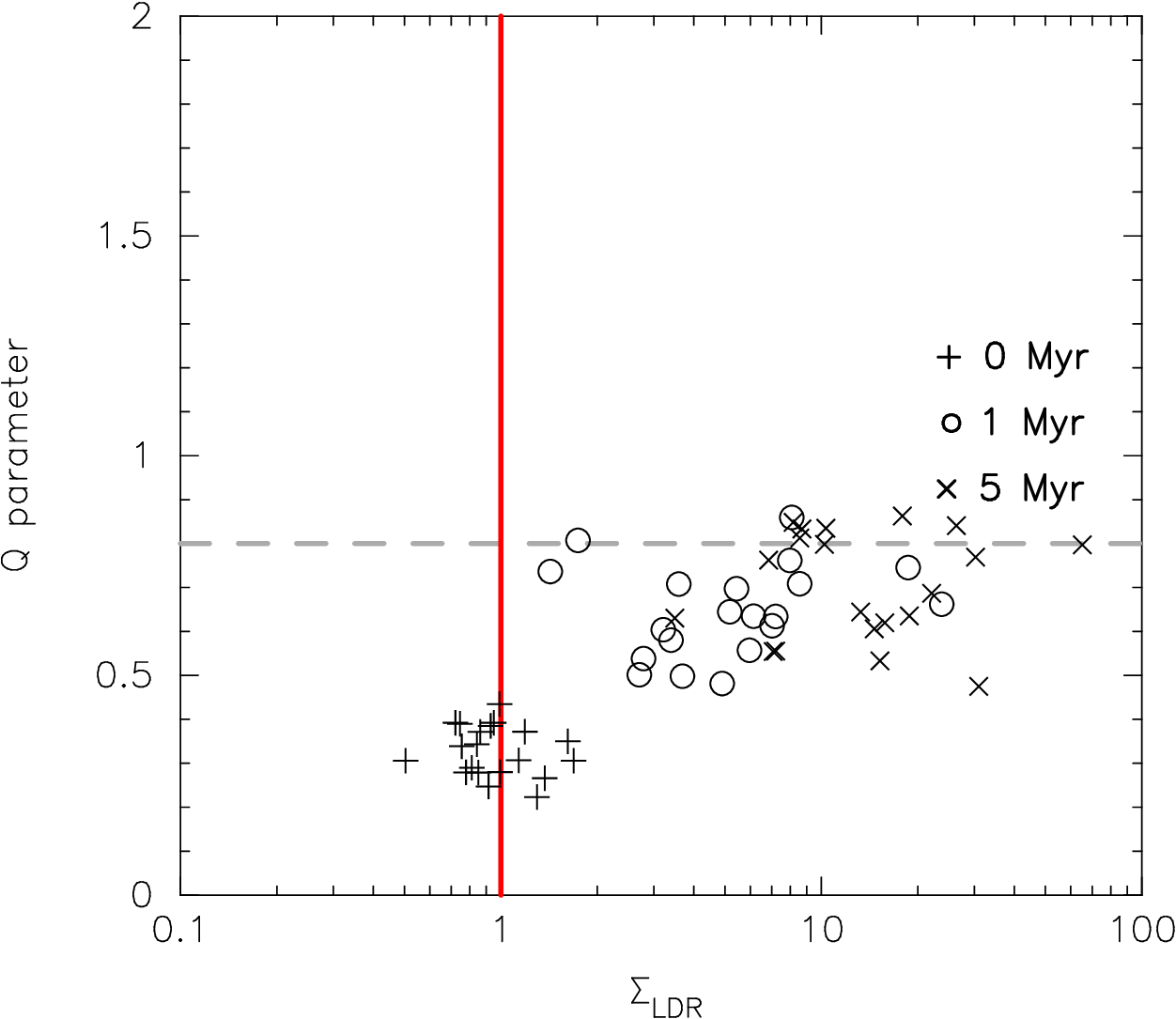}}}
\hspace*{0.1cm} 
\subfigure[$\alpha_{\rm vir} = 1.5, D = 2.0$]{\label{Q_Sig-h}\rotatebox{270}{\includegraphics[scale=0.28]{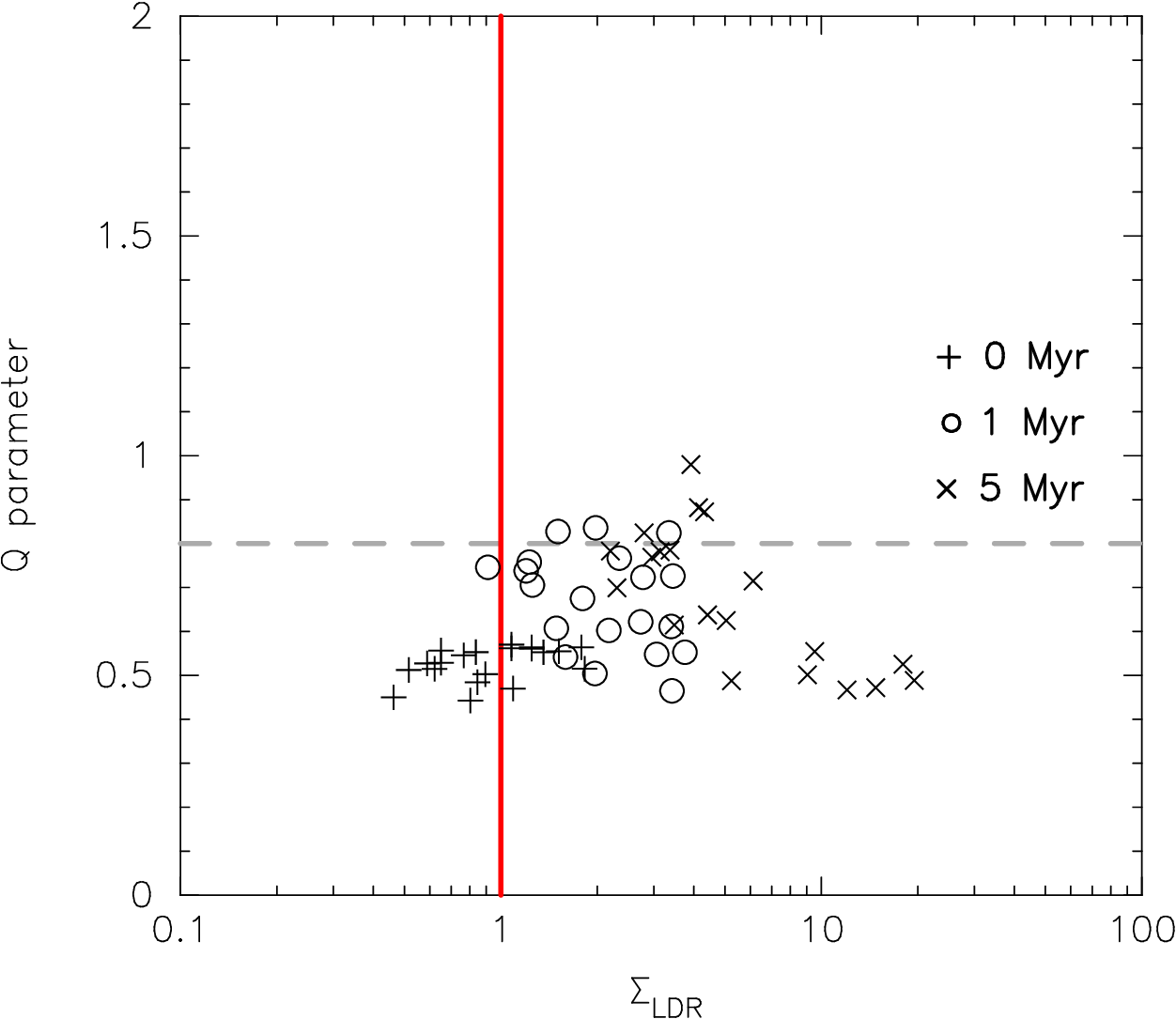}}}
\hspace*{0.1cm} 
\subfigure[$\alpha_{\rm vir} = 1.5, D = 3.0$]{\label{Q_Sig-i}\rotatebox{270}{\includegraphics[scale=0.28]{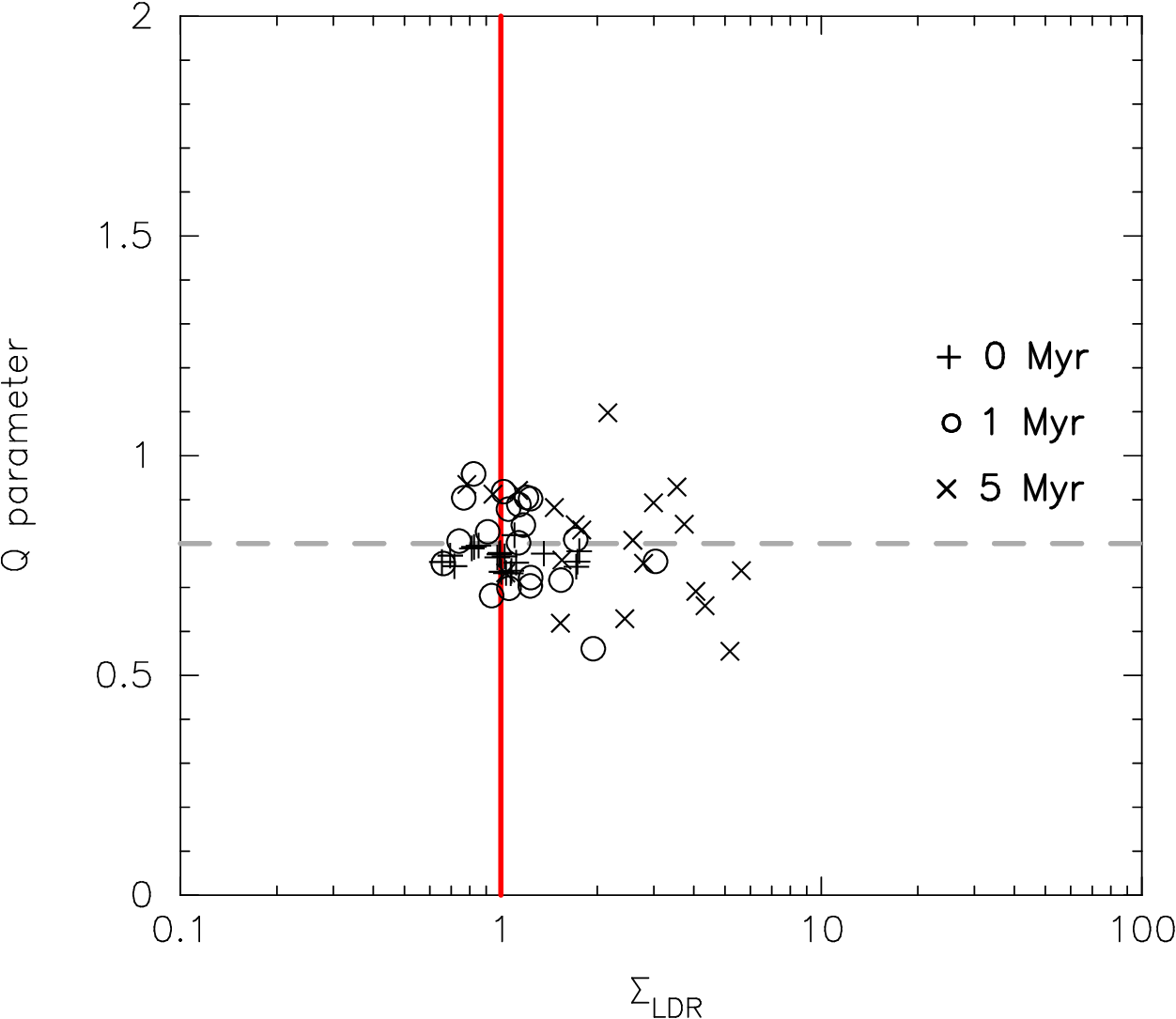}}}
\caption[bf]{The evolution of $\mathcal{Q}$--parameter versus $\Sigma_{\rm LDR}$ with time for the simulations with $N = 1500$ stars. For each simulation we plot the $\mathcal{Q}$--parameter and 
$\Sigma_{\rm LDR}$ at 0\,Myr (the plus signs), 1\,Myr (the open circles) and 5\,Myr (the crosses). The boundary between substructured associations and radially smooth clusters ($\mathcal{Q} = 0.8$) is indicated by the horizontal grey dashed line, 
and $\Sigma_{\rm LDR} = 1$ is shown by the vertical red line.   }
\label{Q_Sig}
  \end{center}
\end{figure*}

As shown in Fig.~\ref{Qpar_all}, the $\mathcal{Q}$--parameter measured at a given time can indicate the likely initial conditions of  a star forming region; something that is centrally concentrated after only 1 -- 2\,Myr is likely to have formed stars with either subvirial, or virial velocities. 
However, the large spread in possible values of $\mathcal{Q}$ means that any further inference of the initial conditions is not possible using $\mathcal{Q}$ alone. However, as pointed out by \citet{Allison10} and demonstrated in Figs.~\ref{msr_med}~and~\ref{sigm_med}, the more 
subvirial and substructured a star forming region is, the more likely mass segregation is to occur within 1\,Myr (with the caveat that any residual gas potential does not strongly affect the dynamical interactions -- see Section~\ref{discuss}). Furthermore, the level of mass segregation (if it occurs) is also higher for subvirial and substructured regions.

In Fig.~\ref{Q_MSR} we plot the $\mathcal{Q}$--parameter against mass segregation ratio, $\Lambda_{\rm MSR}$, for each suite of $N = 1500$ stars simulations. We show the values at 0\,Myr (i.e.\,\,before any dynamical evolution has taken place) by the plus signs, at 1\,Myr (the open circles) and 
at 5\,Myr (the crosses). The boundary between substructured and centrally concentrated morphologies ($\mathcal{Q} = 0.8$) is shown by the dashed grey line, and a mass segregation ratio of unity (i.e.\,\,no mass segregation) is shown by the solid red line.  

For certain initial conditions ($\alpha_{\rm vir} = 0.3, 0.5$ -- panels (a)-(f)), the evolution of the regions can be clearly seen in $\mathcal{Q} - \Lambda_{\rm MSR}$. However, it remains difficult to distinguish between subvirial and virial initial conditions. Furthermore, for the 
clusters with supervirial initial conditions (panels (g)-(i)) the plot is degenerate, as these regions do not mass segregate according to the definition of $\Lambda_{\rm MSR}$ (because the massive stars are unable to group together) and so the $\mathcal{Q} - \Lambda_{\rm MSR}$ values at different ages are overlaid.   

In order to overcome the degeneracy in $\mathcal{Q} - \Lambda_{\rm MSR}$ space for regions with supervirial velocities, we also plot the $\mathcal{Q}$--parameter against the ratio of surface densities $\Sigma_{\rm LDR}$ for the simulations with $N = 1500$ stars in Fig.~\ref{Q_Sig}. In Fig.~\ref{Q_Sig} 
we show the values at 0\,Myr (i.e.\,\,before any dynamical evolution has taken place) by the plus signs, at 1\,Myr (the open circles) and at 5\,Myr (the crosses). The boundary between substructured and centrally concentrated morphologies ($\mathcal{Q} = 0.8$) 
is shown by the dashed grey line, and $\Sigma_{\rm LDR} = 1$ is shown by the solid red line. The  $\mathcal{Q} - \Sigma_{\rm LDR}$ plot shows distinct differences between unevolved regions (the plus signs) and clusters with ages of 5\,Myr (the crosses) 
for most initial conditions, and only becomes degenerate when the initial conditions are supervirial and not substructured (Fig.~\ref{Q_Sig-i}). 

\begin{figure*}
  \begin{center}
\setlength{\subfigcapskip}{10pt}
\vspace*{-0.3cm}
\hspace*{-1.cm}\subfigure[$\alpha_{\rm vir} = 0.3, D = 1.6$]{\label{Q_Sig_lowN-a}\rotatebox{270}{\includegraphics[scale=0.28]{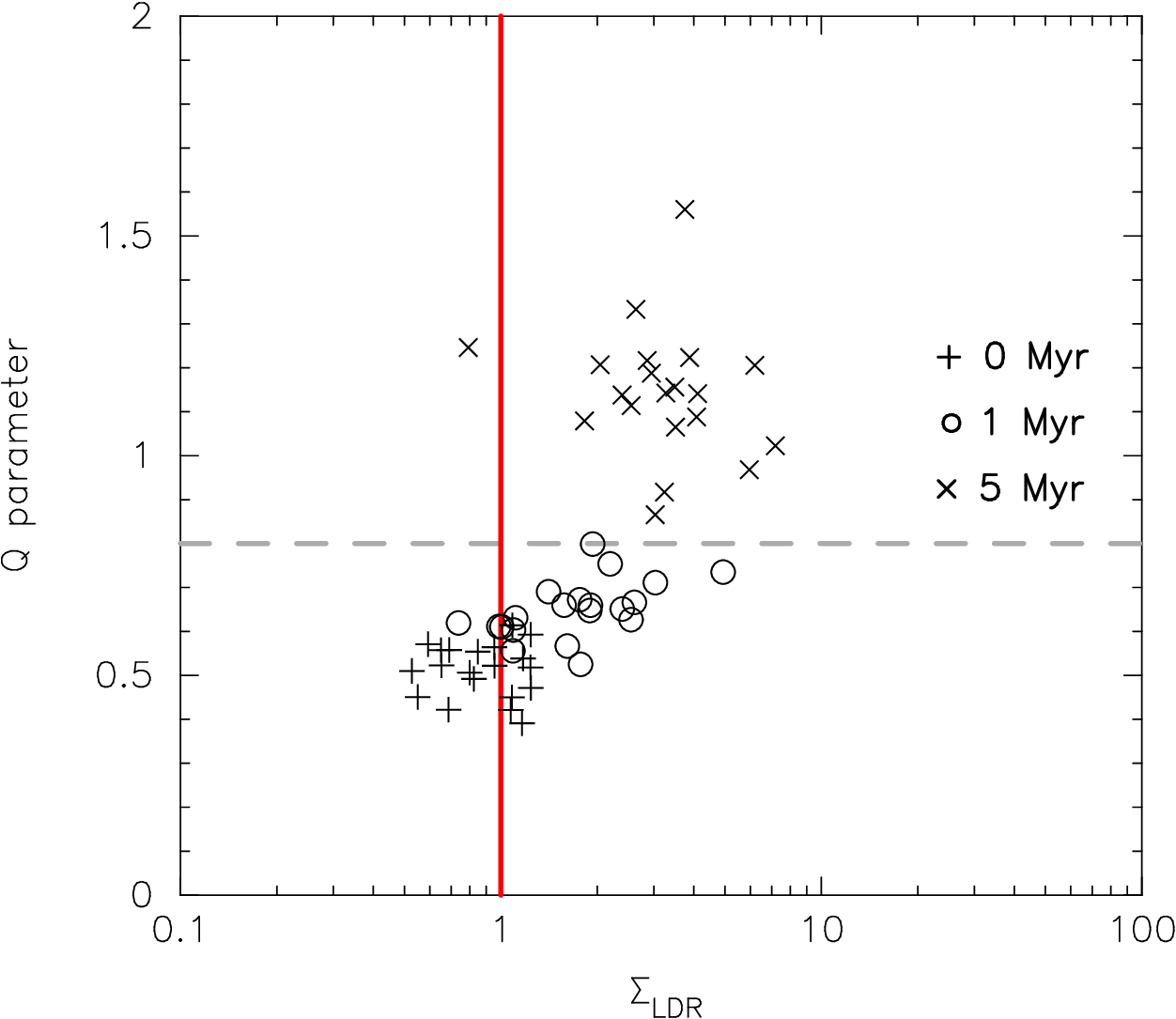}}}
\hspace*{0.1cm} 
\subfigure[$\alpha_{\rm vir} = 0.3, D = 2.0$]{\label{Q_Sig_lowN-b}\rotatebox{270}{\includegraphics[scale=0.28]{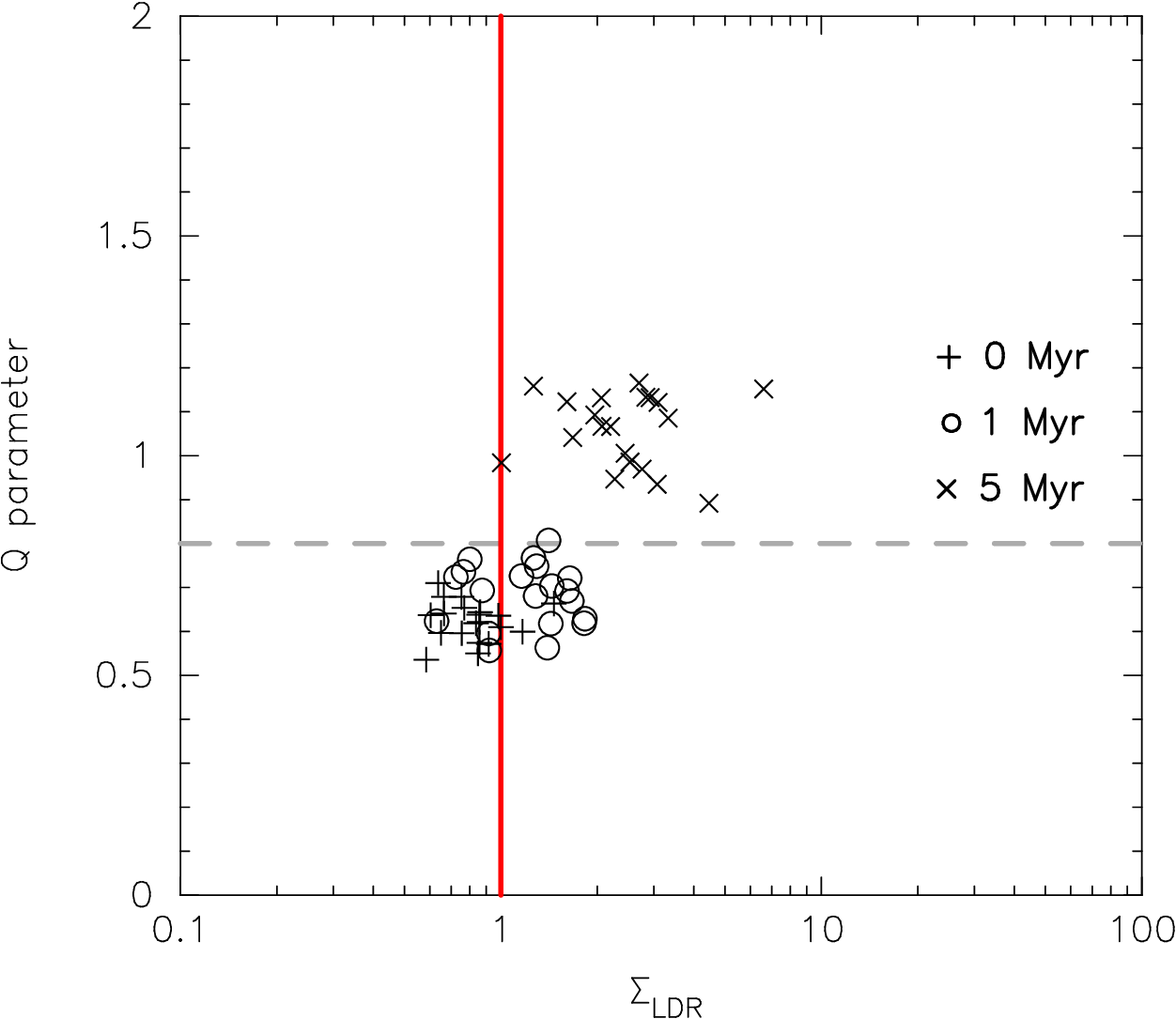}}}
\hspace*{0.1cm} 
\subfigure[$\alpha_{\rm vir} = 0.3, D = 3.0$]{\label{Q_Sig_lowN-c}\rotatebox{270}{\includegraphics[scale=0.28]{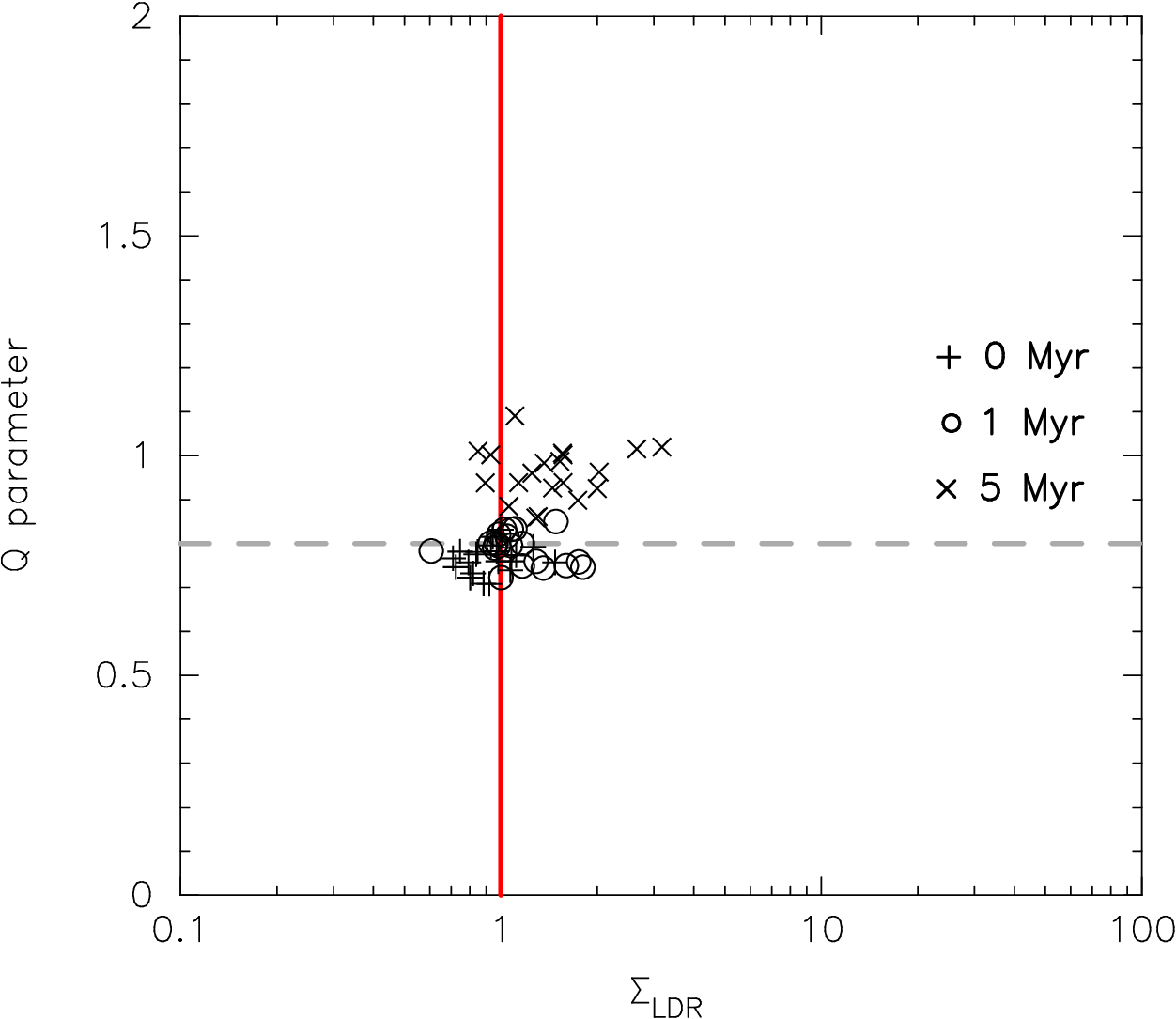}}}
\hspace*{-1.cm}
\subfigure[$\alpha_{\rm vir} = 0.5, D = 1.6$]{\label{Q_Sig_lowN-d}\rotatebox{270}{\includegraphics[scale=0.28]{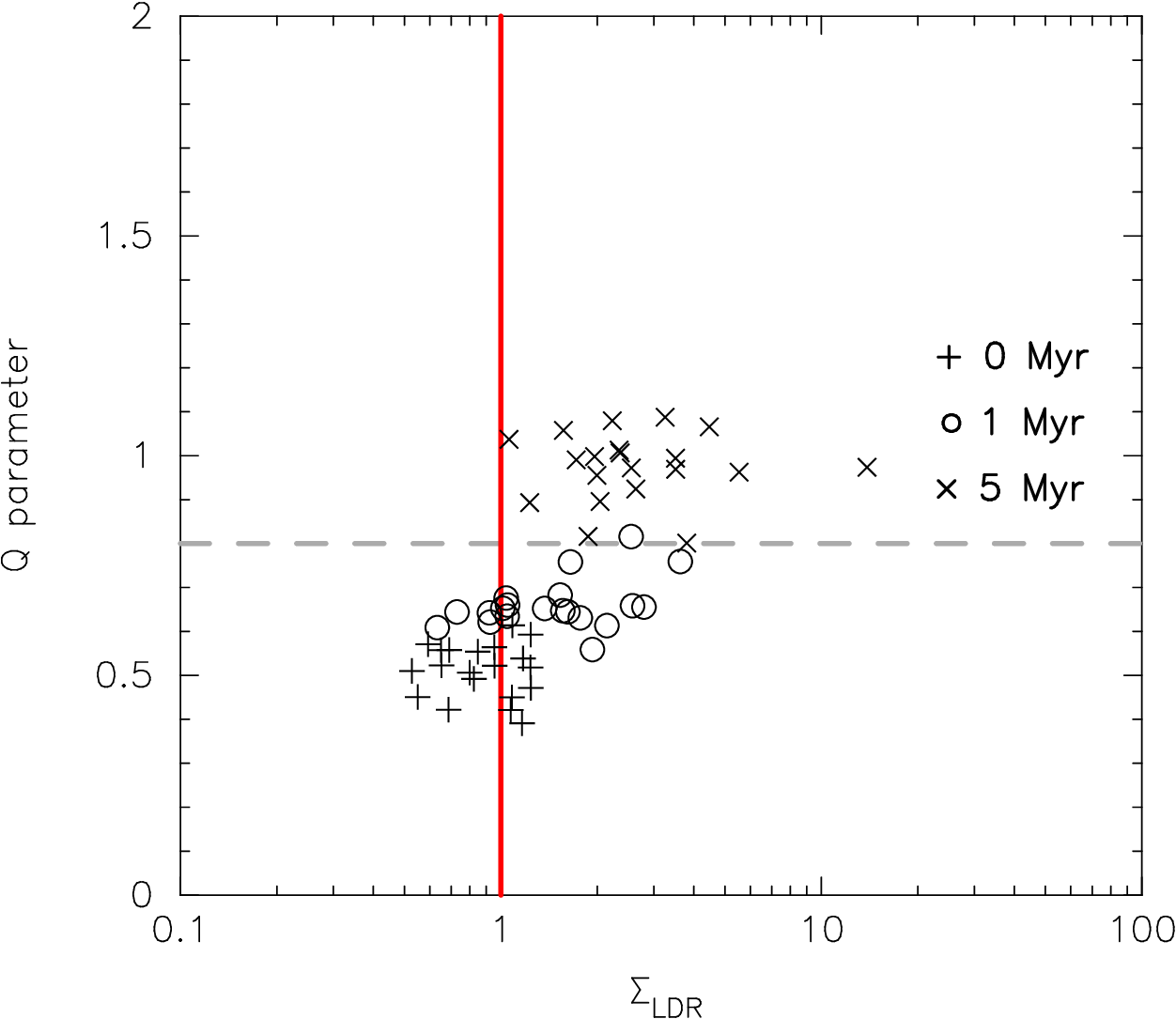}}}
\hspace*{0.1cm} 
\subfigure[$\alpha_{\rm vir} = 0.5, D = 2.0$]{\label{Q_Sig_lowN-e}\rotatebox{270}{\includegraphics[scale=0.28]{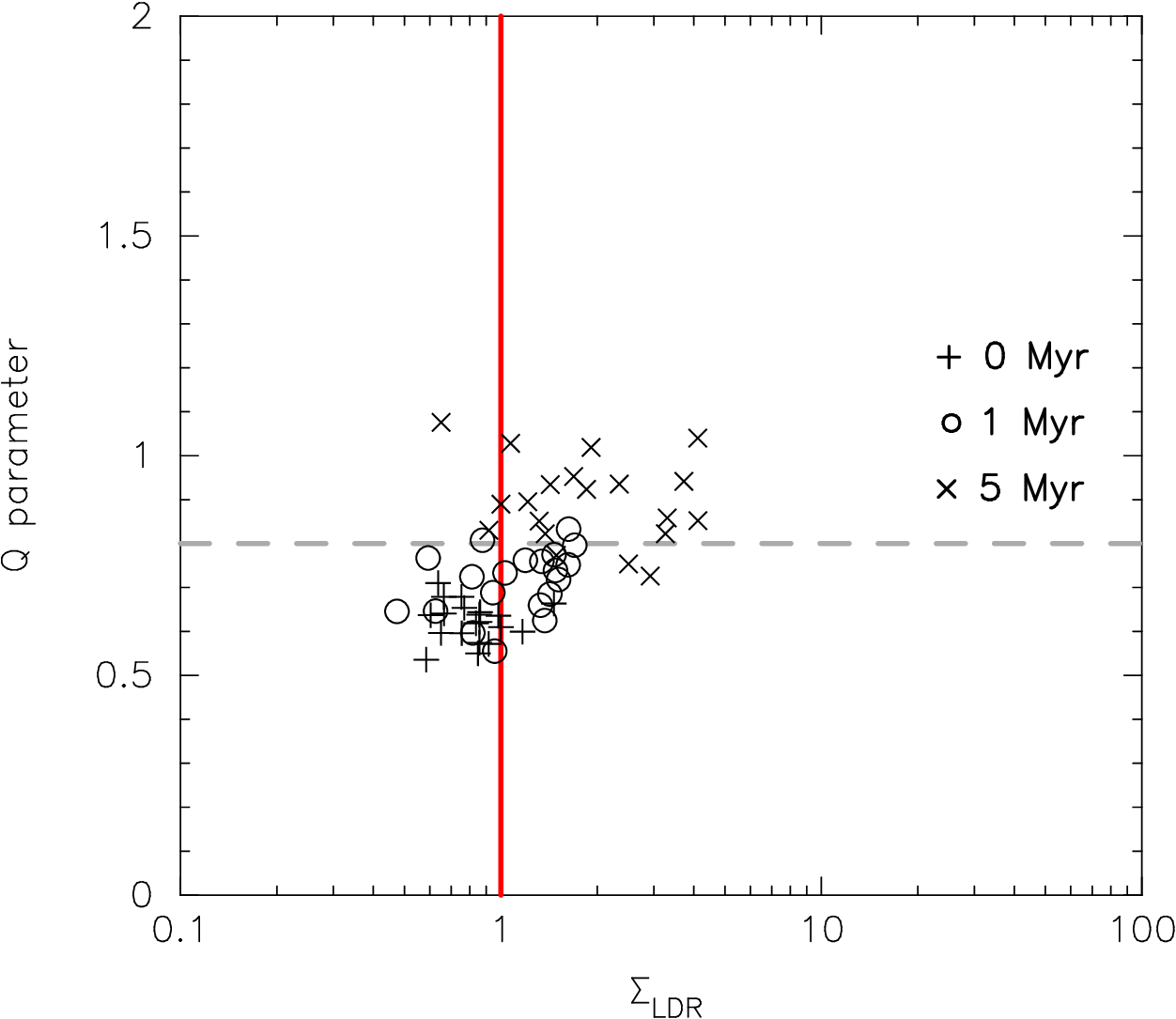}}}
\hspace*{0.1cm} 
\subfigure[$\alpha_{\rm vir} = 0.5, D = 3.0$]{\label{Q_Sig_lowN-f}\rotatebox{270}{\includegraphics[scale=0.28]{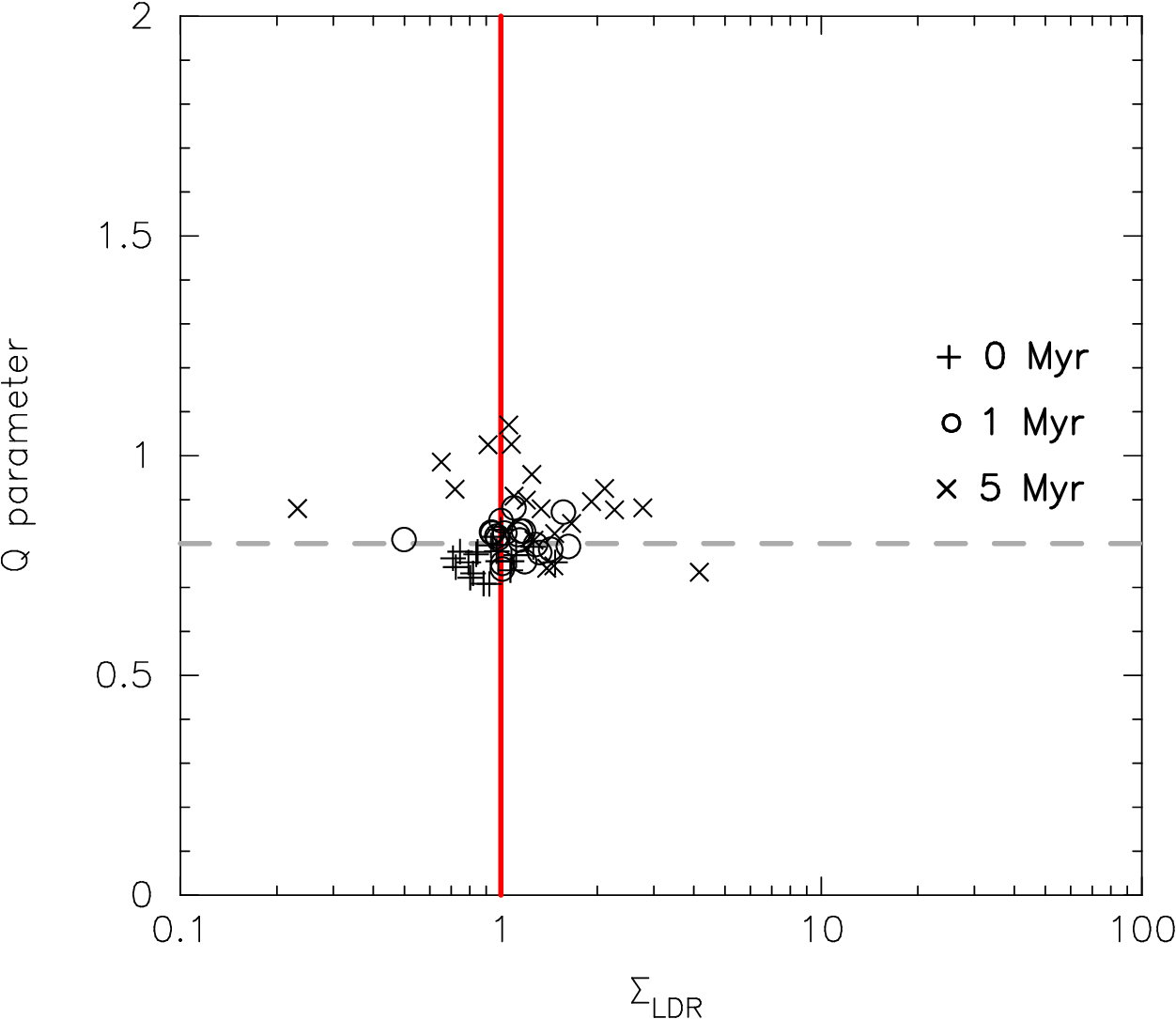}}}
\hspace*{-1.cm}
\subfigure[$\alpha_{\rm vir} = 1.5, D = 1.6$]{\label{Q_Sig_lowN-g}\rotatebox{270}{\includegraphics[scale=0.28]{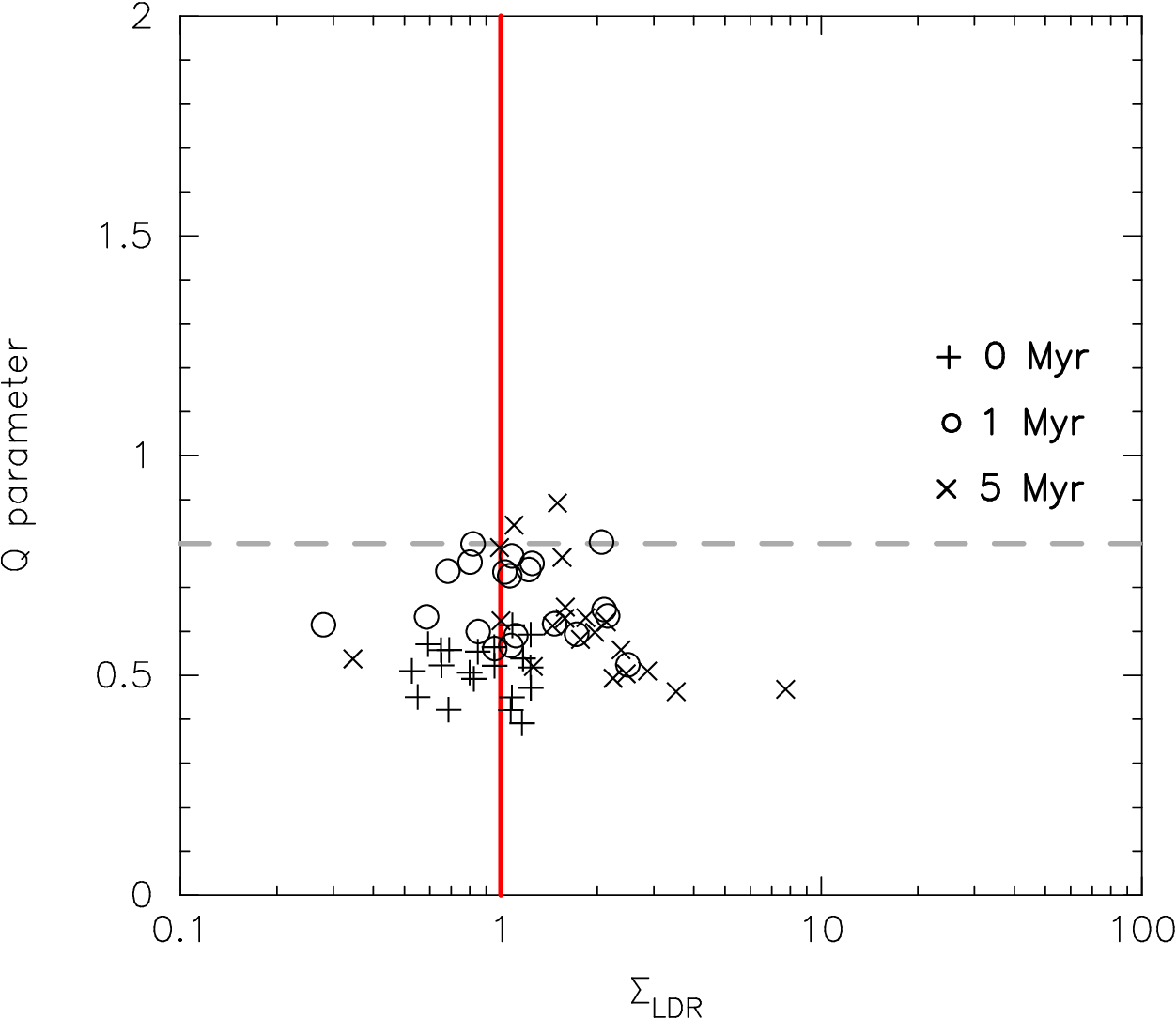}}}
\hspace*{0.1cm} 
\subfigure[$\alpha_{\rm vir} = 1.5, D = 2.0$]{\label{Q_Sig_lowN-h}\rotatebox{270}{\includegraphics[scale=0.28]{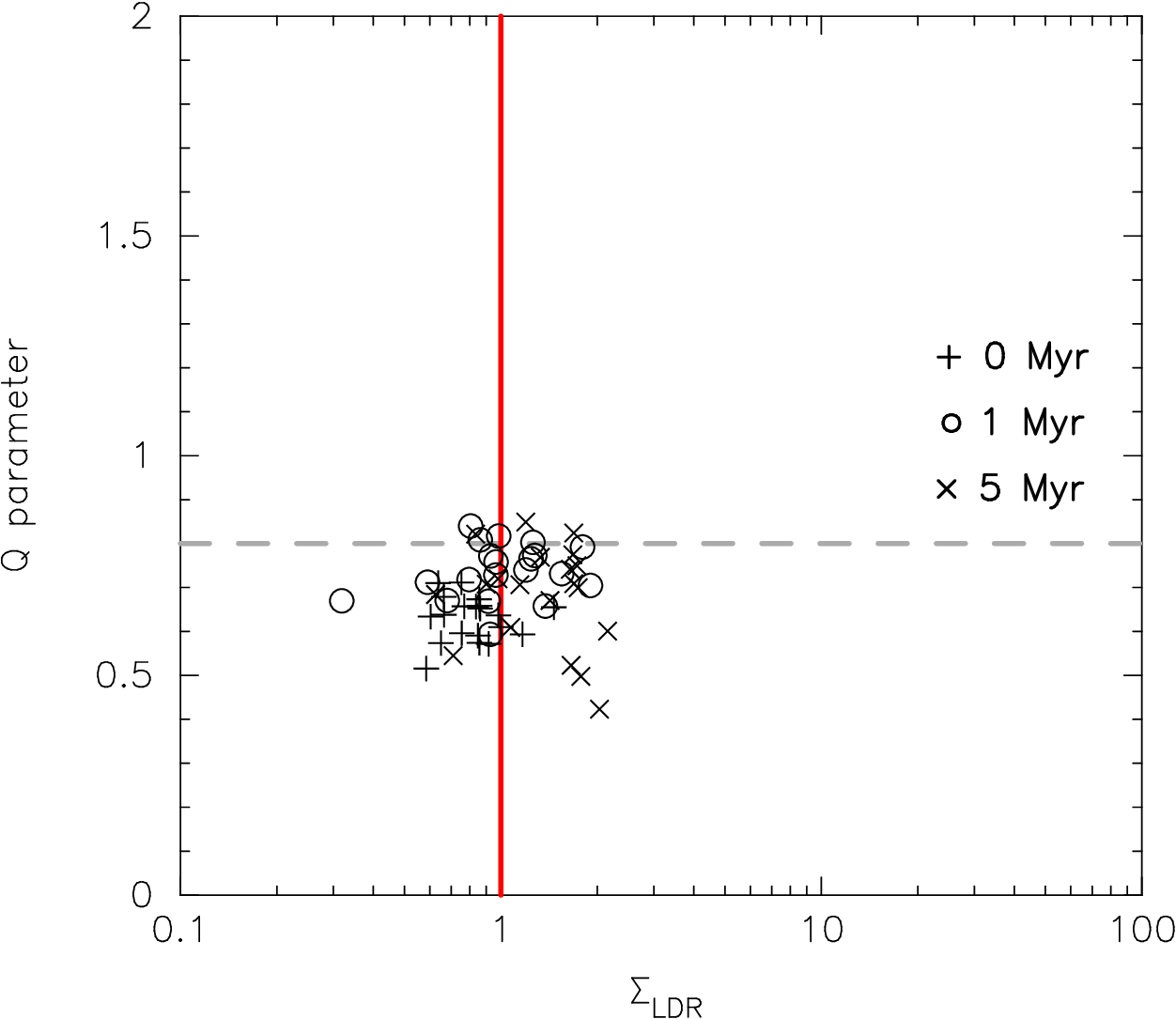}}}
\hspace*{0.1cm} 
\subfigure[$\alpha_{\rm vir} = 1.5, D = 3.0$]{\label{Q_Sig_lowN-i}\rotatebox{270}{\includegraphics[scale=0.28]{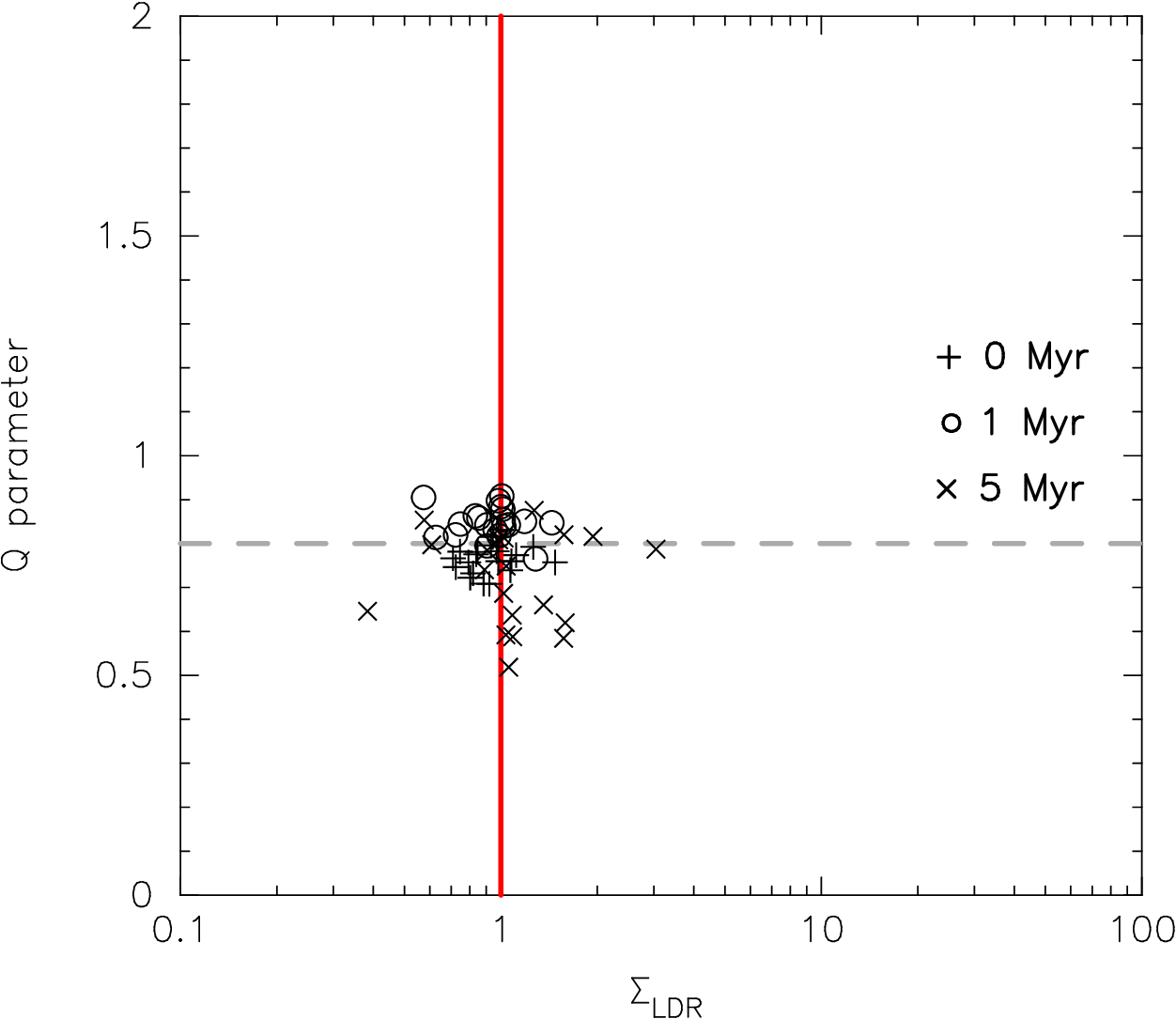}}}
\caption[bf]{The evolution of $\mathcal{Q}$--parameter versus
  $\Sigma_{\rm LDR}$ with time for the low-density ($N = 150$ stars) simulations. For each simulation we plot the $\mathcal{Q}$--parameter and 
$\Sigma_{\rm LDR}$ at 0\,Myr (the plus signs), 1\,Myr (the open circles) and 5\,Myr (the crosses). The boundary between substructured associations and radially smooth clusters ($\mathcal{Q} = 0.8$) is indicated by the horizontal grey dashed line, 
and $\Sigma_{\rm LDR} = 1$ is shown by the vertical red line.   }
\label{Q_Sig_lowN}
  \end{center}
\end{figure*}

Finally, we note that the initial densities of our simulations with $N = 1500$ stars may be higher than those observed in $\sim$50\,per cent of nearby star forming regions \citep{Bressert10,Parker12d}. We therefore show the $\mathcal{Q} - \Sigma_{\rm LDR}$ plot 
for the low-density regions (initial median surface densities
$\sim$100\,stars\,pc$^{-2}$) in Fig.~\ref{Q_Sig_lowN}. 

The dynamical evolution of these $N=150$ regions is not as dramatic as
in the higher-$N$ simulations.  The reason for this is three-fold.
Firstly, lower-$N$ results in lower-number statistics and so the
quantitative measures we are calculating are less significant.
Secondly, lower-$N$ results in fewer stars to form retinues around the
higher-mass stars (and those higher-mass stars are less likely to be
significantly more massive than the average due to random sampling
from the IMF).  Finally, the lower-$N$ clusters have longer dynamical
timescales (for the same radius systems) than the larger-$N$ clusters,
so that for the same physical age they will be dynamically less
evolved (note that two-body relaxation is generally unimportant except
in dynamically mass segregating regions that become dense, so the
$N$-dependence of two-body relaxation is unimportant).

For these reasons we can only readily distinguish between different 
epochs (0 and 5\,Myr) in the $\mathcal{Q} - \Sigma_{\rm LDR}$ plot
when the simulations are
substructured and subvirial ($\alpha_{\rm vir} = 0.3; D = 1.6, 2.0$ --
panels (a) and (b) in Fig.~\ref{Q_Sig_lowN});  i.e.\,\,when the evolution
has been dramatic.

\section{Discussion}
\label{discuss}


In the simulations presented in Section~\ref{results},  we have seen that there are some trends (and some lack of trends) in
the evolution of the quantitative structure parameters  $\mathcal{Q}$,
$\Lambda_{\rm MSR}$, and $\Sigma_{\rm LDR}$ with time.  How these
parameters evolve depends on the initial substructure present in a
region (modelled by the fractal dimension, $D$, of the initial
conditions), and the global dynamical `temperature' of the region
(modelled by the global virial ratio, $\alpha_{\rm vir}$).

It is very interesting to uncover the
dynamics at work in the systems we have simulated.  In common with past work we have
found that substructure is erased in cool and tepid regions, but
retained in hot regions \citep[e.g.][]{Goodwin04a,Parker12d}.  Also
following past work we have found that the massive stars rapidly dynamically
segregate in cool and tepid regions, but that the clusters so formed
can evolve in different ways, and even destroy themselves
\citep[e.g.][]{Allison09b,Allison10,Allison11}. 

The parameter space covered by \citeauthor{Allison10} did not include
regions that initially expand (i.e.\,\,supervirial velocities). When
regions have supervirial velocities, the massive stars usually have a
similar spatial concentration to the average stars,  and so the
$\Lambda_{\rm MSR}$ technique does not find these associations to be
mass segregated. 

When using the
$\Sigma - m$ method with its corresponding $\Sigma_{\rm LDR}$ local ratio, in nearly \emph{all} regions the massive stars 
find themselves in regions of higher density than the median value in the cluster,
irrespective of the initial amount of substructure, or virial
ratio. Indeed,  even in the regions which start as uniform spheres
(Figs.~\ref{sigm_med-c},~\ref{sigm_med-f}~and~\ref{sigm_med-i}) the
ten most massive stars have significantly higher surface densities
than the average. 

We attribute this to gravitational focusing from the massive
  stars, which act as 
potential wells and effectively `sweep up' a retinue of low mass stars as the
region evolves (note the concentration of low-mass stars (shown by
the black points) around the high-mass stars (the red triangles) in
Figs.~\ref{hot-b}~and~\ref{hot-c}).  This means that the local surface density around high-mass
stars almost always increases with time.

The evolution of spatial structure, as measured by the
$\mathcal{Q}$--parameter, follows a similar pattern to the measures of
mass segregation and local density. As noted by \citet{Parker12d}, subvirial,
substructured regions lose their structure  much faster, and become
more centrally concentrated, than virialised regions with smooth
initial conditions (compare the $D = 1.6, \alpha_{\rm vir} = 0.3$
region in Fig.~\ref{Qpar-a} with the $D = 3.0, \alpha_{\rm vir} = 0.5$
region  in Fig.~\ref{Qpar-f}). Supervirial regions tend to erase some
substructure, but remain substructured for the duration of the
simulation as the stars in these models never fully mix
together. However, a low $\mathcal{Q}$-parameter   does not
necessarily imply that the region is dynamically young, or that it has
always been substructured to some degree. In Fig.~\ref{Qpar-i} we see
that several models that began as uniform spheres have developed
substructure during their evolution.

Whilst an old (10\,Myr) region with substructure may imply that the
initial conditions were supervirial, or that the region is dynamically
young, it is more informative to combine the $\mathcal{Q}$--parameter
with the measures  of mass segregation and local density to decide whether a
star-forming region has undergone dynamical evolution. In
Figs.~\ref{Q_MSR}~and~\ref{Q_Sig} we show the evolution of
$\mathcal{Q}$ against $\Lambda_{\rm MSR}$, and $\mathcal{Q}$ against
$\Sigma_{\rm LDR}$, respectively.  The $\mathcal{Q}-\Lambda_{\rm MSR}$
plot (Fig.~\ref{Q_MSR}) shows that, for cool and tepid regions,
structure is erased as the level of mass segregation
increases. However, as dynamical evolution leads to massive stars
attaining higher local densities than  low-mass stars, the
$\mathcal{Q}-\Sigma_{\rm LDR}$ plot (Fig.~\ref{Q_Sig}) enables us to 
determine whether dynamical evolution has taken place, and if so, to
distinguish between initially cool/tepid and hot regions
(i.e.\,\,bound clusters versus unbound associations).


The most obvious question to ask is to what extent these results can
be applied to real observations of a single snapshot in the evolution
of a real star forming region?

Before we do this, it is worth quickly dicussing the difference
between physical and dynamical ages.  Two systems with the same
physical age (say, 1~Myr), can have very different dynamical ages.
The dynamical age is a measure of how much the system can have relaxed
into a rough equilibirum.  To first order, the dynamical age is the
number of crossing times old the system is.  \citet{Gieles11} 
define a `cluster' as distinct from an `association' from the number
of crossing times old a system is.  An unbound system is, by
definition, always less than a crossing time old (they define this as
an association).  A bound system can be more than one crossing time
old (a cluster).  However, in the very young systems we are
considering, even if a system is bound it may still be dynamically
young in that it is only 1 or 2 crossing times old.

The evolution of $\mathcal{Q}$ we find above is a proxy of dynamical
age.  Relaxation will increase $\mathcal{Q}$ as substructure is
erased, therefore with increasing dynamical age we have increasing
$\mathcal{Q}$.  

\subsection{Comparison with observations}

What might we say from real observations of $\mathcal{Q}$, 
$\Lambda_{\rm MSR}$, and $\Sigma_{\rm LDR}$? The key parameter is $\mathcal{Q}$, which provides an upper limit on
the initial $\mathcal{Q}$, and hence the initial degree of
substructure in the region. Here, we discuss three different regimes of $\mathcal{Q}$:\\


{\bf Low-$\mathcal{Q}$} ($\mathcal{Q} < 0.8$ or $1$).  If 
a region has a low-$\mathcal{Q}$ then {\em it must be dynamically young}.  It has
not managed to erase its substructure, and is not well-mixed. 

In dynamically young regions $\Lambda_{\rm MSR}$ provides a measure of
how well-separated the massive stars were at birth.  In our
simulations we randomly place the massive stars and so for
low-$\mathcal{Q}$ we always find $\Lambda_{\rm MSR} \sim 1$.  If we observed 
a low-$\mathcal{Q}$ but a $\Lambda_{\rm MSR}$ value significantly above or below unity this would
provide information on the formation of massive stars relative to
low-mass stars.

Even if a region is globally dynamically young, locally (especially around massive
stars) it might be dynamically older.  We find that $\Sigma_{\rm LDR}$
increases with time as the massive stars gain a retinue of low-mass
stars.  Therefore, the observed value of $\Sigma_{\rm LDR}$ is an
upper limit on the initial $\Sigma_{\rm LDR}$.  

High values of $\Sigma_{\rm LDR}$ with low-$\mathcal{Q}$ are probably indicative of
a high degree of initial substructure acting as seeds for massive
stars to gain a retinue (see panels (g)--(i) of Fig.~\ref{Q_Sig}).\\

{\bf Moderate-$\mathcal{Q}$} ($\mathcal{Q} \sim 1$).  Regions with
moderate-$\mathcal{Q}$ may have formed with moderate-$\mathcal{Q}$ and
be dynamically young (i.e.\,\,not have changed their structure much since
birth).  Alternatively, they may have formed with low-$\mathcal{Q}$ and have
undergone a small degree of violent relaxation.  They cannot be
globally dynamically old.

In our simulations with no initial mass segregation we find that when
$\mathcal{Q} \sim 1$ that $\Lambda_{\rm MSR}$ is around 1 -- 2
(i.e.\,\,hardly significant).  An observation of a significant
$\Lambda_{\rm MSR}$ in a moderate-$\mathcal{Q}$ region would strongly
suggest that the massive stars formed with a significant-$\Lambda_{\rm
  MSR}$ (i.e.\,\,they were initially mass segregated).

However, we do find a wide range in $\Sigma_{\rm LDR}$ for
moderate-$\mathcal{Q}$ (see again panels (g)--(i) of
Fig.~\ref{Q_Sig}).  Again, high values of $\Sigma_{\rm MSR}$ 
with moderate-$\mathcal{Q}$ are probably indicative of
a high degree of initial substructure.\\

{\bf High-$\mathcal{Q}$} ($\mathcal{Q} > 1.5$).  Regions with
high-$\mathcal{Q}$ could be dynamically very old and have erased their
substructure (i.e.\,\,the 5~Myr old systems in panels (a)--(f) of
Figs.~\ref{Q_MSR} and~\ref{Q_Sig}).  Or they may have formed with a
high-$\mathcal{Q}$.  After only 1~Myr cool regions have erased their
substructure and are indistinguishable (by $\mathcal{Q}$) from regions
that formed centrally concentrated.

With no initial mass segregation high-$\mathcal{Q}$ regions rapidly
increase both $\Lambda_{\rm MSR}$ and $\Sigma_{\rm LDR}$ meaning they
are indistinguishable from systems that started with high 
$\Lambda_{\rm MSR}$ and $\Sigma_{\rm LDR}$.  And by 5~Myr the
evolution is such that almost any value of $\Lambda_{\rm MSR}$ and
$\Sigma_{\rm LDR}$ is associated with high-$\mathcal{Q}$ regions.
Typically, $\Sigma_{\rm LDR}$ is high at late times in 
high-$\mathcal{Q}$, but there are outliers in which it is not.

\subsubsection{$\rho$~Oph, Taurus and Cyg~OB2}

There are three young regions for which we have information in the
literature on $\mathcal{Q}$, $\Lambda_{\rm MSR}$, and $\Sigma_{\rm LDR}$.

\citet{Parker12c} analyse $\rho$~Oph and find for this $\sim 1$~Myr
region that $\Lambda_{\rm MSR} \sim 1$, and
$\Sigma_{\rm LDR} \sim 1$ (no evidence for any mass segregation or massive stars residing in over-densities).  From \citet{Cartwright04} we have
$\mathcal{Q} = 0.85$ for $\rho$~Oph.

Examination of Fig.~\ref{Q_Sig} shows that several sets of initial
conditions have $\Sigma_{\rm LDR} \sim 1$ and $\mathcal{Q} = 0.85$ at
1~Myr.  These tend to have tepid or hot initial conditions, and
relatively high fractal dimensions.  

The best-bet for $\rho$~Oph is that it formed with a $\mathcal{Q}$
similar to what we see now, and is dynamically young.  It could be
globally bound or unbound, and without dynamical information it is
impossible to tell.  But we are probably seeing something with a
global structure not too dissimilar to that with which it formed.

However, it should be noted that $\rho$~Oph only contains $\sim 250$
members \citep[e.g.][]{Alves12} meaning that any quantitative measure, and especially $\Lambda_{\rm MSR}$, and
$\Sigma_{\rm LDR}$ will be rather noisy (see Section~\ref{structms}).

For Taurus, \citet{Cartwright04} find $\mathcal{Q} = 0.47$
(extremely substructured).  \citet{Parker11b} find that
$\Lambda_{\rm MSR} \sim 0.7$ (it is inversely mass segregated).  
\citet{Kirk10} perform an analysis of subgroups of Taurus which is not
too dissimilar to $\Sigma_{\rm LDR}$ and find {\em local} mass
segregation.

The very low global value of $\mathcal{Q}$ shows that globally Taurus
is dynamically young (an unsurprising result given 
the roughly 20~pc extent of
this 1~Myr old region).  As discussed by \citet{Parker11b}, the
$\Lambda_{\rm MSR} \sim 0.7$ is therefore probably primordial as
dynamics have had no opportunity to change the global structure.
However, the local mass segregation in groups could suggest that
subclusters are dynamically old (given their sub-pc sizes this is very
likely).  

In a recent paper, \citet{Wright13} find that the massive association Cyg~OB2 has a low
$\mathcal{Q}$--parameter ($\mathcal{Q} \sim 0.4 - 0.5$) and no evidence for 
mass segregation or massive stars residing in over-densities
($\Lambda_{\rm MSR} = 1.14$ and $\Sigma_{\rm LDR} = 1.44$), based
  on data collated in \citet{Wright10}. This strongly suggests 
that  Cyg~OB2 has not undergone any significant dynamical evolution, and probably formed with a similar morphology 
and density to that currently observed; i.e. a sparse \citep[$\tilde{\Sigma} = 19$\,stars\,pc$^{-2}$,][]{Wright13}, 
unbound association.

\subsection{Caveats and assumptions}

Our main result is that the $\mathcal{Q}-\Sigma_{\rm LDR}$ plot (Fig.~\ref{Q_Sig}) enables us
determine whether dynamical evolution has taken place, and if so, to
distinguish between initially cool/tepid and hot regions
(i.e.\,\,bound clusters versus unbound associations).

We find that the massive stars only acquire a retinue of low mass stars if the initial density of the region is relatively
high. Our model fractals have initial median local surface densities of
$\sim$5000\,stars\,pc$^{-2}$ (the blue dashed lines in
Figs.~\ref{cool-g}~and~\ref{hot-g}).  If we plot the evolution of
$\mathcal{Q}-\Sigma_{\rm LDR}$  for fractals with initial median
surface densities of $\sim$100\,stars\,pc$^{-2}$, then little
dynamical evolution occurs and the $\mathcal{Q}-\Sigma_{\rm LDR}$ plot
becomes degenerate in time (Fig.~\ref{Q_Sig_lowN}). This (surface) density threshold of 100\,stars\,pc$^{-2}$ 
corresponds to a volume density threshold of 100\,stars\,pc$^{-3}$ because the depth scale of the simulations is of order 1\,pc.

A recent analysis of simulations of high-mass star formation by
\citet{Parker13a} did not find any evidence for primordial mass
segregation in regions which form in hydrodynamical simulations, nor
did they find evidence for subsequent accumulation of retinues 
according to the  $\Sigma -m$ method. In those simulations
the initial median surface density was similar to that of the low-density regions modelled here,  and this result is consistent with 
the lack of high $\Sigma_{\rm LDR}$ ratios in Fig.~\ref{Q_Sig_lowN}.  This returns us to the
distinction between physical and dynamical ages.  The relatively
higher initial (surface) densities of our initial conditions reduce
the dynamical (crossing) times of our simulations relative to lower
densities -- 5~Myr of physical time in Fig. ~\ref{Q_Sig_lowN}
corresponds to less dynamical time than 5~Myr of physical time in
Fig.~\ref{Q_Sig}.  

\begin{figure}
\begin{center}
\rotatebox{270}{\includegraphics[scale=0.3]{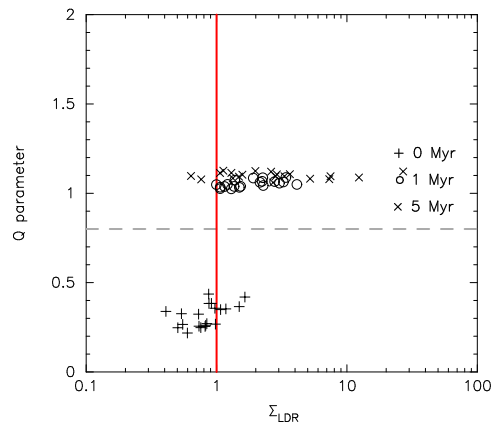}}
\end{center}
\caption[bf]{As Fig.~\ref{Q_Sig-g}, but for regions where the velocities of stars are not correlated. All substructure is erased, and fewer models show over-densities around massive stars according to the $\Sigma_{\rm LDR}$ method, than in the case of correlated velocities.}
\label{Q_Sig_uncorr}
\end{figure}

Aside from the initial density, another assumption in our simulations
is that the velocities of stars are correlated by distance. If we
remove any correlation, and simply draw velocities from a Gaussian
distribution and scale to the required virial ratio, we remove any
initial substructure on  very fast ($<$1\,Myr) timescales
(Fig.~\ref{Q_Sig_uncorr}). Therefore, in order to distinguish between
bound clusters and unbound associations using the
$\mathcal{Q}-\Sigma_{\rm LDR}$ plot, we require the (reasonable)
assumption that the velocities of stars are initially correlated on
local scales \citep{Larson81}.

The competitive accretion theory of star formation predicts that the
most massive stars should be primordially mass segregated \citep[e.g.][]{Zinnecker82,Bonnell03}, and such
behaviour is seen in \emph{some} simulations of massive star formation
\citep{Maschberger11}, but not in others
\citep{Girichidis12,Dale12a,Dale13,Parker13a}. Whilst mass segregation
is observed in some clusters \citep{Hillenbrand98,Allison09a,Sana10,Pang13},
it may occur purely due to dynamical interactions \citep{Allison10} as it is not
observed in some clusters and associations which are dynamically young
\citep{Parker11b,Parker12c,Delgado13,Wright13}. Alternatively, the observed levels 
of mass segregation (when it is present) in clusters could be a combination of some primordial 
segregation, with later dynamical segregation \citep{Moeckel09a}. 

In our simulations we see that $\Sigma_{\rm LDR}$ always increases
with time.  We started with no mass segregation or high local densities around massive stars, but even if we had
started with $\Sigma_{\rm LDR} > 1$, then $\Sigma_{\rm LDR}$ should
still tend to increase.  But if we make the assumption of no {\em initial}
mass segregation, then observed values of $\mathcal{Q}$ and
$\Sigma_{\rm LDR}$ can be used to estimate the
initial density of the region.

The dynamical evolution of substructured regions is highly
stochastic. Two regions with (statistically) identical initial
conditions can exhibit very different degrees of mass segregation,
which can occur at different times in the regions'
evolution. Similarly,  \citet{Parker12d} find a large scatter in the
evolution of $\mathcal{Q}$--parameter and median surface density; and
\citet{Parker12b} find a large scatter in the evolution of binary star
orbitial properties, in substructured regions. Using the
$\mathcal{Q}$--parameter  in isolation is not enough to determine
whether or not a star forming region has undergone dynamical
evolution, and should be coupled with the $\Sigma_{\rm LDR}$ ratio of
mass-weighted local density. Recent work by \citet{Delgado13} considered both the
$\mathcal{Q}$--parameter  and $\Lambda_{\rm MSR}$, but adopted the
median MST length. This is in some ways analagous to using the $\Sigma_{\rm LDR}$
ratio, and these authors argued for different initial conditions 
for the clusters Berkeley~94 and Berkeley~96 (the former
likely to have undergone warm expansion,  and the latter cool
collapse).

Our simulations are pure $N$-body models, and as such do not
  include the effects of gas left over from the star formation
  process. If star formation is inefficient, then previous studies have suggested that the rapid removal of this gas from a star forming region 
 will dominate its subsequent evolution \citep[e.g.][and many others]{Tutukov78,Whitworth79,Lada84,Goodwin97,Goodwin06,Baumgardt07,Parmentier13}.

However, these (and other) studies usually assume that the stars and
gas are well coupled; i.e.\,\,that the spatial distribution of the stars
and gas is similar, and that the stars are in equilibrium with the gas.
It is highly unclear if either of these assumptions are true.

\citet{Goodwin09} showed that the the dynamical state of the stars is
crucial to how they react to gas expulsion, slow-moving stars will
`feel' the effects of gas expulsion far less than fast-moving stars
(relative to their equilibrium values).  Recent work 
analysing hydrodynamical simulations of star formation has shown that when the stars and gas are decoupled \citep{Offner09}, then the regions where stars form tend to be gas-poor 
and so the influence of gas expulsion on the cluster's evolution is minimal \citep{Kruijssen12a}. In such a scenario, the subsequent dynamical evolution of the cluster is then dominated by 
two-body interactions, rather than gas removal \citep[e.g.][]{Smith11,Moeckel12,Gieles12}. This appears to be the case in some observed young massive clusters, which are gas-free, but still 
(sub-)virial, implying that gas expulsion has had little, or no effect
\citep{Rochau10,Cottaar12}.  \citet{Smith11} also showed that
substructure in the stellar distribution during gas expulsion can be
extremely important in the response of a system to gas expulsion (the albeit simple
case of a smooth external gas potential).

This does not, of course, mean that we should completely neglect the effects of gas on the evolution of star forming regions, even if they are modest. Recent advances in code development have enabled 
a better treatment of gas to be included in $N$-body simulations \citep[e.g.][]{Moeckel11a,Pelupessy12,Fujii13,Hubber13}, and such codes will be used in future studies.

Finally, we note that in our simulations we have access to the full,
three dimensional spatial data. However, even if we use only the 2D data 
and remove stars that lie outside two half-mass radii, the results do not change by much and we are still able to distinguish 
between bound and unbound star formation using the $\mathcal{Q} - \Sigma_{\rm LDR}$ 
plot. 

\subsection{Kinematics and the influence of binaries}

In this paper, we have deliberately refrained from presenting information on the velocities of stars in the simulations, such as the velocity dispersion as a function of stellar mass. Our reasons for doing this are two-fold; firstly, 
although velocity dispersions are available for some clusters \citep[e.g.][]{Bosch01,Gieles10,Rochau10,Cottaar12,Henault12} and associations \citep[e.g.][]{Steenbrugge03,Kiminki07}, the data are often restricted to a narrow stellar mass range 
 \citep[e.g.][]{Cottaar12}, making detailed comparisons with simulations difficult. Here it should be noted that ongoing spectroscopic surveys, such as the ESO VLT/FLAMES programme \citep{Randich12}, and the APOGEE survey \citep{Zasowski13} could soon remedy this issue. Secondly, 
the development of other quantitative measures of the dynamical state of a cluster using radial velocity measurements is currently in its infancy. \Citet{Cartwright09b} adapted the $\mathcal{Q}$-parameter for use with radial velocity measurements, 
but found that the use of the third spatial dimension instead of radial velocities gave better results. However, in future work we will analyse our simulations in greater detail to search for observational diagnostics in velocity space.

We note that the \emph{Gaia} satellite (and associated spectroscopic surveys) have the potential to distinguish between field stars and cluster members on the periphery of embedded and young open clusters, which could facilitate a direct comparison 
with runaway stars in simulations \citep{Allison12}, and even trace back individual field stars to their natal regions \citep{Moyano13}.

For simplicity, the simulations presented here do not contain any primordial binaries. Whilst the binary populations in star-forming regions are not as well constrained as in the Galactic field \citep{King12a,King12b,Duchene13}, the semi-major axis 
distributions are, to zeroth order, similar to the field \citep{King12b} and the distributions of mass ratios also are consistent with the field \citep{Metchev09,Reggiani11a,Reggiani13}. 

The presence of primordial binaries is likely to influence the degree 
to which a system mass segregates, although this needs to be tested fully as all simulations have so far neglected binaries \citep[e.g.][]{Allison10,Olczak11,Yu11}. Due to their increased system mass, massive star binaries could in principle facilitate 
a higher degree of mass segregation because the mass segregation timescale is a function of relative stellar mass \citep{Spitzer69}, although this may be balanced by an increased frequency of ejections of massive stars from unstable Trapezium-like 
systems \citep{Allison11}. Recently, \citet{Geller13b} showed that the binary fraction as a function of distance from the cluster centre could be an indicator of the amount of dynamical mass segregation that has taken place in the cluster. We plan to make a full assessment 
of the impact of primordial binaries on the $\mathcal{Q}$--parameter, $\Lambda_{\rm MSR}$ and $\Sigma_{\rm LDR}$ in future studies.

\section{Conclusions}
\label{conclude} 

In this paper, we have modelled the dynamical evolution of star
forming regions with $N = 1500$ or $N=150$ stars and varied the amount of initial
substructure and the initial bulk motion of the stars. We have searched for mass segregation, and increases in the local density around massive stars, and
compared these to the  evolution of the spatial structure of stars
over a 10\,Myr timeframe. Our conclusions are the following:\\

(i) The level of substructure in a region, as measured by the
$\mathcal{Q}$--parameter generally stays the same or 
increases with (dynamical) age.  Low values of $\mathcal{Q}$ show that
a region is dynamically young (see also Parker \& Meyer 2012).

(ii) The surface density around massive stars, as measured by the
$\Sigma - m$ technique generally increases with time.  This is due to
massive stars collecting a retinue of low-mass stars.

(iii) The relative closeness of massive stars, as measured 
by the $\Lambda_{\rm  MSR}$ method, stays the same or increases at
first, but can evolve in many different ways according to the details
of the dynamics in any situation (see also Allison et al. 2009).

We have introduced the $\mathcal{Q}-\Sigma_{\rm LDR}$ plot, which
traces the dynamical evolution of a star forming and removes (some of)
the degeneracies of using the $\mathcal{Q}$--parameter in
isolation.  Combining $\mathcal{Q}$ and $\Sigma_{\rm LDR}$
(and $\Lambda_{\rm  MSR} $ can certainly help) can provide
information on the initial energy (boundness), dynamical age, initial 
structure, initial density and initial degree of mass segregation of star forming
regions from the instantaneous projected positions and masses of the
stars.  

Finally, we note that the upcoming \emph{Gaia} space telescope, and associated ground-based spectroscopic surveys, will soon add a wealth of information on stellar velocities in star forming regions, clusters and associations.
If used in tandem with the  analysis of spatial distributions such as the $\mathcal{Q}-\Sigma_{\rm LDR}$ plot, we will be able to characterise the dynamical state (and hence initial conditions) of star forming regions, clusters and associations.

\section*{Acknowledgements}

We thank the anonymous referee for their comments and suggestions,
which improved the original manuscript. NJW acknowledges a Royal Astronomical Society Research Fellowship. The simulations in this work were performed on the \texttt{BRUTUS}
computing cluster at ETH Z{\"u}rich.  Part of this work was developed during an
 International Team programme at the International Space Science
Institute, Bern, Switzerland.

\bibliographystyle{mn2e}
\bibliography{general_ref}

\begin{thebibliography}{}

\bibitem[\protect\citeauthoryear{Allison}{Allison}{2012}]{Allison12}
Allison R.~J.,  2012, MNRAS, 421, 3338

\bibitem[\protect\citeauthoryear{Allison \& Goodwin}{Allison \&
  Goodwin}{2011}]{Allison11}
Allison R.~J.,  Goodwin S.~P.,  2011, MNRAS, 415, 1967

\bibitem[\protect\citeauthoryear{Allison, Goodwin, Parker, de Grijs, {Portegies
  Zwart} \& Kouwenhoven}{Allison et~al.}{2009}]{Allison09b}
Allison R.~J.,  Goodwin S.~P.,  Parker R.~J.,  de Grijs R.,  {Portegies Zwart}
  S.~F.,    Kouwenhoven M. B.~N.,  2009, ApJ, 700, L99

\bibitem[\protect\citeauthoryear{Allison, Goodwin, Parker, {Portegies Zwart} \&
  de Grijs}{Allison et~al.}{2010}]{Allison10}
Allison R.~J.,  Goodwin S.~P.,  Parker R.~J.,  {Portegies Zwart} S.~F.,    de
  Grijs R.,  2010, MNRAS, 407, 1098

\bibitem[\protect\citeauthoryear{Allison, Goodwin, Parker, {Portegies Zwart},
  de Grijs \& Kouwenhoven}{Allison et~al.}{2009}]{Allison09a}
Allison R.~J.,  Goodwin S.~P.,  Parker R.~J.,  {Portegies Zwart} S.~F.,  de
  Grijs R.,    Kouwenhoven M. B.~N.,  2009, MNRAS, 395, 1449

\bibitem[\protect\citeauthoryear{{Alves de Oliveira}, {Moraux}, {Bouvier} \&
  {Bouy}}{{Alves de Oliveira} et~al.}{2012}]{Alves12}
{Alves de Oliveira} C.,  {Moraux} E.,  {Bouvier} J.,    {Bouy} H.,  2012, A\&A,
  539, A151

\bibitem[\protect\citeauthoryear{{Andr{\'e}}, {Men'shchikov}, {Bontemps},
  {K{\"o}nyves}, {Motte}, {Schneider}, {Didelon}, {Minier}, {Saraceno},
  {Ward-Thompson} \& {et al.}}{{Andr{\'e}} et~al.}{2010}]{Andre10}
{Andr{\'e}} P.,  {Men'shchikov} A.,  {Bontemps} S.,  {K{\"o}nyves} V.,  {Motte}
  F.,  {Schneider} N.,  {Didelon} P.,  {Minier} V.,  {Saraceno} P.,
  {Ward-Thompson} D.,    {et al.} 2010, A\&A, 518, L102

\bibitem[\protect\citeauthoryear{Bastian, Covey \& Meyer}{Bastian
  et~al.}{2010}]{Bastian10}
Bastian N.,  Covey K.~R.,    Meyer M.~R.,  2010, ARA\&A, 48, 339

\bibitem[\protect\citeauthoryear{Bastian, Gieles, Ercolano \&
  Gutermuth}{Bastian et~al.}{2009}]{Bastian09}
Bastian N.,  Gieles M.,  Ercolano B.,    Gutermuth R.,  2009, MNRAS, 392, 868

\bibitem[\protect\citeauthoryear{Bate}{Bate}{2012}]{Bate12}
Bate M.~R.,  2012, MNRAS, 419, 3115

\bibitem[\protect\citeauthoryear{{Baumgardt} \& {Kroupa}}{{Baumgardt} \&
  {Kroupa}}{2007}]{Baumgardt07}
{Baumgardt} H.,  {Kroupa} P.,  2007, MNRAS, 380, 1589

\bibitem[\protect\citeauthoryear{Bonnell, Bate, Clarke \& Pringle}{Bonnell
  et~al.}{2001}]{Bonnell01}
Bonnell I.~A.,  Bate M.~R.,  Clarke C.~J.,    Pringle J.~E.,  2001, MNRAS, 323,
  785

\bibitem[\protect\citeauthoryear{Bonnell, Bate \& Vine}{Bonnell
  et~al.}{2003}]{Bonnell03}
Bonnell I.~A.,  Bate M.~R.,    Vine S.~G.,  2003, MNRAS, 343, 413

\bibitem[\protect\citeauthoryear{Bonnell, Clark \& Bate}{Bonnell
  et~al.}{2008}]{Bonnell08}
Bonnell I.~A.,  Clark P.~C.,    Bate M.~R.,  2008, MNRAS, 389, 1556

\bibitem[\protect\citeauthoryear{Bonnell \& Davies}{Bonnell \&
  Davies}{1998}]{Bonnell98}
Bonnell I.~A.,  Davies M.~B.,  1998, MNRAS, 295, 691

\bibitem[\protect\citeauthoryear{{Bosch}, {Selman}, {Melnick} \&
  {Terlevich}}{{Bosch} et~al.}{2001}]{Bosch01}
{Bosch} G.,  {Selman} F.,  {Melnick} J.,    {Terlevich} R.,  2001, A\&A, 380,
  137

\bibitem[\protect\citeauthoryear{Bressert, Bastian, Gutermuth, Megeath, Allen,
  {Evans, II}, Rebull, Hatchell, Johnstone, Bourke, Cieza, Harvey, Merin, Ray
  \& Tothill}{Bressert et~al.}{2010}]{Bressert10}
Bressert E.,  Bastian N.,  Gutermuth R.,  Megeath S.~T.,  Allen L.,  {Evans,
  II} N.~J.,  Rebull L.~M.,  Hatchell J.,  Johnstone D.,  Bourke T.~L.,  Cieza
  L.~A.,  Harvey P.~M.,  Merin B.,  Ray T.~P.,    Tothill N. F.~H.,  2010,
  MNRAS, 409, L54

\bibitem[\protect\citeauthoryear{{Carpenter}, {Meyer}, {Dougados}, {Strom} \&
  {Hillenbrand}}{{Carpenter} et~al.}{1997}]{Carpenter97}
{Carpenter} J.~M.,  {Meyer} M.~R.,  {Dougados} C.,  {Strom} S.~E.,
  {Hillenbrand} L.~A.,  1997, AJ, 114, 198

\bibitem[\protect\citeauthoryear{Cartwright}{Cartwright}{2009}]{Cartwright09b}
Cartwright A.,  2009, MNRAS, 400, 1427

\bibitem[\protect\citeauthoryear{Cartwright \& Whitworth}{Cartwright \&
  Whitworth}{2004}]{Cartwright04}
Cartwright A.,  Whitworth A.~P.,  2004, MNRAS, 348, 589

\bibitem[\protect\citeauthoryear{{Cartwright} \& {Whitworth}}{{Cartwright} \&
  {Whitworth}}{2009}]{Cartwright09a}
{Cartwright} A.,  {Whitworth} A.~P.,  2009, MNRAS, 392, 341

\bibitem[\protect\citeauthoryear{Casertano \& Hut}{Casertano \&
  Hut}{1985}]{Casertano85}
Casertano S.,  Hut P.,  1985, ApJ, 298, 80

\bibitem[\protect\citeauthoryear{Chabrier}{Chabrier}{2003}]{Chabrier03}
Chabrier G.,  2003, PASP, 115, 763

\bibitem[\protect\citeauthoryear{Chabrier}{Chabrier}{2005}]{Chabrier05}
Chabrier G.,  2005 Vol.~327 of {Astrophysics and Space Science Library}, The
  {I}nitial {M}ass {F}unction: from {S}alpeter 1955 to 2005.
p.~41

\bibitem[\protect\citeauthoryear{{Cottaar}, {Meyer}, {Andersen} \&
  {Espinoza}}{{Cottaar} et~al.}{2012}]{Cottaar12}
{Cottaar} M.,  {Meyer} M.~R.,  {Andersen} M.,    {Espinoza} P.,  2012, A\&A,
  539, A5

\bibitem[\protect\citeauthoryear{{Dale}, {Ercolano} \& {Bonnell}}{{Dale}
  et~al.}{2012}]{Dale12a}
{Dale} J.~E.,  {Ercolano} B.,    {Bonnell} I.~A.,  2012, MNRAS, 424, 377

\bibitem[\protect\citeauthoryear{{Dale}, {Ercolano} \& {Bonnell}}{{Dale}
  et~al.}{2013}]{Dale13}
{Dale} J.~E.,  {Ercolano} B.,    {Bonnell} I.~A.,  2013, MNRAS, 430, 234

\bibitem[\protect\citeauthoryear{{Delgado}, {Djupvik}, {Costado} \&
  {Alfaro}}{{Delgado} et~al.}{2013}]{Delgado13}
{Delgado} A.~J.,  {Djupvik} A.~A.,  {Costado} M.~T.,    {Alfaro} E.~J.,  2013,
  MNRAS, 435, 429

\bibitem[\protect\citeauthoryear{{Duch{\^e}ne} \& {Kraus}}{{Duch{\^e}ne} \&
  {Kraus}}{2013}]{Duchene13}
{Duch{\^e}ne} G.,  {Kraus} A.,  2013, ARA\&A, 51, 269

\bibitem[\protect\citeauthoryear{Elmegreen \& Elmegreen}{Elmegreen \&
  Elmegreen}{2001}]{Elmegreen01}
Elmegreen B.~G.,  Elmegreen D.~M.,  2001, AJ, 121, 1507

\bibitem[\protect\citeauthoryear{{Fujii} \& {Portegies Zwart}}{{Fujii} \&
  {Portegies Zwart}}{2013}]{Fujii13}
{Fujii} M.~S.,  {Portegies Zwart} S.,  2013, ArXiv e-prints

\bibitem[\protect\citeauthoryear{{Geller}, {de Grijs}, {Li} \&
  {Hurley}}{{Geller} et~al.}{2013}]{Geller13b}
{Geller} A.~M.,  {de Grijs} R.,  {Li} C.,    {Hurley} J.~R.,  2013, ApJ, in
  press (arXiv: 1310.1085)

\bibitem[\protect\citeauthoryear{Gieles, Moeckel \& Clarke}{Gieles
  et~al.}{2012}]{Gieles12}
Gieles M.,  Moeckel N.,    Clarke C.~J.,  2012, MNRAS, 426, L11

\bibitem[\protect\citeauthoryear{Gieles \& {Portegies Zwart}}{Gieles \&
  {Portegies Zwart}}{2011}]{Gieles11}
Gieles M.,  {Portegies Zwart} S.~F.,  2011, MNRAS, 410, L6

\bibitem[\protect\citeauthoryear{{Gieles}, {Sana} \& {Portegies
  Zwart}}{{Gieles} et~al.}{2010}]{Gieles10}
{Gieles} M.,  {Sana} H.,    {Portegies Zwart} S.~F.,  2010, MNRAS, 402, 1750

\bibitem[\protect\citeauthoryear{{Girichidis}, {Federrath}, {Allison},
  {Banerjee} \& {Klessen}}{{Girichidis} et~al.}{2012}]{Girichidis12}
{Girichidis} P.,  {Federrath} C.,  {Allison} R.,  {Banerjee} R.,    {Klessen}
  R.~S.,  2012, MNRAS, 420, 3264

\bibitem[\protect\citeauthoryear{Goodwin}{Goodwin}{1997}]{Goodwin97}
Goodwin S.~P.,  1997, MNRAS, 286, 669

\bibitem[\protect\citeauthoryear{{Goodwin}}{{Goodwin}}{2009}]{Goodwin09}
{Goodwin} S.~P.,  2009, Ap\&SS, 324, 259

\bibitem[\protect\citeauthoryear{Goodwin \& Bastian}{Goodwin \&
  Bastian}{2006}]{Goodwin06}
Goodwin S.~P.,  Bastian N.,  2006, MNRAS, 373, 752

\bibitem[\protect\citeauthoryear{Goodwin \& Whitworth}{Goodwin \&
  Whitworth}{2004}]{Goodwin04a}
Goodwin S.~P.,  Whitworth A.~P.,  2004, A\&A, 413, 929

\bibitem[\protect\citeauthoryear{Gouliermis, Keller, Kontizas, Kontizas \&
  {Bellas-Velidis}}{Gouliermis et~al.}{2004}]{Gouliermis04}
Gouliermis D.,  Keller S.~C.,  Kontizas M.,  Kontizas E.,    {Bellas-Velidis}
  I.,  2004, A\&A, 416, 137

\bibitem[\protect\citeauthoryear{{Gouliermis}, {de Grijs} \&
  {Xin}}{{Gouliermis} et~al.}{2009}]{Gouliermis09}
{Gouliermis} D.~A.,  {de Grijs} R.,    {Xin} Y.,  2009, ApJ, 692, 1678

\bibitem[\protect\citeauthoryear{{H{\'e}nault-Brunet}, {Evans}, {Sana},
  {Gieles}, {Bastian}, {Ma{\'{\i}}z Apell{\'a}niz}, {Markova}, {Taylor},
  {Bressert}, {Crowther} \& {van Loon}}{{H{\'e}nault-Brunet}
  et~al.}{2012}]{Henault12}
{H{\'e}nault-Brunet} V.,  {Evans} C.~J.,  {Sana} H.,  {Gieles} M.,  {Bastian}
  N.,  {Ma{\'{\i}}z Apell{\'a}niz} J.,  {Markova} N.,  {Taylor} W.~D.,
  {Bressert} E.,  {Crowther} P.~A.,    {van Loon} J.~T.,  2012, A\&A, 546, A73

\bibitem[\protect\citeauthoryear{Hillenbrand \& Hartmann}{Hillenbrand \&
  Hartmann}{1998}]{Hillenbrand98}
Hillenbrand L.~A.,  Hartmann L.~W.,  1998, ApJ, 492, 540

\bibitem[\protect\citeauthoryear{{Hubber}, {Allison}, {Smith} \&
  {Goodwin}}{{Hubber} et~al.}{2013}]{Hubber13}
{Hubber} D.~A.,  {Allison} R.~J.,  {Smith} R.,    {Goodwin} S.~P.,  2013,
  MNRAS, 430, 1599

\bibitem[\protect\citeauthoryear{{Kiminki}, {Kobulnicky}, {Kinemuchi}, {Irwin},
  {Fryer}, {Berrington}, {Uzpen}, {Monson}, {Pierce} \& {Woosley}}{{Kiminki}
  et~al.}{2007}]{Kiminki07}
{Kiminki} D.~C.,  {Kobulnicky} H.~A.,  {Kinemuchi} K.,  {Irwin} J.~S.,  {Fryer}
  C.~L.,  {Berrington} R.~C.,  {Uzpen} B.,  {Monson} A.~J.,  {Pierce} M.~J.,
  {Woosley} S.~E.,  2007, ApJ, 664, 1102

\bibitem[\protect\citeauthoryear{King, Parker, Patience \& Goodwin}{King
  et~al.}{2012a}]{King12a}
King R.~R.,  Parker R.~J.,  Patience J.,    Goodwin S.~P.,  2012a, MNRAS, 421,
  2025

\bibitem[\protect\citeauthoryear{King, Goodwin, Parker \& Patience}{King
  et~al.}{2012b}]{King12b}
King R.~R.,  Goodwin S.~P.,  Parker R.~J.,    Patience J.,  2012b, MNRAS, 427,
  2636

\bibitem[\protect\citeauthoryear{Kirk \& Myers}{Kirk \& Myers}{2011}]{Kirk10}
Kirk H.,  Myers P.~C.,  2011, ApJ, 727, 64

\bibitem[\protect\citeauthoryear{{Korchagin}, {Girard}, {Borkova}, {Dinescu} \&
  {van Altena}}{{Korchagin} et~al.}{2003}]{Korchagin03}
{Korchagin} V.~I.,  {Girard} T.~M.,  {Borkova} T.~V.,  {Dinescu} D.~I.,    {van
  Altena} W.~F.,  2003, AJ, 126, 2896

\bibitem[\protect\citeauthoryear{Kruijssen}{Kruijssen}{2012}]{Kruijssen12b}
Kruijssen J. M.~D.,  2012, MNRAS, 426, 3008

\bibitem[\protect\citeauthoryear{Kruijssen, Maschberger, Moeckel, Clarke,
  Bastian \& Bonnell}{Kruijssen et~al.}{2012}]{Kruijssen12a}
Kruijssen J. M.~D.,  Maschberger T.,  Moeckel N.,  Clarke C.~J.,  Bastian N.,
   Bonnell I.~A.,  2012, MNRAS, 419, 841

\bibitem[\protect\citeauthoryear{Lada \& Lada}{Lada \& Lada}{2003}]{Lada03}
Lada C.~J.,  Lada E.~A.,  2003, ARA\&A, 41, 57

\bibitem[\protect\citeauthoryear{Lada, Margulis \& Dearborn}{Lada
  et~al.}{1984}]{Lada84}
Lada C.~J.,  Margulis M.,    Dearborn D.,  1984, ApJ, 285, 141

\bibitem[\protect\citeauthoryear{Larson}{Larson}{1981}]{Larson81}
Larson R.~B.,  1981, MNRAS, 194, 809

\bibitem[\protect\citeauthoryear{Maschberger}{Maschberger}{2013}]{Maschberger1%
3}
Maschberger T.,  2013, MNRAS, 429, 1725

\bibitem[\protect\citeauthoryear{Maschberger \& Clarke}{Maschberger \&
  Clarke}{2011}]{Maschberger11}
Maschberger T.,  Clarke C.~J.,  2011, MNRAS, 416, 541

\bibitem[\protect\citeauthoryear{Metchev \& Hillenbrand}{Metchev \&
  Hillenbrand}{2009}]{Metchev09}
Metchev S.~A.,  Hillenbrand L.~A.,  2009, ApJS, 181, 62

\bibitem[\protect\citeauthoryear{{Moeckel} \& {Bonnell}}{{Moeckel} \&
  {Bonnell}}{2009a}]{Moeckel09a}
{Moeckel} N.,  {Bonnell} I.~A.,  2009a, MNRAS, 396, 1864

\bibitem[\protect\citeauthoryear{{Moeckel} \& {Bonnell}}{{Moeckel} \&
  {Bonnell}}{2009b}]{Moeckel09b}
{Moeckel} N.,  {Bonnell} I.~A.,  2009b, MNRAS, 400, 657

\bibitem[\protect\citeauthoryear{{Moeckel} \& {Clarke}}{{Moeckel} \&
  {Clarke}}{2011}]{Moeckel11a}
{Moeckel} N.,  {Clarke} C.~J.,  2011, MNRAS, 410, 2799

\bibitem[\protect\citeauthoryear{Moeckel, Holland, Clarke \& Bonnell}{Moeckel
  et~al.}{2012}]{Moeckel12}
Moeckel N.,  Holland C.,  Clarke C.~J.,    Bonnell I.~A.,  2012, MNRAS, 425,
  450

\bibitem[\protect\citeauthoryear{{Moyano Loyola} \& {Hurley}}{{Moyano Loyola}
  \& {Hurley}}{2013}]{Moyano13}
{Moyano Loyola} G.~R.~I.,  {Hurley} J.~R.,  2013, MNRAS, 434, 2509

\bibitem[\protect\citeauthoryear{Offner, Hansen \& Krumholz}{Offner
  et~al.}{2009}]{Offner09}
Offner S. S.~R.,  Hansen C.~E.,    Krumholz M.~R.,  2009, ApJ, 704, L124

\bibitem[\protect\citeauthoryear{Olczak, Spurzem \& Henning}{Olczak
  et~al.}{2011}]{Olczak11}
Olczak C.,  Spurzem R.,    Henning T.,  2011, A\&A, 532, 119

\bibitem[\protect\citeauthoryear{{Pang}, {Grebel}, {Allison}, {Goodwin},
  {Altmann}, {Harbeck}, {Moffat} \& {Drissen}}{{Pang} et~al.}{2013}]{Pang13}
{Pang} X.,  {Grebel} E.~K.,  {Allison} R.~J.,  {Goodwin} S.~P.,  {Altmann} M.,
  {Harbeck} D.,  {Moffat} A.~F.~J.,    {Drissen} L.,  2013, ApJ, 764, 73

\bibitem[\protect\citeauthoryear{Parker, Bouvier, Goodwin, Moraux, Allison,
  Guieu \& G{\"u}del}{Parker et~al.}{2011}]{Parker11b}
Parker R.~J.,  Bouvier J.,  Goodwin S.~P.,  Moraux E.,  Allison R.~J.,  Guieu
  S.,    G{\"u}del M.,  2011, MNRAS, 412, 2489

\bibitem[\protect\citeauthoryear{Parker \& Dale}{Parker \&
  Dale}{2013}]{Parker13a}
Parker R.~J.,  Dale J.~E.,  2013, MNRAS, 432, 986

\bibitem[\protect\citeauthoryear{Parker \& Goodwin}{Parker \&
  Goodwin}{2012}]{Parker12b}
Parker R.~J.,  Goodwin S.~P.,  2012, MNRAS, 424, 272

\bibitem[\protect\citeauthoryear{Parker, Maschberger \& {Alves de
  Oliveira}}{Parker et~al.}{2012}]{Parker12c}
Parker R.~J.,  Maschberger T.,    {Alves de Oliveira} C.,  2012, MNRAS, 426,
  3079

\bibitem[\protect\citeauthoryear{Parker \& Meyer}{Parker \&
  Meyer}{2012}]{Parker12d}
Parker R.~J.,  Meyer M.~R.,  2012, MNRAS, 427, 637

\bibitem[\protect\citeauthoryear{{Parmentier} \& {Pfalzner}}{{Parmentier} \&
  {Pfalzner}}{2013}]{Parmentier13}
{Parmentier} G.,  {Pfalzner} S.,  2013, A\&A, 549, A132

\bibitem[\protect\citeauthoryear{{Pelupessy} \& {Portegies Zwart}}{{Pelupessy}
  \& {Portegies Zwart}}{2012}]{Pelupessy12}
{Pelupessy} F.~I.,  {Portegies Zwart} S.,  2012, MNRAS, 420, 1503

\bibitem[\protect\citeauthoryear{{Portegies Zwart}, McMillan, Hut \&
  Makino}{{Portegies Zwart} et~al.}{2001}]{Zwart01}
{Portegies Zwart} S.~F.,  McMillan S. L.~W.,  Hut P.,    Makino J.,  2001,
  MNRAS, 321, 199

\bibitem[\protect\citeauthoryear{{Portegies Zwart}, Makino, McMillan \&
  Hut}{{Portegies Zwart} et~al.}{1999}]{Zwart99}
{Portegies Zwart} S.~F.,  Makino J.,  McMillan S. L.~W.,    Hut P.,  1999,
  A\&A, 348, 117

\bibitem[\protect\citeauthoryear{Prim}{Prim}{1957}]{Prim57}
Prim R.~C.,  1957, Bell Syst. Tech. J., 36, 1389

\bibitem[\protect\citeauthoryear{{Randich}}{{Randich}}{2012}]{Randich12}
{Randich} S.,  2012, in Chemical Evolution of the Milky Way {The Gaia-ESO
  Survey}

\bibitem[\protect\citeauthoryear{{Reggiani} \& {Meyer}}{{Reggiani} \&
  {Meyer}}{2011}]{Reggiani11a}
{Reggiani} M.~M.,  {Meyer} M.~R.,  2011, ApJ, 738, 60

\bibitem[\protect\citeauthoryear{{Reggiani} \& {Meyer}}{{Reggiani} \&
  {Meyer}}{2013}]{Reggiani13}
{Reggiani} M.~M.,  {Meyer} M.~R.,  2013, A\&A, 553, A124

\bibitem[\protect\citeauthoryear{{Rochau}, {Brandner}, {Stolte}, {Gennaro},
  {Gouliermis}, {Da Rio}, {Dzyurkevich} \& {Henning}}{{Rochau}
  et~al.}{2010}]{Rochau10}
{Rochau} B.,  {Brandner} W.,  {Stolte} A.,  {Gennaro} M.,  {Gouliermis} D.,
  {Da Rio} N.,  {Dzyurkevich} N.,    {Henning} T.,  2010, ApJL, 716, L90

\bibitem[\protect\citeauthoryear{Sabbi, Sirianni, Nota, Tosi, Gallagher, Smith,
  Angeretti, Meixner, Oey, Walterbos \& Pasquali}{Sabbi et~al.}{2008}]{Sabbi08}
Sabbi E.,  Sirianni M.,  Nota A.,  Tosi M.,  Gallagher J.,  Smith L.~J.,
  Angeretti L.,  Meixner M.,  Oey M.~S.,  Walterbos R.,    Pasquali A.,  2008,
  AJ, 135, 173

\bibitem[\protect\citeauthoryear{Salpeter}{Salpeter}{1955}]{Salpeter55}
Salpeter E.~E.,  1955, ApJ, 121, 161

\bibitem[\protect\citeauthoryear{Sana, Momany, Gieles, Carraro, Beletsky,
  Ivanov, {De Silva} \& James}{Sana et~al.}{2010}]{Sana10}
Sana H.,  Momany Y.,  Gieles M.,  Carraro G.,  Beletsky Y.,  Ivanov V.~D.,  {De
  Silva} G.,    James G.,  2010, A\&A, 515, A26

\bibitem[\protect\citeauthoryear{S{\'a}nchez \& Alfaro}{S{\'a}nchez \&
  Alfaro}{2009}]{Sanchez09}
S{\'a}nchez N.,  Alfaro E.~J.,  2009, ApJ, 696, 2086

\bibitem[\protect\citeauthoryear{Scally \& Clarke}{Scally \&
  Clarke}{2002}]{Scally02}
Scally A.,  Clarke C.,  2002, MNRAS, 334, 156

\bibitem[\protect\citeauthoryear{Schmeja}{Schmeja}{2011}]{Schmeja11}
Schmeja S.,  2011, AN, 332, 172

\bibitem[\protect\citeauthoryear{{Schmeja} \& {Klessen}}{{Schmeja} \&
  {Klessen}}{2006}]{Schmeja06}
{Schmeja} S.,  {Klessen} R.~S.,  2006, A\&A, 449, 151

\bibitem[\protect\citeauthoryear{Smith, Fellhauer, Goodwin \& Assmann}{Smith
  et~al.}{2011}]{Smith11}
Smith R.,  Fellhauer M.,  Goodwin S.,    Assmann P.,  2011, MNRAS, 414, 3036

\bibitem[\protect\citeauthoryear{{Spitzer} Jr.}{{Spitzer}}{1969}]{Spitzer69}
{Spitzer} Jr. L.,  1969, ApJL, 158, L139

\bibitem[\protect\citeauthoryear{{Steenbrugge}, {de Bruijne}, {Hoogerwerf} \&
  {de Zeeuw}}{{Steenbrugge} et~al.}{2003}]{Steenbrugge03}
{Steenbrugge} K.~C.,  {de Bruijne} J.~H.~J.,  {Hoogerwerf} R.,    {de Zeeuw}
  P.~T.,  2003, A\&A, 402, 587

\bibitem[\protect\citeauthoryear{Tutukov}{Tutukov}{1978}]{Tutukov78}
Tutukov A.~V.,  1978, A\&A, 70, 57

\bibitem[\protect\citeauthoryear{{Whitworth}}{{Whitworth}}{1979}]{Whitworth79}
{Whitworth} A.,  1979, MNRAS, 186, 59

\bibitem[\protect\citeauthoryear{{Wright}, {Drake}, {Drew} \& {Vink}}{{Wright}
  et~al.}{2010}]{Wright10}
{Wright} N.~J.,  {Drake} J.~J.,  {Drew} J.~E.,    {Vink} J.~S.,  2010, ApJ,
  713, 871

\bibitem[\protect\citeauthoryear{Wright, Parker, Goodwin \& Drake}{Wright
  et~al.}{2013}]{Wright13}
Wright N.~J.,  Parker R.~J.,  Goodwin S.~P.,    Drake J.~J.,  2013, MNRAS,
  submitted, pp~--

\bibitem[\protect\citeauthoryear{{Yu}, {de Grijs} \& {Chen}}{{Yu}
  et~al.}{2011}]{Yu11}
{Yu} J.,  {de Grijs} R.,    {Chen} L.,  2011, ApJ, 732, 16

\bibitem[\protect\citeauthoryear{{Zasowski}, {Johnson}, {Frinchaboy},
  {Majewski}, {Nidever}, {Rocha Pinto}, {Girardi}, {Andrews}, {Chojnowski},
  {Cudworth}, {Jackson}, {Munn}, {Skrutskie}, {Beaton}, {Blake}, {Covey} \& {et
  al.}}{{Zasowski} et~al.}{2013}]{Zasowski13}
{Zasowski} G.,  {Johnson} J.~A.,  {Frinchaboy} P.~M.,  {Majewski} S.~R.,
  {Nidever} D.~L.,  {Rocha Pinto} H.~J.,  {Girardi} L.,  {Andrews} B.,
  {Chojnowski} S.~D.,  {Cudworth} K.~M.,  {Jackson} K.,  {Munn} J.,
  {Skrutskie} M.~F.,  {Beaton} R.~L.,  {Blake} C.~H.,  {Covey} K.,    {et al.}
  2013, AJ, 146, 81

\bibitem[\protect\citeauthoryear{{Zinnecker}}{{Zinnecker}}{1982}]{Zinnecker82}
{Zinnecker} H.,  1982, Annals of the New York Academy of Sciences, 395, 226

\end{thebibliography}

\label{lastpage}

\end{document}